\documentclass{article}

\usepackage{arxiv}

\usepackage[utf8]{inputenc} 
\usepackage[T1]{fontenc}    
\usepackage{hyperref}       
\usepackage{url}            
\usepackage{booktabs}       
\usepackage{amsfonts}       
\usepackage{nicefrac}       
\usepackage{microtype}      
\usepackage{lipsum}		
\usepackage{graphicx}
\usepackage{natbib}
\usepackage{doi}

\usepackage{enumitem}
\usepackage{tikz}
\usetikzlibrary{calc,arrows.meta,positioning}

\usetikzlibrary{shapes,decorations,arrows,calc,arrows.meta,fit,positioning}
\tikzset{
    -Latex,auto,node distance =1 cm and 1 cm,semithick,
    state/.style ={circle, draw, minimum width = 0.5 cm},
    point/.style = {circle, draw, inner sep=0.04cm,fill,node contents={}},
    bidirected/.style={Latex-Latex,dashed},
    el/.style = {inner sep=2pt, align=left, sloped}
}

\makeatletter
\newcommand{\xleftrightarrow}[2][]{\ext@arrow 3359\leftrightarrowfill@{#1}{#2}}
\makeatother

\newcommand{\xdasharrow}[2][->]{
\tikz[baseline=-\the\dimexpr\fontdimen22\textfont2\relax]{
\node[anchor=south,font=\scriptsize, inner ysep=1.5pt,outer xsep=2.2pt](x){#2};
\draw[shorten <=3.4pt,shorten >=3.4pt,dashed,#1](x.south west)--(x.south east);
}
}

\usepackage{longtable}
\usepackage{soul}

\usepackage{amsmath}
\usepackage{amsthm}
\usepackage{amsfonts}
\usepackage{verbatim}

\usepackage{times}
\usepackage{latexsym}
\usepackage{graphicx}
\usepackage{color}
\usepackage{dsfont}

\usepackage{multirow}
\usepackage{float}

\usepackage{enumitem}

\usepackage{booktabs} 

\usepackage{subfig}

\newtheorem{thm}{Theorem}[section]

\newtheorem{prop}[thm]{Proposition}

\newtheorem{defn}[thm]{Definition}

\newcommand{\ind}{\perp\!\!\!\!\perp} 

\newcommand{\field}[1]{\mathbb{#1}}
\DeclareMathOperator{\PR}{\field{P}}     
\DeclareMathOperator{\E}{\field{E}}      

\newcommand{\ls}[1]
  {\dimen0=\fontdimen6\the\font \lineskip=#1\dimen0
  \advance\lineskip.5\fontdimen5\the\font \advance\lineskip-\dimen0
  \lineskiplimit=.9\lineskip \baselineskip=\lineskip
  \advance\baselineskip\dimen0 \normallineskip\lineskip
  \normallineskiplimit\lineskiplimit \normalbaselineskip\baselineskip
  \ignorespaces }

\newcommand{\blot}{\hfill{\vrule height .9ex width .8ex depth -.1ex }}
\newcommand{\EndPf}{\hfill $\blot$ \medskip}     

\title{Average Response Curves for Treatment Time in the Emergency Department}

\author{ {Sebastian A. Alvarez Avendaño}\\
	Department of Industrial and Systems Engineering \\
	University of Wisconsin-Madison\\
	Madison, WI \\
	\texttt{alvarezavend@wisc.edu} \\
	\And
	{Amy L. Cochran} \\
	Department of Population Health Sciences \\and Department of Mathematics\\
	University of Wisconsin-Madison\\
	Madison, WI \\
	\texttt{cochran4@wisc.edu} \\
	\And
	{Keith E. Kocher} \\
	Department of Emergency Medicine\\
	University of Michigan\\
	Ann Arbor, MI\\
	\texttt{kkocher@med.umich.edu} \\
	\And
	{Brian W. Patterson} \\
	BerbeeWalsh Department of Emergency Medicine\\
	University of Wisconsin-Madison\\
	Madison, WI \\
	\texttt{bpatter@medicine.wisc.edu} \\
	\And
	{Gabriel Zayas-Cab\'{a}n}\\ 
	Department of Industrial and Systems Engineering\\
	University of Wisconsin-Madison\\
	Madison, WI \\
	\texttt{zayascaban@wisc.edu} \\
}

\renewcommand{\headeright}{Preprint}
\renewcommand{\undertitle}{Preprint}
\renewcommand{\shorttitle}{Average Response Curves for Treatment Time}

\hypersetup{
pdftitle={Average Response Curves for Treatment Time},
pdfauthor={Sebastian A. Alvarez Avendaño, Amy L. Cochran, Keith E. Kocher, Brian W. Patterson, Gabriel Zayas-Cab\'{a}n},
pdfkeywords={Causal Inference, Continuous Treatment, Unmeasured Confounding, Latent Variable, Emergency Department, Admission Decision, Treatment Time},
}

\begin{document}
\maketitle

\begin{abstract}
	We estimate average responses curves for treatment time in the Emergency Department (ED). Extending treatment time is considered a promising solution for improving admission decisions. Providing empirical support for this solution, however, is difficult because this intervention (treatment) is a continuous time-to-event; is strongly influenced by unmeasured patient health needs and is jointly determined with the admission decision (admit vs. discharge); and may be only modifiable up to a shift in the realized time. We formalize the admission process as a directed acyclic graph and show that average responses curves for treatment time cannot be identified nonparametrically due to unmeasured confounding from patient health needs. We thus use a parametric model that includes a latent variable for health needs and a threshold regression model for the admission process. We fit this model to observational data (n = 28,862) from abdominal pain patients at a large tertiary teaching hospital. We estimate that fixing ED treatment time to 2 hours rather than 1 hour decreases admission rates from 41.6\% (95\% CI: [40.6, 42.7]) to 32.7\% (95\% CI: [32.2, 33.2]) and that increasing the realized treatment time by 30 minutes can reduce admission rates by 1.1\% (95\% CI: [-1.1, 3.2]), with little change to 30-day revisit and readmission rates.
\end{abstract}

\keywords{Causal Inference; Continuous Treatment; Unmeasured Confounding; Latent Variable; Emergency Department; Admission Decision; Treatment Time}

\section{Introduction}

Emergency department (ED) admissions account for two thirds of national health expenditures related to care in the ED, making the decision to admit to inpatient hospital units one of the most critical, routine, and expensive decisions made in health care \citep{sabbatini2014reducing,galarraga2016costs}. While largely informed by clinical factors, the admission decision (admit vs. discharge)  does not completely rest on black and white clinical factors. Rather, it often rests on a provider's judgment regarding the benefits of hospital care versus the risks of discharge. Our goal was to provide empirical support for the role that treatment time plays in the admission decision and its consequences.

Abdominal pain patients provide a good example of patients for whom treatment time is a factor in the admission decision. Patients presenting to the ED with abdominal pain are common and among the most difficult patients to decide whether to admit or discharge \citep{sabbatini2014reducing}. A wrong decision can have life or death consequences for these patients. Yet only a minority of individuals experience such serious diagnoses, creating some uncertainty about how best to evaluate and manage this presentation in the ED.  As a result, these patients are subject to large variability in admission rates within hospitals and nationally. This variability can lead to harm, from not only incorrectly discharging a person home but also from an unnecessary admission. 

A big part of ED care for abdominal pain patients is monitoring and examining patients through laboratory work and imaging (e.g. ultrasound CT scans, MRI). By extending treatment time, an ED provider can gather more information for their admission decision by ordering more tests or monitoring patients longer. Indeed, patients are being increasingly sent to medical short-stay or observation units, to allow for extended evaluation \citep{sun2014randomized,zuckerman2016readmissions,centersmedicare}.  However, whether extending treatment time alters admissions decisions or improves outcomes is an open-ended and hotly-debated topic \citep{noel2015observation,zuckerman2016readmissions}.

Providing evidence to support extending treatment time, however, requires overcoming several methodological challenges. The first is the presence of unmeasured confounding. The admission process is influenced by the health needs of the patient in terms of how challenging their clinical, social, or residential situations may be. That is, patients tend to receive treatment and an admission decision according to indications that are available to a physician but not always recorded in the data. In clinical epidemiology, this type of confounding is referred to as \emph{confounding by indication}.  Without randomizing treatment time, patient health needs confound the relationship between treatment time and outcomes. This could lead to spurious conclusions about whether extending treatment time could improve outcomes.


A second challenge is that treatment time is a continuous intervention. One way to handle continuous outcomes are generalized propensity score (GPS) methods developed by \citet{hirano2004propensity} and \citet{imai2004causal}. GPS methods recover average responses curves of a continuous intervention. These curves afford a causal interpretation within the potential outcomes framework under the usual assumptions of consistency, positivity, exchangeability, and stable unit treatment value (SUTVA). Numerous alternatives or extensions to GPS methods ensued such as methods with different distributional assumptions \citep{guardabascioventura}, longitudinal methods \citep{moodiestephens}, doubly-robust methods \citep{turobust}), non-parametric methods \citep{diasvanderlaan,kennedynonparametric,NIPS2016_6107,hillbart}, and Bayesian methods \citep{PapadogeorgouDominici2020}. GPS methods, like their predecessor: propensity score methods, can only adjust for measured, not unmeasured, confounding, making them less suited for evaluating ED treatment time. 

A third challenge is that treatment time and the admission decision are jointly determined. If we could ignore admission decision, treatment time could be modelled using standard time-to-event or survival analysis regression models (e.g., Cox proportional hazards model, log-normal models) \citep{cox1972regression}. The \emph{event} in these time-to-event models has the same outcome, usually an adverse outcome such as remission or death.  By contrast, treatment time is associated with two outcomes: an admission or a discharge. Which outcome ultimately occurs is tied to treatment time. For example, a short treatment time might reflect a person who needs to be admitted urgently. Competing risk models, which describe a time-to-event and multiple events arising from competing processes (e.g. stroke vs. cancer) \citep{austin2016introduction}, may also be a poor match for the admission process, because of the aforementioned unmeasured confounding and because there is only one event: the admission decision. A more accurate reflection of the admission process is a model that jointly describes the admission decision and treatment time. 

A fourth and final challenge is that average responses curves for a \emph{fixed} treatment time may have limited practical value since it would be difficult to implement an intervention whereby patients are kept in treatment a fixed amount of time. There is too much variability in their care to impose such a restriction. A promising alternative is to estimate average responses curves for a \emph{shift} in the realized treatment time. This translates into to an intervention that encourages a physician to spend additional or less time with each patient. These interventions are called shift interventions, as defined in \citet{sani2020identification}.

In this paper, we pursue a model-based approach for measuring average response curves for ED treatment time. We first formalize assumed causal relationships in a directed acyclic graph (DAG). We provide conditions under which average responses can and cannot be identified using the potential outcomes framework presented in \cite{malinsky2019potential}. We then describe a latent variable model that includes a latent variable to capture \emph{unmeasured} health needs and a joint model of treatment time and the admission decision \citep{cochran2019latent}. This joint model uses a threshold regression model of hitting times of a drift-diffusion process between two boundaries. With a parametric model specified, we then estimate the average causal response of \emph{fixed} treatment time on various outcomes using observational data (n = 28,862) from abdominal pain patients at a large tertiary teaching hospital. We also estimate average responses of shifting the realized treatment time a fixed amount. We then systematically alter assumptions in our approach to examine sensitivity of estimates to alternative approaches. These alternatives include violating a key identifiability assumption, replacing the model of the admission decision process, removing the influence of the latent variable, and obtaining other non-latent variable approaches using the classic generalized propensity score (GPS) \citep{imai2004causal,austin2018assessing}.

\textbf{Organization of the paper.}  The rest of this paper is organized as follows. Section~\ref{sec:causal} defines a causal framework for evaluating our intervening variable: treatment time. Section~\ref{sec:identification} presents a parametric model for estimating average potential responses. Section~\ref{sec:results} presents results from applying these methods to electronic health records (EHR) from a large tertiary hospital. Concluding remarks are given in Section~\ref{sec:conclusion}.

\section{Causal inference framework} \label{sec:causal}

\subsection{Notation and definitions}

Consider the following random variables:
\begin{itemize}[topsep=2pt]
\setlength\itemsep{0pt}
\item $X$ --- patient characteristics known prior to the start of treatment such as age and sex. 
\item $H$ --- binary latent health state, representing patient's unknown needs for hospital resources.  
\item $Z$ --- (noisy) observations of health needs taken prior to the start of treatment.
\item $T$ --- realized time from when treatment begins to admission decision is made, i.e. treatment time.  
\item $A$ --- binary admission decision of admit or discharge.
\item $Y$ --- outcome of interest (e.g., ED revisits, subsequent admission/readmission, death). 
\end{itemize}

\subsection{Directed acyclic graph (DAG)}\label{sec:dag}
We assume that each ED visit yields an iid observation of~$(X, H, Z, A, T, Y).$  
These variables capture the following decision-making scenario. Each patient has an underlying health needs $H$ just before the treatment starts, reflecting the information prior to treatment that indicates whether the patient requires hospitalization or not. In most cases, $H$ is unobserved and a confounder between the transfer process $(A,T)$ and $Y$. The patient undergoes treatment/testing that yield (noisy) observations $Z$ of the health needs $H$. In other words, $Z$, which is measured, serves as proxy for $H$, which is not measured. Observations $Z$ might be acuity level or other information, such as vitals, obtained at baseline. We remark that the latent health state $H$ is binary to simplify subsequent estimation and interpretation of results and to reflect that the latent health state $H$ is largely an indicator of whether or not a person should be admitted. One could also consider an model in which $H$ takes on a finite number of ordered values.

\begin{figure}[h!]
\centering
\begin{tikzpicture}
    \node (xi) at (-4,0.75) [label=above:$X$,point];
    \node[state] (hi) at (-2,-2) {$H$};
    \node (zi) at (-2,0) [label=above:$Z$,point];
    \node (ti) at (0,0) [label=above:$T$,point];
    \node (ai) at (0,-2) [label=below:$A$,point];    
    \node (yi) at (2,-2) [label=above:$Y$,point];
    \path (hi) edge (zi);
    \path (hi) edge (ti);
    \path (zi) edge (ti);
    \path (zi) edge (ai);
    \path (ti) edge (ai);
    \path (hi) edge (ai);
    \path (ai) edge (yi);
    \path (hi) edge[out= 315, in= 225] (yi);
    \path (xi) edge ($ (xi) !.32! (ti) $);
    \path (xi) edge ($ (xi) !.3! (hi) $);
    \path (xi) edge ($ (xi) !.25! (ai) $);
    \path (xi) edge ($ (xi) !.2! (yi) $);
\end{tikzpicture}
\vspace{-1em}   
\caption{DAG for admission decision process.} \label{fig:dag1}
\vspace{-1em}
\end{figure}
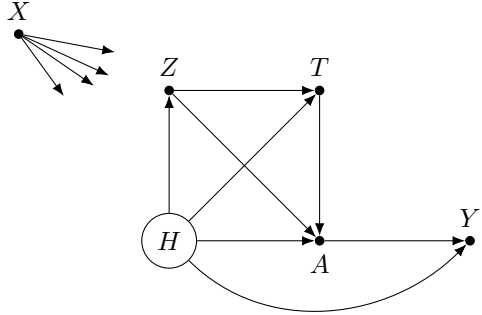

The care provider treats and evaluates the patient until a random amount of time $T$, at which point they make decision $A$ denoting the decision to admit or not. Once the physician decides to admit or discharge a patient, a patient no longer undergoes treatment/testing and waits to be transferred. Outcomes $Y$ after the transfer decisions are then measured. 
Baseline characteristics $X$ (e.g. age or sex) can influence any of the variables. Observations $Z$ and treatment time $T$, however, are assumed to be related to downstream outcomes, only insofar as they inform both the admission decision $A$ and the underlying latent health state $H$. Our model of the admission process yields the DAG depicted in Fig~\ref{fig:dag1}. Observed variables are expected to only partly account for the impact of patient health needs on the transfer decision process and outcomes. Latent health needs $H$ will be needed to control for confounding, which we formalize below.

\subsection{Average potential responses}

We define average potential responses using the potential outcomes framework presented in \cite{malinsky2019potential}, which generalizes Pearl's do-calculus framework \citep{pearl2009causality} and is compatible with the Neyman-Rubin potential outcome framework \citep{rubin1974estimating}. A causal model $\mathcal{M}$ can be defined based on the DAG depicted in Figure~\ref{fig:dag1}. This causal model consists of a finite set of (not necessarily observed) exogenous variables $\{H\}$; a set of endogenous observed variables $\{X,Z,A,T,Y\}$; a set of arrows, given by the DAG, with each arrow mapping a variable to an endogenous variable; and a probability measure on the variables, which is induced by the random vector $(X,H,Z,A,T,Y)$. 

We are interested average potential responses of intervening on the treatment time $T$ on both the admission decision $A$ and patient outcomes $Y$. This intervention involves modifying the treatment time $T$ to artificially attain the value $t$. The potential response of the intervention is denoted in the Neyman-Rubin framework (\cite{rubin1974estimating}) with $A(t)$ and $Y(t)$. A more formal definition is provided in Appendix \ref{appendix:proof}.  We are also interested in average potential responses of shifting the admission decision by a given length of time. This intervention involves specifying a policy function $f$ on our probability space, artificially shifting ED treatment time $T$ to $f(T)$ time units, and evaluating the potential response of this shift on admissions, i.e., $A(f(T))$, and outcomes, i.e., $Y(f(T))$. A formal definition is provided in Appendix \ref{appendix:proof_shift}.  Our goal is to recover the following:

\begin{defn}[Treatment time on admission decision]
To evaluate the causal impact of treatment time on the admission decision, we introduce 
 \begin{align*}
    \theta(t) := \E\left[ A(t) \right] = \PR\left( A(t)=1 \right)
\end{align*}
to reflect the average potential admission decision when the timing of the admission decision equals $t$. 
\end{defn}

\begin{defn}[Treatment time on outcomes]
To evaluate the causal impact of treatment time on patient outcomes, we introduce 
 \begin{align*}
    \gamma(t) := \E\left[ Y(t) \right] = \PR\left( Y(t)=1 \right)
\end{align*}
to reflect the average potential outcome when the timing of the admission decision equals $t$. 
\end{defn}

\begin{defn}[Shift intervention $f$ on admission decision]
To evaluate the causal impact of shifting treatment time to $f$ on the admission decision, we introduce 
 \begin{align*}
    \hat{\theta}(f) := \E\left[ A(f(T)) \right]
\end{align*}
to reflect the average potential admission decision when the timing of the admission decision equals $f(T)$. 
\end{defn}

\begin{defn}[Shift intervention $f$ on outcomes]
To evaluate the causal impact of shifting treating time to $f$ on patient outcomes, we introduce 
 \begin{align*}
    \hat{\gamma}(f) := \E\left[ Y(f(T)) \right]
\end{align*}
to reflect the average potential outcome when the timing of the admission decision equals $f(T)$. 
\end{defn}

\subsection{Identification}

There is a difference between $\E\left[ Y(t) \right]$ and $\E[Y|T=t]$. The former term is the average potential response on outcome $Y$ upon intervening on treatment time $T$ and setting it to $t$, while the latter is an average of the outcome $Y$ among those visits whose treatment time is $t$. So while $\E[Y|T=t]$ can be identified from observed variables, the average potential response $\E\left[ Y (t)\right]$ cannot generally be identified, i.e. expressed in terms of observed variables.  Similarly, the value of the shift interventions $A(f(T))$ and $Y(f(T))$ for a given function $f$ applied to $T$ might not be observed unless $f(T) = T$, and requires assumptions to identify.  We proceed to formulate necessary and sufficient conditions for identifying $\theta(t)$, $\gamma(t)$, $\hat{\theta}(f)$, and $\hat{\gamma}(f)$ from the observed variables.

\begin{prop} \label{thm:identifiability}
 Based on the DAG~\ref{fig:dag1}, average potential responses of treatment time on the admission decision and outcomes, i.e. $\theta(t)$ and $\gamma(t)$, can be nonparametrically identified if and only if health needs $H$ is observed. 
\end{prop}

\begin{prop}\label{thm:identifiability_shift}
 Based on the DAG \ref{fig:dag1}, average potential responses of shift intervention $f$ on the admission decision and outcomes, i.e. $\hat{\theta}(f)$ and $\hat{\gamma}(f)$, can be nonparametrically identified if and only if health needs $H$ is observed.
\end{prop}

We leave the proofs to Appendices~\ref{appendix:proof} and \ref{appendix:proof_shift}. Two key insights can be garnered from Propositions~\ref{thm:identifiability} and~\ref{thm:identifiability_shift}. First, if some unobserved health needs $H$ confounds the admission process, then it is generally not possible to identify the average potential responses of treatment time, and efforts to do so without adjusting for this latent variable are prone to bias. Second, since we do assume unobserved health needs $H$ confounds the admission process, then we will need to rely on \emph{parametric} assumptions to identify the causal effect. These assumptions are detailed in the next section.

\section{Parametric estimation of average potential responses}\label{sec:identification}


\subsection{Parametric model of the admission process}

We require a joint parametric model for the final decision $A$ and treatment time $T$ to parametrically identify average potential responses. Treatment time $T$ is a time-to-event that culminates in two possible outcomes reflected by the binary variable $A$. To capture such a time-to-event model, we rely on a threshold regression model introduced in \citep{cochran2019latent} to evaluate the impact of the admission decision and applied in \citep{cochran2019latent} to learn how the admission decision impacts chest-pain patients. Threshold regression is a popular framework for jointly describing a continuous time-to-event and an event's outcome~\citep{10.2307/27645791}. The idea is to construct a joint model of admission decision $A$ and treatment time $T$ (conditional on $X$, $H$, and $Z$) from first-passage locations and times of Brownian motion $B_t$. The Brownian motion $B_t$ of interest is a continuous-time stochastic process with stationary and independent increments $B_{t+s} - B_{t}$ which are normally-distributed with mean $d(H)b(X)s$ and variance $\sigma^2 s$. Treatment time $T$ is modelled as the first-passage time of $B_t$ out of an open interval $(0,b(X))$ starting at some point $B_0 := c(X,Z)b(X)$ :
\begin{align*}
    T   := \inf\{ t > 0 : B_t \notin \left(0,b(X)\right) \}.
\end{align*}
The admission decision $A$ captures which boundary $B_t$ exits through:
\begin{align*}
    A   := \begin{cases}
     0 & B_T = 0 \\
     1 & B_T = b(X). 
     \end{cases}
\end{align*}
This joint model relies on three parameter functions: the boundary $b(X)$, the relative starting point of the Brownian motion $c(X,Z) \in (0,1)$, and the drift rate $d(H)$.  Without loss of generality, we set $\sigma^2= 1$ with units 1/time, since we can scale $B_t$, $\sigma$, and $b(X)$ without changing the distribution of $(A,T)$ given $X,H,Z$.  These assumptions define a joint probability density function for $(A,T)=(a,t)$:
\begin{align*}
 \PR\left[ (T,A)=(a,t) | X, H, Z \right] := g\left(a,t|b(X),c(X,Z),d(H)\right).
\end{align*}
Here, the function $g(a,t|b,c,d)$ denotes the joint density of $(A,T)=(a,t)$ for a given initial point $bc$, drift rate $db$, and upper boundary $b$. Parameter $c$, which determines the initial value of $B_t$, is a function of patient characteristics $X$ and initial observation $Z$ to reflect that care providers have $Z$ and $X$ to initially evaluate the patient. Parameter $d$, which determines the drift rate of the $B_t$, is a function of health needs $H$ to reflect that health needs $H$ determines the information collected by care providers in the ED. Parameter $b$, which determines the boundary for $B_t$, is a function of patient characteristics $X$ to reflect that the level of information required to make a decision depends on patient characteristics. 

\subsection{Specification of parameters} \label{sec:parameter}

Parameter functions $b(X)$, $c(X,Z)$, and $d(H)$ have to be further specified. Since $c$ takes values in $(0,1)$ and $b$ is strictly positive, we assume $\mbox{logit}\,c(X,Z)$ and $\log b(X)$ are linear in their arguments and estimate the set of linear coefficients of these functions. Meanwhile, $\log |d|(H)$ is assumed to be linear in $H$, but restricted in such a way that the drift rate takes only a negative value for the lower health needs ($H=0$) and only positive value for the higher health needs ($H=1$). That way, patients tend to be discharged in the lower health state and admitted in the higher health state. Several other functions need to be specified. The function $\mbox{logit}\,\PR( H=1 | X )$ is linear in its argument, where we use a logistic function since the latent health state $H$ is binary. Similarly, we assume $\mbox{logit}\,\PR( Z_i=1 | X, H)$ is linear in its argument for any binary variable $Z_i$ in the vector of initial observations $Z$. Otherwise, $Z_i$ is assumed to be normal with unknown variance and a mean $\E[Z|X,H]$ that is linear in its arguments. All entries $Z_i$ in $Z$ are assumed to be mutually independent given $X$ and $H$. Lastly, we need a model for patient outcomes $Y$. We let
\begin{align*}
    \PR( Y=1 | X,Z,T,A=a,H) &:= \mu_1 + (\mu_2-\mu_1) H + (\mu_3 -\mu_1) a + (\mu_4-\mu_2-\mu_3+\mu_1) H a.
\end{align*} 
For simplicity, we omit $X$ in the above expression to reduce the number of parameters and to reflect that patient characteristics, acuity, and treatment time might not have a direct effect on potential outcomes when controlling for patient health needs. However, variables $X$ may be included if desired.

\subsection{Estimation of parameters} \label{sec:parameter_est}

We perform maximum likelihood estimation (MLE) using the expectation-maximization (EM) algorithm \citep{dempster1977maximum} to estimate the linear coefficients and unknown variances associated with the functions that specify a parametric model for $(X,H,Z,T,A,Y)$. Data is collected on $n$ visits, where each visit $i$ ($i =1,2,3, \ldots,n$) contributes a vector of patient characteristics $x_n \in \mathbb{R}^k$; initial observations of health needs $z_n \in \mathbb{R}^{p}$; admission decision $a_n \in \{0,1\}$; treatment time $t_n \in (0,\infty)$; and an outcome $y_n \in \{0,1\}$. Confidence intervals are estimated using a numerical approximation to Oakes Identity \citep{oakes1999direct}. This approximation is used in EM methods to estimate Fisher's information matrix, which when inverted, yields estimates of sampling variances for each parameter. These variances were then used to construct 95\% confidence intervals assuming parameter estimates are normally-distributed. Confidence intervals for functions of the estimates were estimated using the delta method. 

\subsection{Sensitivity analyses}

A fundamental limitation of causal inference is that estimates of average potential responses depend on several unverifiable assumptions. This limitation is addressed by presenting alternative estimates to check the sensitivity of estimates to various assumptions. To that end, we recovered alternative estimates for the average potential responses. Briefly, we first systematically violated one of the key assumptions embedded in the DAG depicted in Figure~\ref{fig:dag1}, which is that $Y(a)$ is independent from $A$ upon conditioning on $X$, $H$, and $Z$. Next, we switched the parametric model of the admission process $(A,T)$ from one based on Brownian motion to a log-normal model for $T|Z,X,H$ and logistic regression model for $A|T,H,X,Z$. Last, we examined the importance of adjusting for latent health needs $H$ by recovering estimates of average potential responses from three models that do not include $H$. The first is recovered by removing the influence of $H$ from our main model. The second is recovered by removing the influence of $H$ from the log-normal version of the model. The third is involves adapting a GPS approach to our causal diagram. These alternatives are detailed in Appendix~\ref{appendix:sensitivitydetails}.

\subsection{Data available for estimation}

ED visits were analyzed using EHR from the University of Wisconsin (UW) Health Innovation Program (HIP), which is curated from UW Health, a network of primary and specialty care clinics throughout south-central Wisconsin which share a common EHR instance.  The analyzed population consists of UW Hospital ED encounters between January 1, 2013, and September 27, 2018. Visits were included in the present analysis if the patient's chief complaint was abdominal pain. Visits were excluded sequentially for the following reasons: the patient arrived to the ED within the last 45 days of data collection period since their 45-day outcomes after their visit are not observed ($n=110$); was assigned acuity level or emergency severity index (ESI) \citep{gilboy2011emergency} of 1, 4, or 5 in order to focus on encounters with the highest degree of medical uncertainty with respect to needs for longer term acute care need in an inpatient hospital  ($n=703$); was diagnosed with trauma ($n=19$); was transferred to the ED from another hospital or hospice (i.e., they were triaged outside of the study ED) ($n=7$); did not have information regarding their diabetes status ($n=14$); or did not have a final admission decision to admit or discharge ($n=710$). Remaining records ($n=28,862$) were analyzed.

Patient characteristics $X$ available at baseline included age, sex, race/ethnicity, insurance, income, mode of arrival, CMS Hierarchical Condition Category (HCC; a risk-adjusted score for estimating future health care costs \citep{pope2000diagnostic}), indicator of diabetes with and without complications, indicator of congestive hearth failure, hypertension, and indicator of obesity. Initial observations $Z$ of the latent health needs included blood pressure, respiration rate, temperature, and ESI level. The admission process was described by treatment time $T$, which is the duration between when treatment starts and ends via admission decision, and the admission decision $A$, a binary variable specifying whether patient was admitted or discharged. Two patient outcomes were analyzed: revisits (whether a patient returns to the ED) and readmissions (whether a patient is admitted in the ED to an inpatient unit) within 30 days of being discharged of any care unit, including the ED. Missing values of continuous variables were imputed with the median value.
After imputation, continuous variables were standardized prior to estimation. Visits with missing health insurance ($n=883$) were placed into the ``Unknown"  category. For analysis, we defined binary insurance as 1 if patient had commercial insurance or worker compensation benefits, and 0 otherwise, including Medicare, Medicaid, Self-Paid, and Unknown. For race and ethnicity, we define 0 if White, and 1 otherwise.

We also explored how the operational context of the ED may influence average potential responses of treatment time. To this end, we worked with congestion, which was calculated for each visit in our dataset as the ED census at the time of arrival. The ED census included everyone in the ED, regardless of who they are or where they are in the ED (e.g., waiting room) or in their treatment pathway (e.g., post-admission decision). We also worked with physician workload, which is the ratio of patients in the ED to providers, i.e. the ED census at the time of arrival averaged over physicians divided by the number of providers in the ED at the time of arrival. We point out that the variables congestion and physician workload use only information available prior to being assigned a patient flow model (as opposed to the workload of any physician assigned to the visit), which allows us to treat this variable as a moderator. 

\section{Results}\label{sec:results}

\subsection{Description of the analyzed sample}

Sample characteristics are summarized in Table \ref{tab:sample_characteristics}. Briefly, the sample of patients presenting with abdominal pain was predominantly white (78\%) and female (60.5\%). Average age was 39.7 years. A majority of patients had commercial insurance or worker compensation benefits (57.9\%). Patients had an average income (at census block group level based on geocoded address) of $\$61,000$, and arrived largely via self means or family (84.3\%). We note that the mean (SD) of treatment time was 4.4 (2.1) hours, whereas 34.2\% of patients were admitted to the hospital, 16.6\% revisited the ED within 30 days of their discharge, and 6.4\% were admitted to the hospital within 30 days of their discharge.

\begin{table}[ht!]
\centering \small
\begin{tabular}{l l c c}
\toprule
\multicolumn{2}{c}{\textbf{Variable}} & \multicolumn{1}{c}{\textbf{Value}} & \multicolumn{1}{c}{\textbf{Missing}} \\
\midrule        
\multicolumn{4}{c}{\textbf{\ul{Patient characteristics $X$}}} \\
\multicolumn{2}{l}{Age in years, mean (SD)} & 39.7 (21.0) & 0 \\
\multicolumn{2}{l}{Female,  n (\%)}  & 17450 (60.5)  & 0 \\
\multicolumn{2}{l}{Race/Ethnicity,    n (\%)}   && 182\\
        & White  & 22379 (78.0) \\
        & Black  & 2878 (10.0)  \\
        & Asian  & 995 (3.5)   \\
        & American Indian/Alaskan  & 127 (0.4) \\
        & Other & 318 (1.1) \\
        & Hispanic/Latino & 1983 (6.9) \\
\multicolumn{2}{l}{Insurance,  n (\%)} && 883 \\
        & Medicaid & 4778 (17.1)\\
        & Medicare & 5728 (20.5)\\
        & Commercial/Workers Comp & 16198 (57.9) \\
        & Self pay & 1275 (4.6) & \\
\multicolumn{2}{l}{Income in \$K, mean   (SD)} & 61.0 (16.7) & 4102 \\
\multicolumn{2}{l}{Arrival by EMS or police,    n (\%)} & 4535 (15.7) & 0 \\
\multicolumn{2}{l}{CMS Hierarchical Condition Category HCC,   mean (SD)} & 0.9 (1.3) & 0\\
\multicolumn{2}{l}{Diabetes with complications,    n (\%)}        & 1118 (3.9)    & 0  \\
\multicolumn{2}{l}{Diabetes without complications,    n (\%)}      & 1201 (4.2)    & 0  \\
\multicolumn{2}{l}{Congital heart failure,    n (\%)}      & 586 (2.0)     & 0  \\
\multicolumn{2}{l}{Hypertension,    n (\%)}      &  5339 (18.5)   & 0  \\
\multicolumn{2}{l}{Obesity,  n (\%)}     &  2475 (8.6)    & 0  \\
\midrule 
\multicolumn{4}{c}{\textbf{\ul{Observations $Z$ of health needs} }} \\
\multicolumn{2}{l}{Acuity 2,  n (\%)}    & 2612 (9.0)    & 0  \\
\multicolumn{2}{l}{Heart rate in bpm,   mean (SD)}    & 87.7 (19.6)   & 41 \\
\multicolumn{2}{l}{Systolic blood   pressure in mmHg, mean (SD)}      & 131.0 (23.5)  & 1454       \\
\multicolumn{2}{l}{Respiration rate in   bpm, mean (SD)}      & 18.6 (3.9)    & 435        \\
\multicolumn{2}{l}{Temperature in   $^\circ$F, mean (SD)}     & 97.8 (1.2)    & 18    \\
\midrule
\multicolumn{4}{c}{\textbf{\ul{Admission process $(A,Z)$}}} \\
\multicolumn{2}{l}{Treatment time in   hours, mean (SD)}      & 4.4 (2.1)     & 0          \\
\multicolumn{2}{l}{Admitted,  n (\%)}    & 9881 (34.2)           & 0          \\
\midrule
\multicolumn{4}{c}{\textbf{\ul{Outcomes of interest $Y$}}}   \\
\multicolumn{2}{l}{Revisited within 30 days,  n (\%)}   & 4782 (16.6)           & 0          \\
\multicolumn{2}{l}{Readmitted within 30 days,    n (\%)}        & 1835 (6.4)            & 0          \\
\bottomrule
\end{tabular}
\caption{Characteristics of analyzed sample of ED visits ($n=28,862$), grouped based on corresponding model variables}\label{tab:sample_characteristics}
\end{table}

\subsection{Average potential response of treatment time on admission decisions}

\subsubsection{Overall estimates.} 
\begin{figure}[htb]
\centering
\begin{tikzpicture}
\node at (-8,0)
{\includegraphics[width=0.49\textwidth]{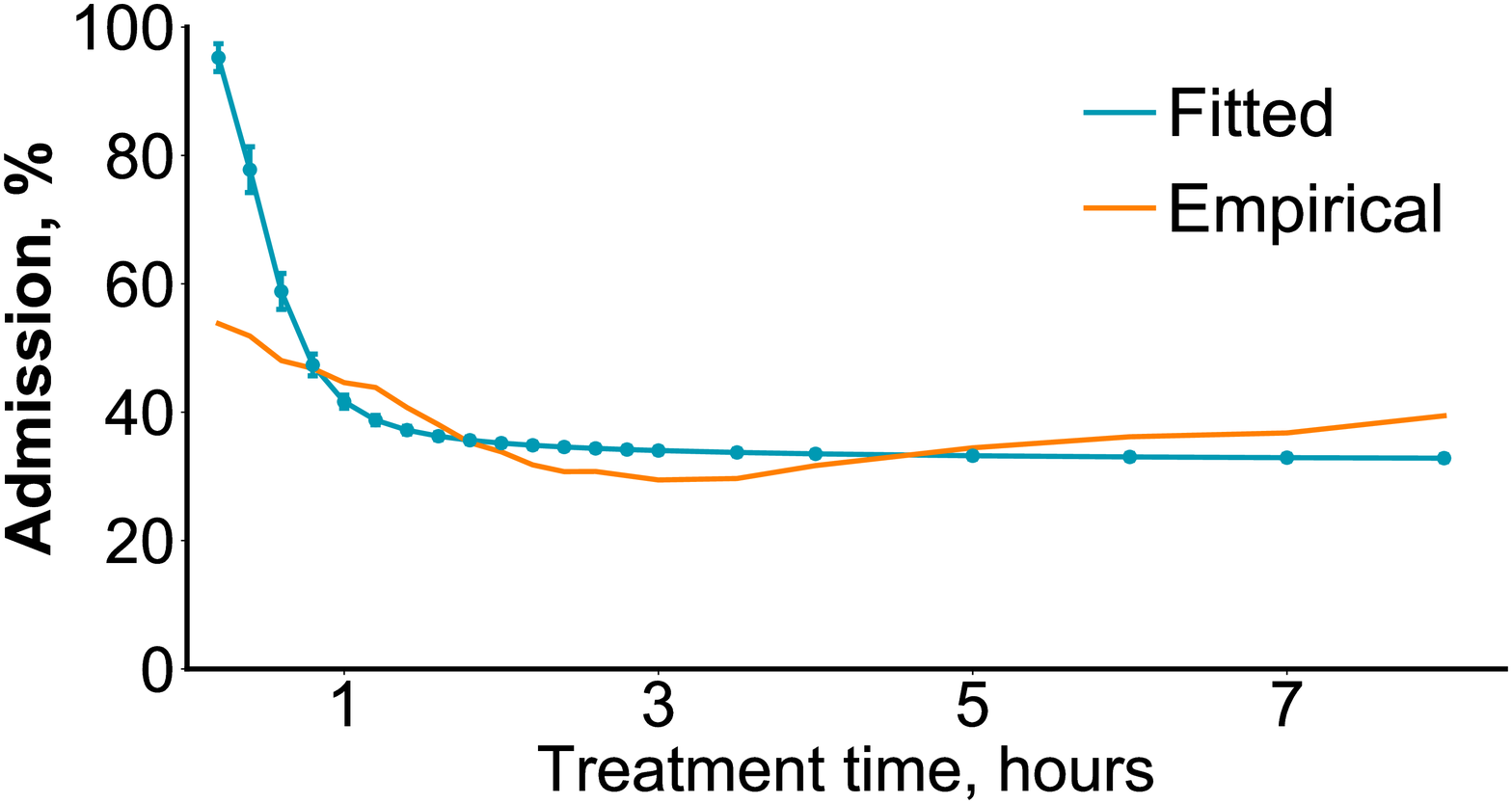}};
\node at (0,0)
{\includegraphics[width=0.49\textwidth]{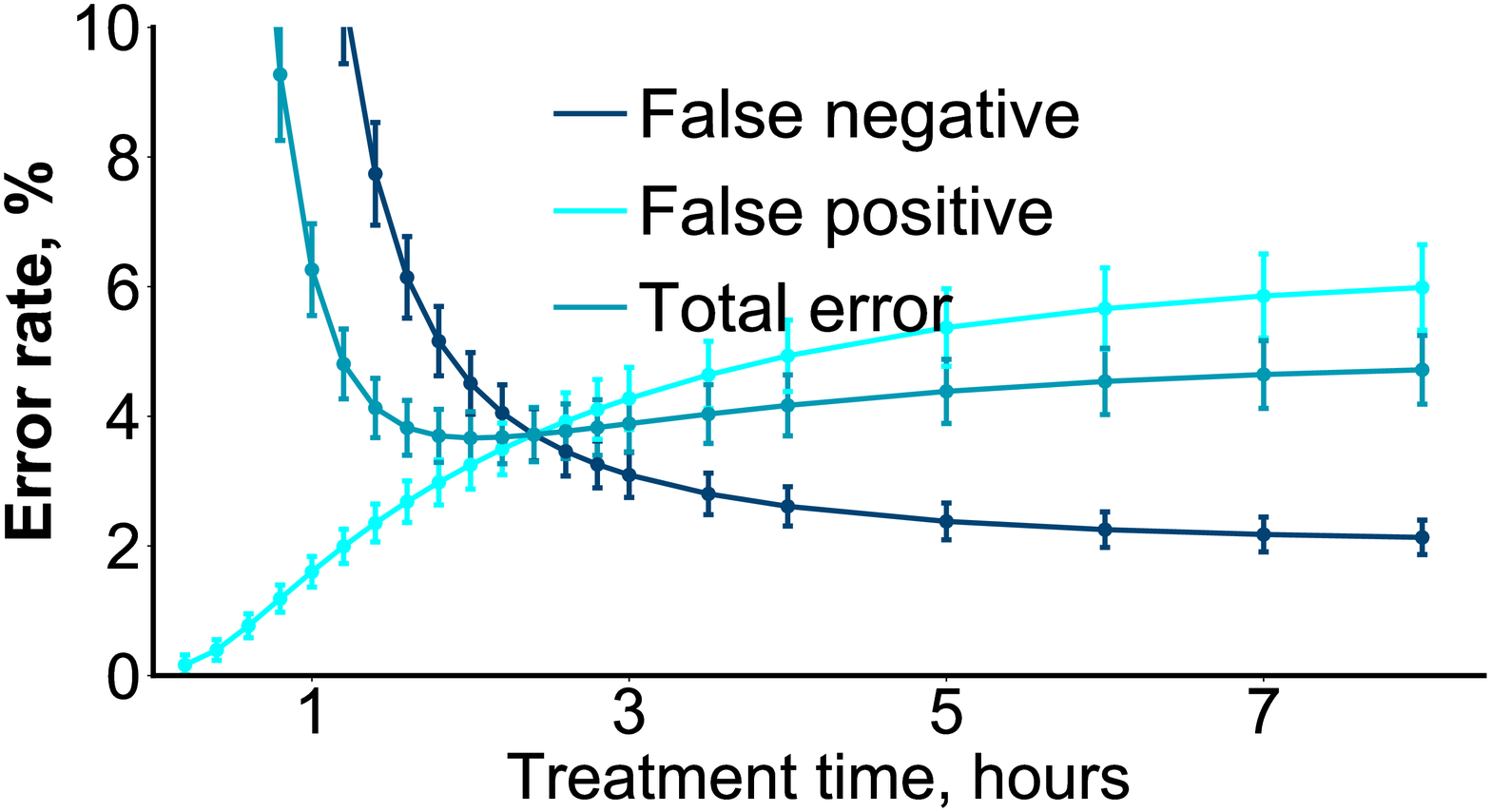}};
\node at (-11.75,2.5) {\large \textsf{\textbf{A}}};
\node at (-3.75,2.5) {\large \textsf{\textbf{B}}};
\end{tikzpicture}
\caption{Average potential responses of treatment time on (A) admission decisions and (B) admission errors based on adjusted and model-based estimates. Admission errors can only be recovered from the model, since it includes the latent health state $H$.}
\label{fig:brown_rev_error_rates}
\end{figure}

We first examine average potential response curves if each patient presenting with abdominal pain is kept in treatment for a fixed length of time $t$.  Theses curves are depicted in Figure~\ref{fig:brown_rev_error_rates}A. Our model suggests that keeping a patient longer in treatment (all else being equal) decreases their chance of being admitted. If patients were kept for one hour, for example, then the admission rate is an estimated 41.6\% (95\% CI: [40.6, 42.7]). By comparison, if patients were kept for 3 hours instead, then this rate drops to 34.0\% (95\% CI: [33.5, 34.5]). Other parameter estimates are presented and checked against empirical quantities in Appendix~\ref{appendix:validity}. 

The most significant reduction occurs within the first hour of ED treatment time, after which admission rates plateaus at roughly 35\%. This suggests the first hour of treatment is critical for reducing admission rates with providers admitting most patients who are treated less than an hour. Meanwhile, unadjusted estimates of admission rates by treatment time provide a different picture: patients experience a more modest drop within the first hour of treatment time and then continue to drop with the second and third hour of treatment, upon which they then start to increase. Unadjusted estimates would thus suggest the first three hours of treatment, as opposed to just the first hour, are critical for reducing admission rates.

\subsubsection{Stratified by latent health needs.} We can stratify estimates by the latent variable $H$ to offer insight into why the first hour might be critical.  In particular, we can calculate $err(t,h) := \PR( A^t \neq H | H=h)$, which can be roughly interpreted as admission error rates when $H$ is the underlying latent health needs of a patient. We call $err(t,0)$ the \emph{false positive} risk of a treatment time time $t$ in terms of admitting someone with lower health needs at time $t$ and call $err(t,1)$ the \emph{false negative} risk at time $t$ in terms of discharging someone with high health needs. We also calculate the total aggregated error defined as $\PR( A^t \neq H )$ for comparison. These risks by treatment time are depicted in Figure~\ref{fig:brown_rev_error_rates}B. 

The false negative rate is a decreasing function and drops dramatically in the first two hours of treatment. It then plateaus at around $2\%$ for longer ED treatment times. For example, the false negative rate is an estimated 15.8\% (95\% CI: [14.1,17.4]) if patients were kept for one hour vs. 3.1\% (95\% CI: [2.8,3.4]) if patients were kept 3 hours. The false positive rate is an increasing function and noticeable higher for patients with ED treatment times greater than 3 hours. It then plateaus at around $6\%$ for longer treatment times. For example, the false positive rate is an estimated 1.6\% (95\% CI: [1.4, 1.8]) if patients were kept for one hour vs. 4.3\% (95\% CI: [3.8, 4.8]) if patients were kept 3 hours. Taken together, these findings can be interpreted as follows: keeping patients who present with abdominal pain for at least one hour, if not two, dramatically reduces admission rates and subsequently the risk of admitting low needs patients with diminishing returns beyond two hours, but does so at the expense of discharging higher needs patients.

\subsubsection{Stratified by patient characteristics.} We also examined how these findings depend on patient characteristics. See Table~\ref{tab:admission_baseline} for average potential responses by baseline characteristics for treatment times fixed at $1/2$, $1$, $2$, and $3$ hours. Several types of individuals have high estimated admission rates if treatment times are fixed at 1/2 hours: individuals who are older ($>$ 39.7 years) followed by individuals who are obese and have high comorbidity ($>$ 0.85 HCC values). Further, women sees the most pronounced drop in admission rates when kept for 1 hour rather than 1/2 hour: starting at 77.5\% (95\% CI: [73.1\%, 81.8\%]) for 1/2-hour treatment times and dropping to 38.7\% (95\% CI: [37.4\%, 40.0\%]) for 1-hour treatment times. 

\begin{table}[t!]
\centering
\begin{tabular}{l l c c c c }
\toprule
\multicolumn{2}{c}{} & \multicolumn{4}{c}{\textbf{Estimate, \%}} \\
\cmidrule(lr){3-6}
\multicolumn{2}{c}{\textbf{Variable}} &
\multicolumn{1}{c}{\textbf{1/2 hr}} & \multicolumn{1}{c}{\textbf{1 hr}} & \multicolumn{1}{c}{\textbf{2 hrs}} & \multicolumn{1}{c}{\textbf{3 hrs}} \\
\midrule
&& \multicolumn{4}{c}{\ul{\textbf{Patient characteristics:}}} \\
Age & $\leq39.67$ & 60.8 (55.4,66.2) & 29.0 (27.9,30.1) & 24.6 (24.0,25.2) & 23.8 (23.2,24.4) \\ 
     & $>39.67$ & 88.5 (86.1,90.8) & 54.3 (53.0,55.6) & 46.5 (45.8,47.2) & 45.5 (44.7,46.2) \\ 
    Sex & Male & 68.5 (64.6,72.5) & 44.4 (43.3,45.6) & 39.8 (39.0,40.6) & 38.8 (38.0,39.6) \\ 
     & Female & 77.5 (73.1,81.8) & 38.7 (37.4,40.0) & 31.8 (31.2,32.5) & 31.0 (30.4,31.6) \\ 
    Race & White & 77.0 (73.6,80.4) & 43.9 (42.9,45.0) & 37.6 (37.0,38.1) & 36.6 (36.0,37.2) \\ 
     & Non-White & 63.3 (56.0,70.6) & 30.9 (29.1,32.6) & 26.1 (25.2,27.1) & 25.4 (24.4,26.4) \\ 
    Insurance & Other & 79.5 (75.5,83.5) & 45.5 (44.1,46.9) & 39.0 (38.2,39.8) & 38.2 (37.5,39.0) \\ 
     & Workers Comp & 70.1 (65.8,74.3) & 37.9 (36.7,39.0) & 32.2 (31.6,32.9) & 31.2 (30.6,31.8) \\ 
    Income & $\leq61.01$ & 76.0 (72.1,79.8) & 41.9 (40.7,43.0) & 35.8 (35.2,36.3) & 34.9 (34.3,35.5) \\ 
     & $>61.01$ & 71.2 (67.2,75.2) & 39.8 (38.7,41.0) & 34.0 (33.3,34.6) & 33.0 (32.3,33.6) \\ 
    Comorbidity & $\leq0.85$ & 72.5 (68.8,76.2) & 39.0 (38.0,40.0) & 33.1 (32.6,33.6) & 32.2 (31.7,32.7) \\ 
     & $>0.85$ & 81.6 (78.9,84.3) & 51.5 (50.4,52.5) & 45.0 (44.4,45.7) & 44.2 (43.6,44.8) \\ 
    Hypertension & No & 63.6 (58.7,68.5) & 33.5 (32.5,34.6) & 28.6 (28.0,29.1) & 27.6 (27.0,28.1) \\ 
     & Yes & 74.8 (71.3,78.2) & 41.6 (40.6,42.6) & 35.5 (35.0,36.0) & 34.6 (34.1,35.1) \\ 
    Obesity & No & 73.5 (69.9,77.1) & 40.4 (39.4,41.4) & 34.4 (33.9,34.9) & 33.5 (33.0,34.0) \\ 
     & Yes & 85.1 (82.1,88.0) & 55.4 (53.5,57.4) & 48.8 (47.6,50.0) & 48.3 (47.1,49.5)\\
     
&& \multicolumn{4}{c}{\ul{\textbf{Operational factors:}}} \\
    Congestion & Low & 73.0 (68.1,78.0) & 41.7 (39.8,43.6) & 34.9 (34.0,35.7) & 33.8 (33.0,34.7) \\ 
     & Medium & 71.3 (65.9,76.7) & 44.2 (41.2,47.2) & 35.0 (34.0,35.9) & 33.4 (32.5,34.4) \\ 
     & High & 90.4 (86.7,94.0) & 46.6 (44.1,49.2) & 35.7 (34.9,36.6) & 34.2 (33.3,35.1) \\ 
    APP workload & Low & 77.6 (73.7,81.5) & 45.1 (43.2,47.0) & 35.1 (34.2,36.0) & 33.6 (32.7,34.4) \\ 
     & Medium & 70.3 (64.5,76.2) & 42.3 (40.1,44.5) & 34.7 (33.8,35.7) & 33.4 (32.5,34.3) \\ 
     & High & 70.8 (64.9,76.8) & 39.7 (38.3,41.1) & 35.3 (34.4,36.1) & 34.8 (33.9,35.6)\\
\bottomrule
\end{tabular}
\caption{Estimated average potential responses of treatment times (1 hr, 3 hrs, 10 hrs) on admissions stratified by baseline characteristics and operational factors.}
\label{tab:admission_baseline}
\end{table}

\subsubsection{Stratified by congestion and physicain workload.} Since keeping a patient longer may not always be feasible if the ED is busy, the model was re-fit to a subset of data that included only visits within each tertile of the congestion variable and similarly for APP workload. Table~\ref{tab:admission_baseline} presents these results. Estimates by congestion level or APP workload follow similar trends as estimates of the overall admission rate: keeping a patient for at least one or two hours is found to significantly reduce admission rates, regardless of how busy the ED is terms of its congestion or the workload of ED providers. 


\subsection{Average potential responses of treatment time on 30-day ED revisits and readmissions}

\subsubsection{Overall estimates.} 
\begin{figure}[t!]
\centering
\begin{tikzpicture}
\node at (-8,0)
{\includegraphics[width=0.49\textwidth]{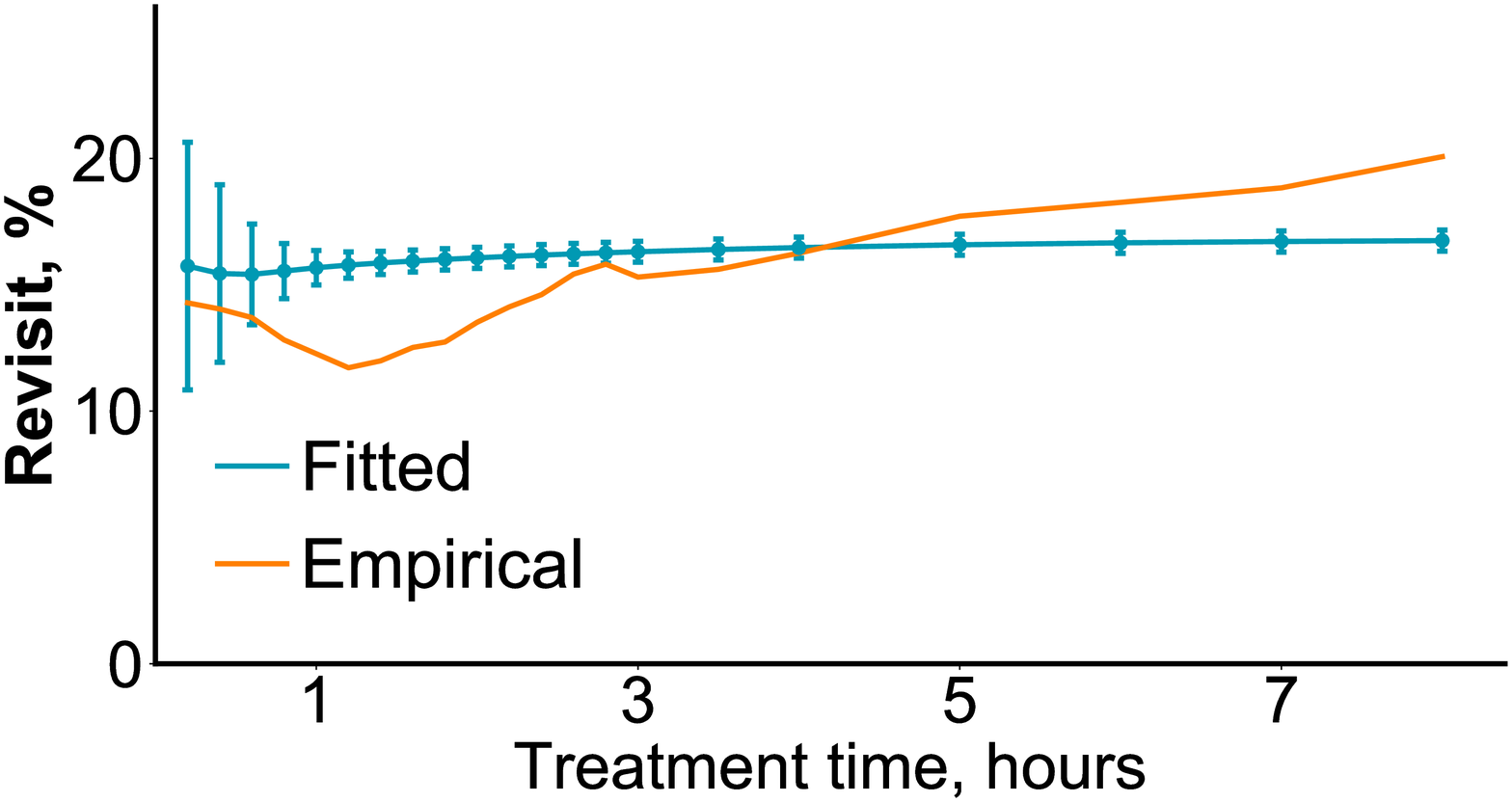}};
\node at (0,0)
{\includegraphics[width=0.49\textwidth]{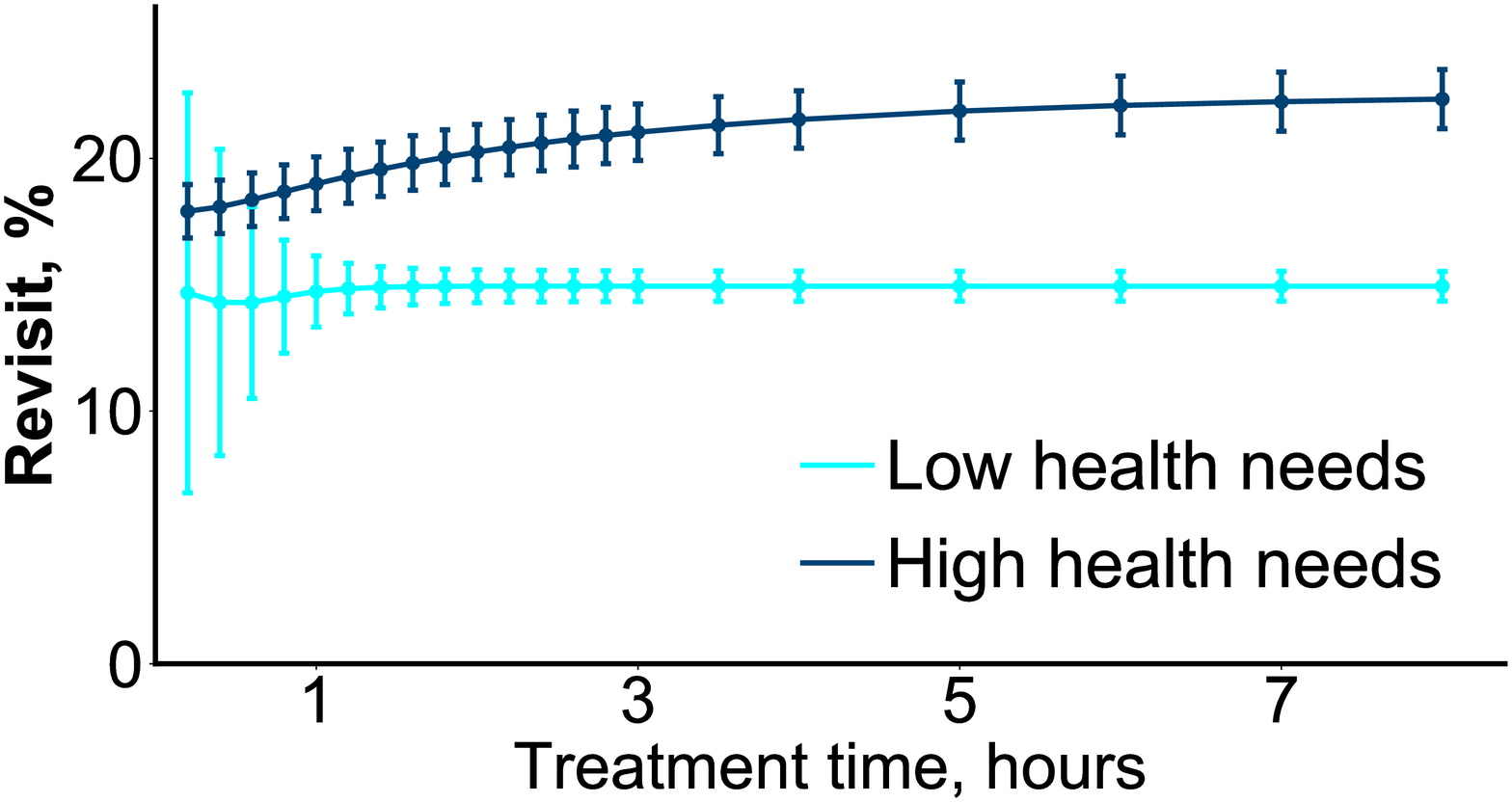}};
\node at (-8,-5)
{\includegraphics[width=0.49\textwidth]{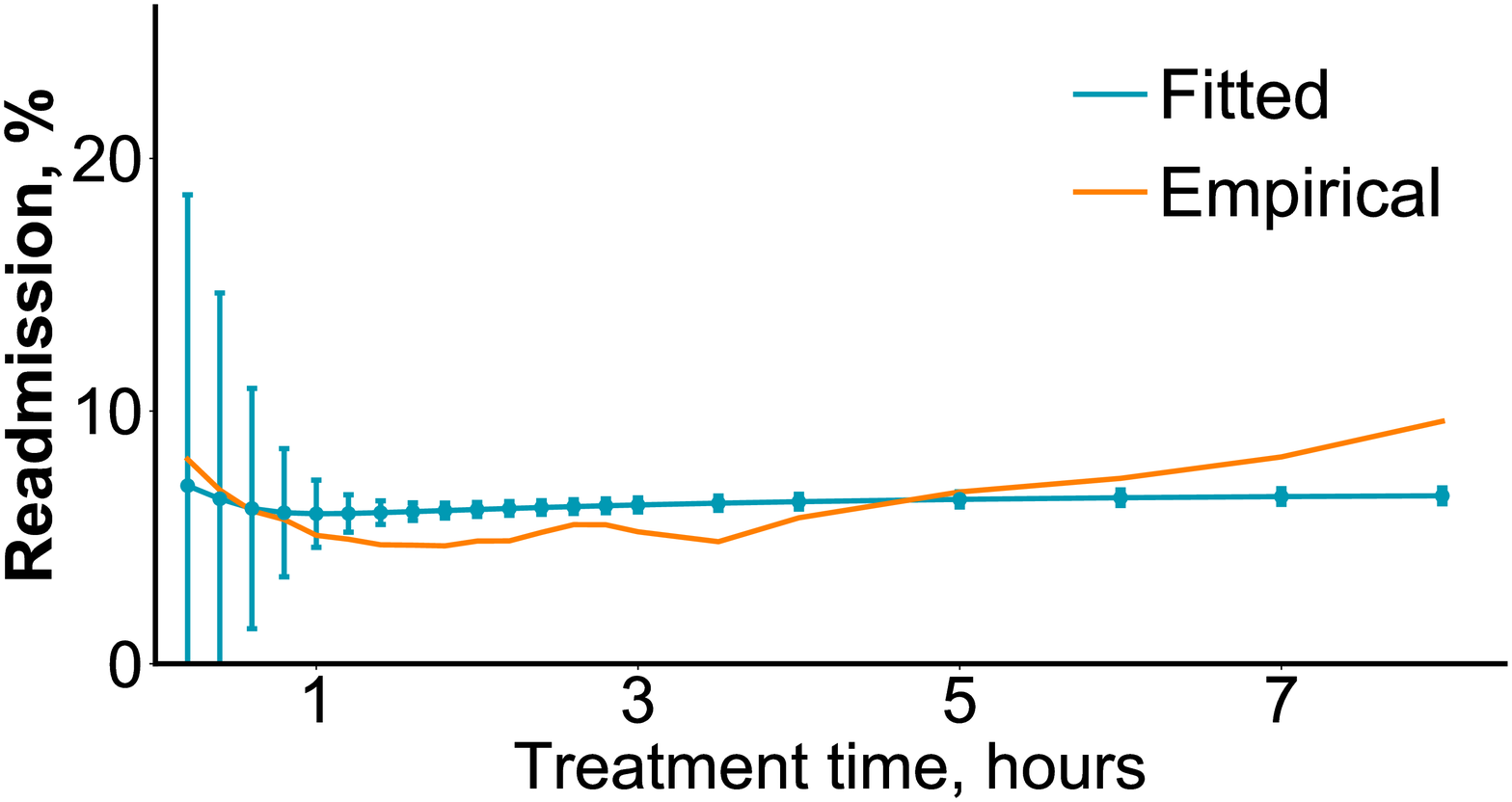}};
\node at (0,-5)
{\includegraphics[width=0.49\textwidth]{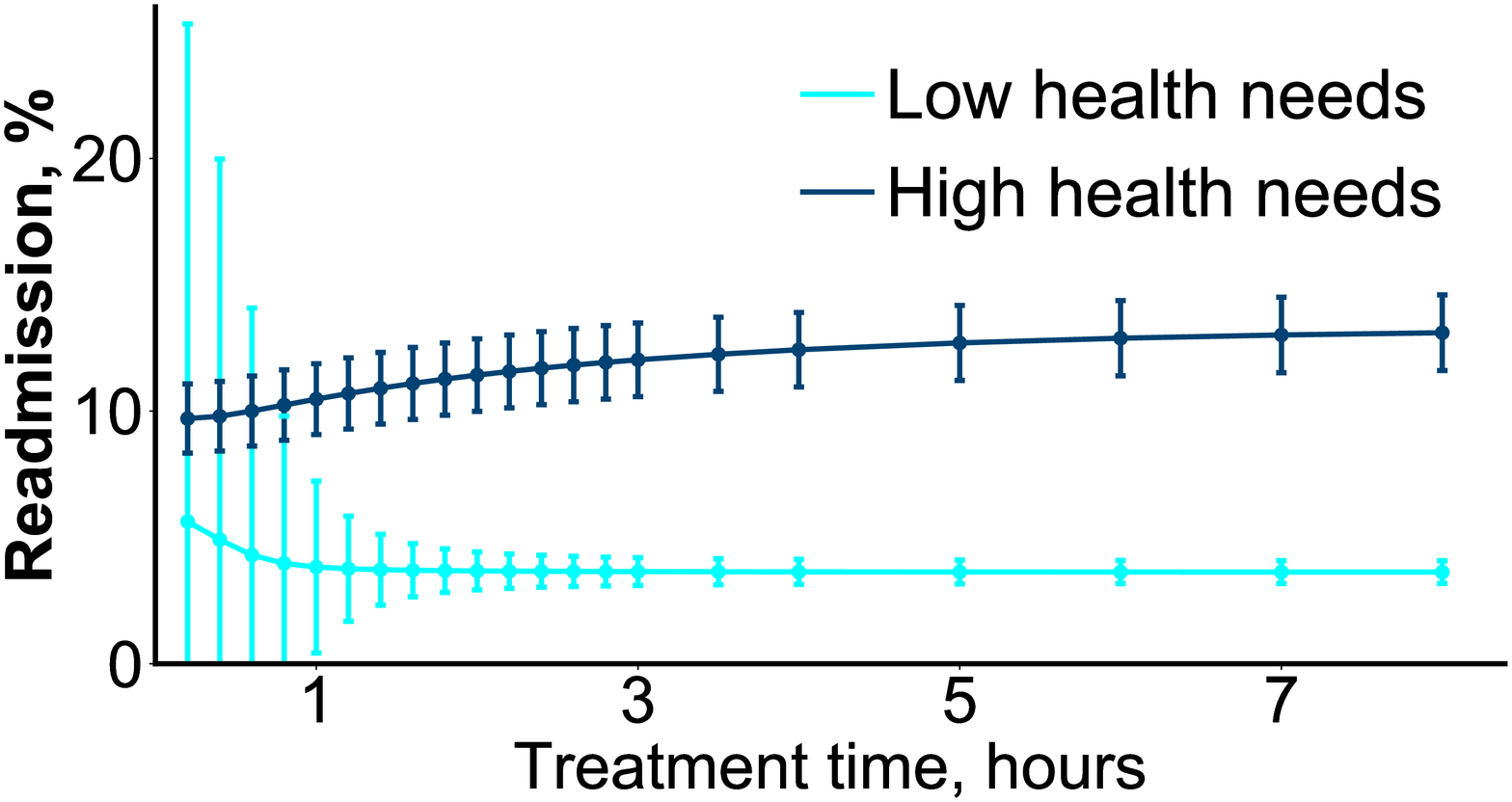}};
\node at (-11.75,2.5) {\large \textsf{\textbf{A}}};
\node at (-3.75,2.5) {\large \textsf{\textbf{B}}};
\node at (-11.75,-2.5) {\large \textsf{\textbf{C}}};
\node at (-3.75,-2.5) {\large \textsf{\textbf{D}}};
\end{tikzpicture}
\caption{Average potential responses of treatment time on 30-day revisits (A) overall and (B) by health needs, and similarly on 30-day readmissions (C) overall and (D) by health needs.}
\label{fig:brown_outcomes} 
\end{figure}

We also explored downstream consequences of keeping a patient presenting with abdominal pain for a fixed time $t$. Figure~\ref{fig:brown_outcomes} depicts estimated average potential responses of treatment time on revisiting the ED or being admitted to the hospital after 30 days of discharge. We find that keeping a patient longer (all else being equal) leads to a slight increase in their risk of revisiting the ED and being readmitting on a subsequent ED visit. If patients were kept for one hour, for example, then the risks of revisit and readmission are an estimated 15.7\% (95\% CI: [15.0, 16.4]) and 5.9\% (95\% CI: [4.6, 7.3]), respectively. By comparison, if patients were kept for 3 hours instead, then these risks increase slightly to an estimated 16.3\% (95\% CI: [15.9, 16.7]) and 6.3\% (95\% CI: [6.0, 6.6]). By contrast, unadjusted estimates are more sensitive to treatment time. For both outcomes, unadjusted estimates decrease in the first hour of treatment time and then steadily increase, which suggests that treatment times that are neither too short or too long is important for reducing revisit and readmission risks.

\subsubsection{Stratified by latent health needs.} As we did with admission rates, we can examine revisit and readmission risk by latent health needs $H$ and by the intervening treatment time (Figure~\ref{fig:brown_outcomes}B, \ref{fig:brown_outcomes}D). For lower health needs patients (i.e. $H=0$), the 30-day risk of revisit and readmission is a relatively constant function of treatment time. For example, the risks of revisit is an estimated 14.7\% (95\% CI: [13.3,16.1]) if patients were kept for one hour vs. 15.0\% (95\% CI: [14.3,15.6]) if patients were kept 3 hours, whereas the risk or readmission is 3.8\% (95\% CI: [0.4,7.2]) if patients were kept 1 hour vs. 3.7\% (95\% CI: [3.1,4.2]) if patients were kept 3 hours. In other words, keeping patients longer appears to impose minimal downstream risk to low needs patients, despite reducing their admission rates dramatically.

For higher health needs patients, the 30-day risk of revisit and readmission is an increasing function that increases quickly in the first four hours of treatment. It then plateaus around $21\%$ for longer ED treatment times. For example, the risks of revisit is an estimated 19.0\% (95\% CI: [17.9, 20.1]) if patients were kept for one hour vs. 21.0\% (95\% CI: [19.9, 22.2]) if patients were kept 3 hours, whereas the risk or readmission is 10.5\% (95\% CI: [9.1, 11.9]) if patients were kept 1 hour vs. 12.0\% (95\% CI: [10.6, 13.5]) if patients were kept 3 hours. Therefore, keeping a patient longer appears to impose a modest risk to higher needs patients, as a result of the greater risk of discharging these patients. 


\subsubsection{Stratified by patient characteristics.}  Table~\ref{tab:outcome_baseline_rev} reports average potential responses of treatment time on 30-day revisits by baseline characteristics when ED treatment times equal $1/2$, $1$, $2$, and $3$ hours. For most people, increasing treatment time appears to slightly increase the 30-day risk of a revisit. The underlying risk, however, differs between each patient group, with individuals who are obese carrying a significantly higher revisit risk than other individuals. Further, individuals who are obese also appear to experience the largest increase in revisit risk from longer treatment times, with the estimated revisit risk increasing from 28.5\% (95\% CI: [24.0, 33.1]) if treated for 1/2 hour to 30.2\% (95\% CI: [18.2,19.3]) if treated for 1 hour. Since individuals who are obese are also more likely to be admitted, these individuals should have disproportionately higher needs; keeping them longer could increase their risk of being incorrectly discharged and a subsequent ED revisit.

\begin{table}[t!]
\centering
\begin{tabular}{l l c c c c }
\toprule
\multicolumn{2}{c}{} & \multicolumn{3}{c}{\textbf{Estimate, \%}} \\
\cmidrule(lr){3-6}
\multicolumn{2}{c}{\textbf{Variable}} &
\multicolumn{1}{c}{\textbf{1/2 hr}} & \multicolumn{1}{c}{\textbf{1 hr}} & \multicolumn{1}{c}{\textbf{2 hrs}}  & \multicolumn{1}{c}{\textbf{3 hrs}} \\
\midrule
&& \multicolumn{4}{c}{\ul{\textbf{Patient characteristics:}}} \\
Age & $\leq39.67$ & 14.0 (10.7,17.4) & 13.7 (13.0,14.3) & 13.9 (13.4,14.4) & 14.1 (13.6,14.6) \\ 
     & $>39.67$ & 17.0 (13.3,20.8) & 17.9 (17.0,18.8) & 18.4 (17.9,19.0) & 18.7 (18.2,19.3) \\ 
    Sex & Male & 15.5 (12.6,18.4) & 15.6 (14.8,16.4) & 16.2 (15.5,16.8) & 16.5 (15.8,17.1) \\ 
     & Female & 15.4 (11.4,19.4) & 15.7 (14.9,16.5) & 16.0 (15.5,16.5) & 16.2 (15.7,16.7) \\ 
    Race & White & 15.1 (11.7,18.5) & 15.4 (14.6,16.1) & 15.8 (15.3,16.3) & 16.1 (15.6,16.6) \\ 
     & Non-White & 16.7 (12.6,20.7) & 16.7 (15.8,17.7) & 16.9 (16.0,17.8) & 17.0 (16.1,17.9) \\ 
    Insurance & Other & 20.1 (15.6,24.5) & 21.0 (20.0,22.1) & 21.5 (20.8,22.2) & 21.7 (21.0,22.5) \\ 
     & Workers Comp & 12.3 (9.3,15.2) & 12.0 (11.3,12.6) & 12.3 (11.8,12.8) & 12.6 (12.1,13.0) \\ 
    Income & $\leq61.01$ & 15.8 (12.0,19.7) & 16.0 (15.2,16.8) & 16.4 (15.9,16.9) & 16.7 (16.2,17.1) \\ 
     & $>61.01$ & 14.9 (11.8,18.0) & 15.2 (14.5,15.9) & 15.6 (15.1,16.1) & 15.8 (15.3,16.4) \\ 
    Comorbidity & $\leq0.85$ & 14.8 (11.3,18.3) & 14.9 (14.2,15.6) & 15.3 (14.9,15.7) & 15.5 (15.1,15.9) \\ 
     & $>0.85$ & 18.9 (15.2,22.6) & 19.8 (18.9,20.6) & 20.3 (19.7,20.8) & 20.5 (20.0,21.1) \\ 
    Hypertension & No & 11.7 (8.7,14.7) & 11.4 (10.7,12.0) & 11.7 (11.2,12.2) & 12.0 (11.5,12.4) \\ 
     & Yes & 15.7 (12.2,19.3) & 16.0 (15.3,16.7) & 16.4 (16.0,16.8) & 16.7 (16.2,17.1) \\ 
    Obesity & No & 14.9 (11.4,18.4) & 15.1 (14.4,15.8) & 15.5 (15.1,15.9) & 15.7 (15.3,16.1) \\ 
     & Yes & 28.5 (24.0,33.1) & 30.2 (28.9,31.5) & 30.6 (29.4,31.8) & 30.8 (29.6,32.0) \\
     && \multicolumn{4}{c}{\ul{\textbf{Operational factors:}}} \\
    Congestion & Low & 15.5 (12.4,18.6) & 15.9 (15.0,16.8) & 16.3 (15.6,17.0) & 16.5 (15.7,17.2) \\ 
     & Medium & 16.7 (10.7,22.8) & 17.0 (15.2,18.9) & 17.3 (16.6,18.1) & 17.4 (16.6,18.1) \\ 
     & High & 17.6 (10.8,24.4) & 16.1 (14.5,17.7) & 15.7 (15.0,16.5) & 15.7 (15.0,16.4) \\ 
    APP workload & Low & 14.1 (10.4,17.8) & 15.5 (14.4,16.6) & 16.3 (15.6,17.0) & 16.6 (15.9,17.3) \\ 
     & Medium & 17.6 (11.6,23.5) & 17.0 (15.4,18.6) & 17.0 (16.2,17.7) & 17.0 (16.2,17.7) \\ 
     & High & 14.3 (11.2,17.3) & 15.6 (14.8,16.4) & 15.9 (15.2,16.6) & 16.0 (15.3,16.7)\\
\bottomrule
\end{tabular}
\caption{Estimated average potential responses of treatment time (1/2 hr, 1 hr, 2 hrs, 3 hrs) on 30-day revisit stratified by baseline characteristics and operational factors.}
\label{tab:outcome_baseline_rev}
\end{table}

Similar to 30-day revisit risk, treatment time also appears to have a slight effect on 30-day readmission risk regardless of patient characteristics. Table~\ref{tab:outcome_baseline_rev} reports average potential responses of treatment time on 30-day readmissions by baseline characteristics when ED treatment times equal $1/2$, $1$, $2$, and $3$ hours. The 30-day risk of readmission actually appears to slightly decrease when lengthening treatment time from 1/2 hour to 1 hour (albeit confidence intervals are wide), but then slightly increase when lengthening treatment time from 1 hour to 3 hours. Again, the underlying risk differs between patient groups, with individuals who are obese having significantly higher readmission risk than other individuals. However, the increase in readmission risk due to longer treatment time is relatively more comparable between groups. So although estimated admission rates drop dramatically with longer treatment times and are accompanied by an estimated  2\% increase in 30-day revisits among individuals who are obese, these changes do not translate into higher estimated 30-day readmission rates.

\begin{table}[t!]
\centering
\begin{tabular}{l l c c c c }
\toprule
\multicolumn{2}{c}{} & \multicolumn{3}{c}{\textbf{Estimate, \%}} \\
\cmidrule(lr){3-6}
\multicolumn{2}{c}{\textbf{Variable}} &
\multicolumn{1}{c}{\textbf{1/2 hr}} & \multicolumn{1}{c}{\textbf{1 hr}} & \multicolumn{1}{c}{\textbf{2 hrs}} & \multicolumn{1}{c}{\textbf{3 hrs}} \\
\midrule
&& \multicolumn{4}{c}{\ul{\textbf{Patient characteristics:}}} \\
Age & $\leq39.67$ & 4.7 (-2.0,11.4) & 4.0 (3.2,4.8) & 4.1 (3.9,4.4) & 4.3 (4.0,4.6) \\ 
     & $>39.67$ & 8.6 (-1.3,18.4) & 8.1 (6.1,10.0) & 8.3 (7.8,8.7) & 8.5 (8.1,9.0) \\ 
    Sex & Male & 6.7 (0.9,12.4) & 6.2 (5.4,7.0) & 6.5 (6.0,7.0) & 6.8 (6.3,7.3) \\ 
     & Female & 6.4 (-3.3,16.2) & 5.8 (4.0,7.5) & 5.8 (5.4,6.2) & 6.0 (5.6,6.3) \\ 
    Race & White & 6.7 (-1.4,14.9) & 6.2 (4.8,7.5) & 6.4 (6.1,6.7) & 6.6 (6.2,6.9) \\ 
     & Non-White & 5.8 (-2.3,14.0) & 5.1 (3.8,6.5) & 5.2 (4.7,5.8) & 5.3 (4.8,5.9) \\ 
    Insurance & Other & 8.7 (-1.6,19.0) & 8.2 (6.5,10.0) & 8.4 (7.9,8.9) & 8.5 (8.0,9.0) \\ 
     & Workers Comp & 5.0 (-1.6,11.7) & 4.3 (3.3,5.4) & 4.5 (4.2,4.9) & 4.7 (4.4,5.1) \\ 
    Income & $\leq61.01$ & 6.8 (-1.8,15.4) & 6.2 (4.8,7.6) & 6.3 (6.0,6.6) & 6.5 (6.1,6.8) \\ 
     & $>61.01$ & 6.2 (-1.4,13.7) & 5.6 (4.3,6.9) & 5.8 (5.5,6.2) & 6.0 (5.7,6.4) \\ 
    Comorbidity & $\leq0.85$ & 6.0 (-1.9,14.0) & 5.4 (4.1,6.7) & 5.6 (5.3,5.8) & 5.8 (5.5,6.0) \\ 
     & $>0.85$ & 9.1 (-0.1,18.4) & 8.8 (7.1,10.5) & 9.0 (8.5,9.4) & 9.2 (8.7,9.6) \\ 
    Hypertension & No & 3.9 (-1.5,9.2) & 3.4 (2.6,4.2) & 3.6 (3.4,3.8) & 3.8 (3.6,4.0) \\ 
     & Yes & 6.7 (-1.6,15.1) & 6.1 (4.8,7.5) & 6.3 (6.0,6.6) & 6.5 (6.2,6.8) \\ 
    Obesity & No & 6.2 (-1.7,14.1) & 5.6 (4.3,6.9) & 5.8 (5.5,6.0) & 5.9 (5.7,6.2) \\ 
     & Yes & 15.3 (1.0,29.6) & 14.8 (11.9,17.7) & 14.8 (13.8,15.8) & 14.9 (13.9,15.9)\\
     
     && \multicolumn{4}{c}{\ul{\textbf{Operational factors:}}} \\
    Congestion & Low & 7.0 (5.8,8.2) & 6.2 (5.7,6.7) & 6.3 (5.8,6.8) & 6.4 (6.0,6.9) \\ 
     & Medium & 7.9 (5.6,10.2) & 6.9 (6.1,7.8) & 6.5 (6.0,7.0) & 6.5 (6.0,6.9) \\ 
     & High & 4.2 (3.8,4.7) & 5.6 (5.2,6.1) & 6.1 (5.6,6.5) & 6.2 (5.7,6.7) \\ 
    APP workload & Low & 6.9 (3.4,10.4) & 6.1 (5.2,7.1) & 6.3 (5.8,6.7) & 6.5 (6.0,7.0) \\ 
     & Medium & 5.9 (4.8,7.0) & 6.4 (5.8,6.9) & 6.6 (6.1,7.1) & 6.7 (6.2,7.1) \\ 
     & High & 6.0 (3.6,8.5) & 5.8 (5.2,6.4) & 5.9 (5.5,6.4) & 6.0 (5.6,6.5)\\
\bottomrule
\end{tabular}
\caption{Estimated average potential responses of treatment time (1/2 hr, 1hr, 2 hrs, 3 hrs) on 30-day readmission stratified by baseline characteristics and operational factors.}
\label{tab:outcome_baseline_read}
\end{table}

\subsubsection{Stratified by congestion and physicain workload.} Tables~\ref{tab:outcome_baseline_rev}--\ref{tab:outcome_baseline_read} also present average potential responses of treatment time on 30-day revisits and readmissions based only on visits within each tertile of the congestion variable and similarly for APP workload. For both potential moderators, trends follow overall 30-day revisit and readmission trends, in that varying the treatment time appears to result in slight changes to these risks. One notable exception is in the first hour of treatment. In this case, the 30-day readmission risk appears to experience a more moderate increase when congestion is high compared to a potential decrease when congestion is low. 

\subsection{From Estimates to Recommendations:  Shift Interventions}

\begin{figure}[t!]
\centering
{\includegraphics[width=0.49\textwidth]{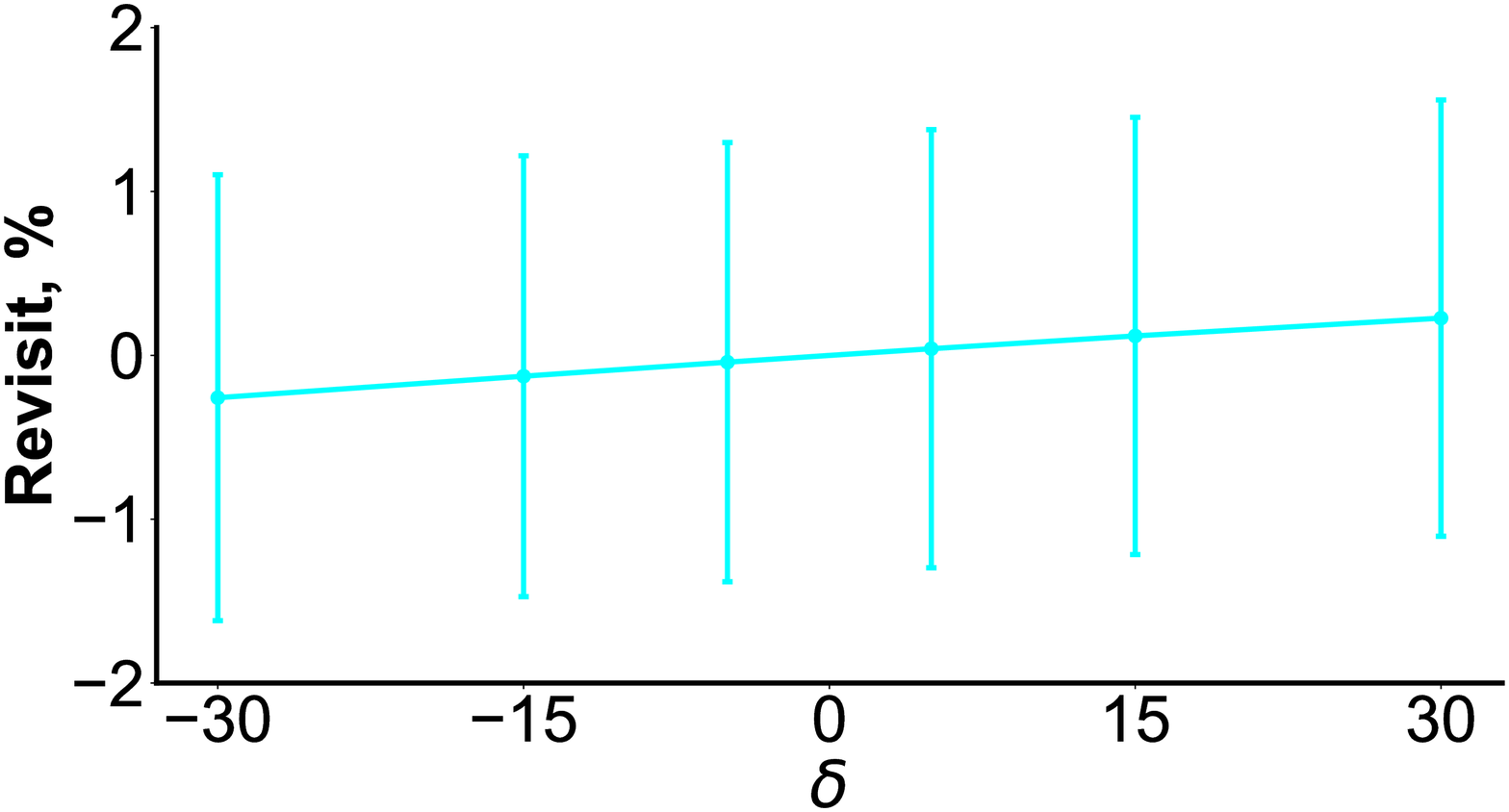}}
{\includegraphics[width=0.49\textwidth]{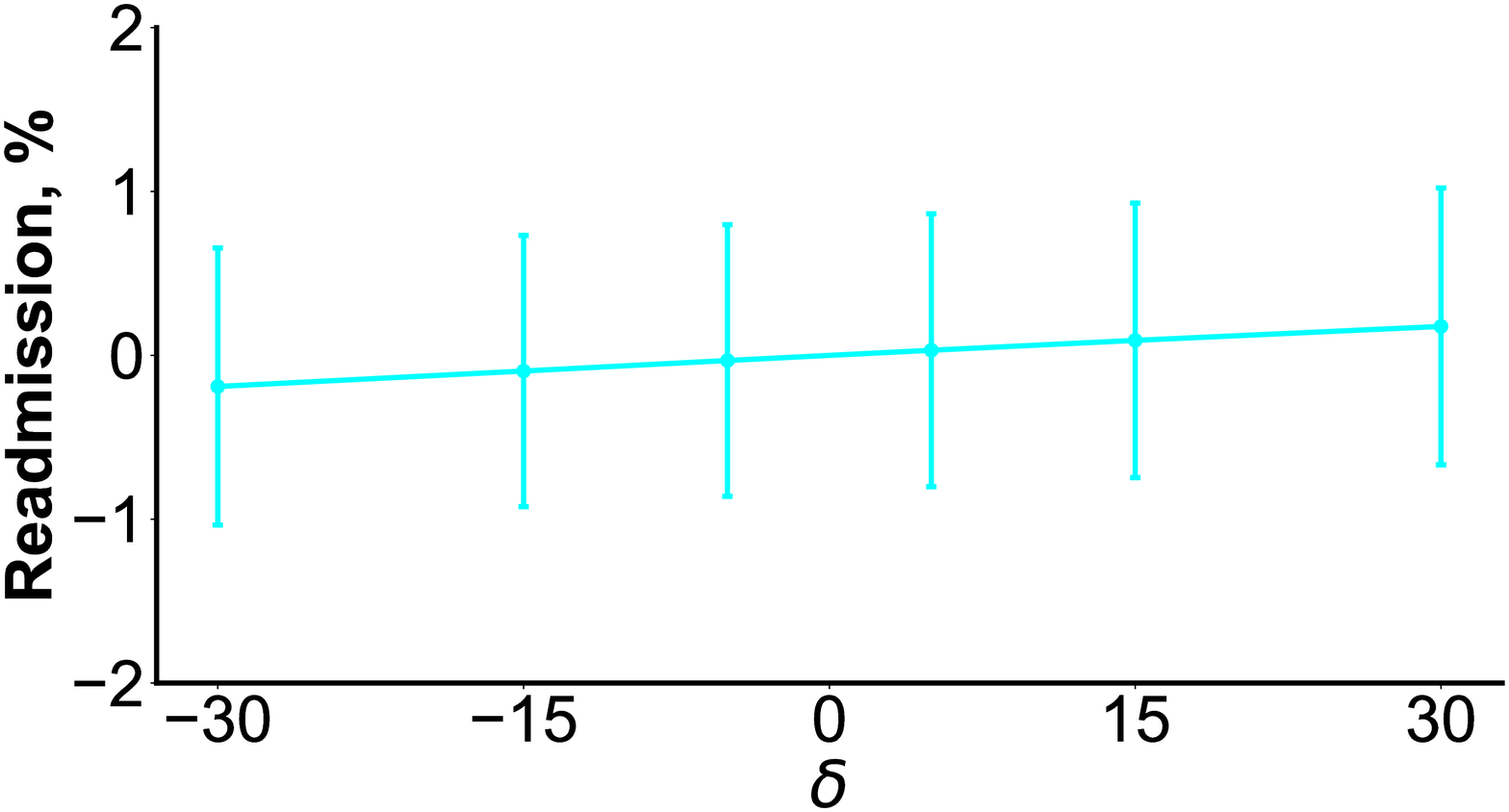}}
{\includegraphics[width=0.49\textwidth]{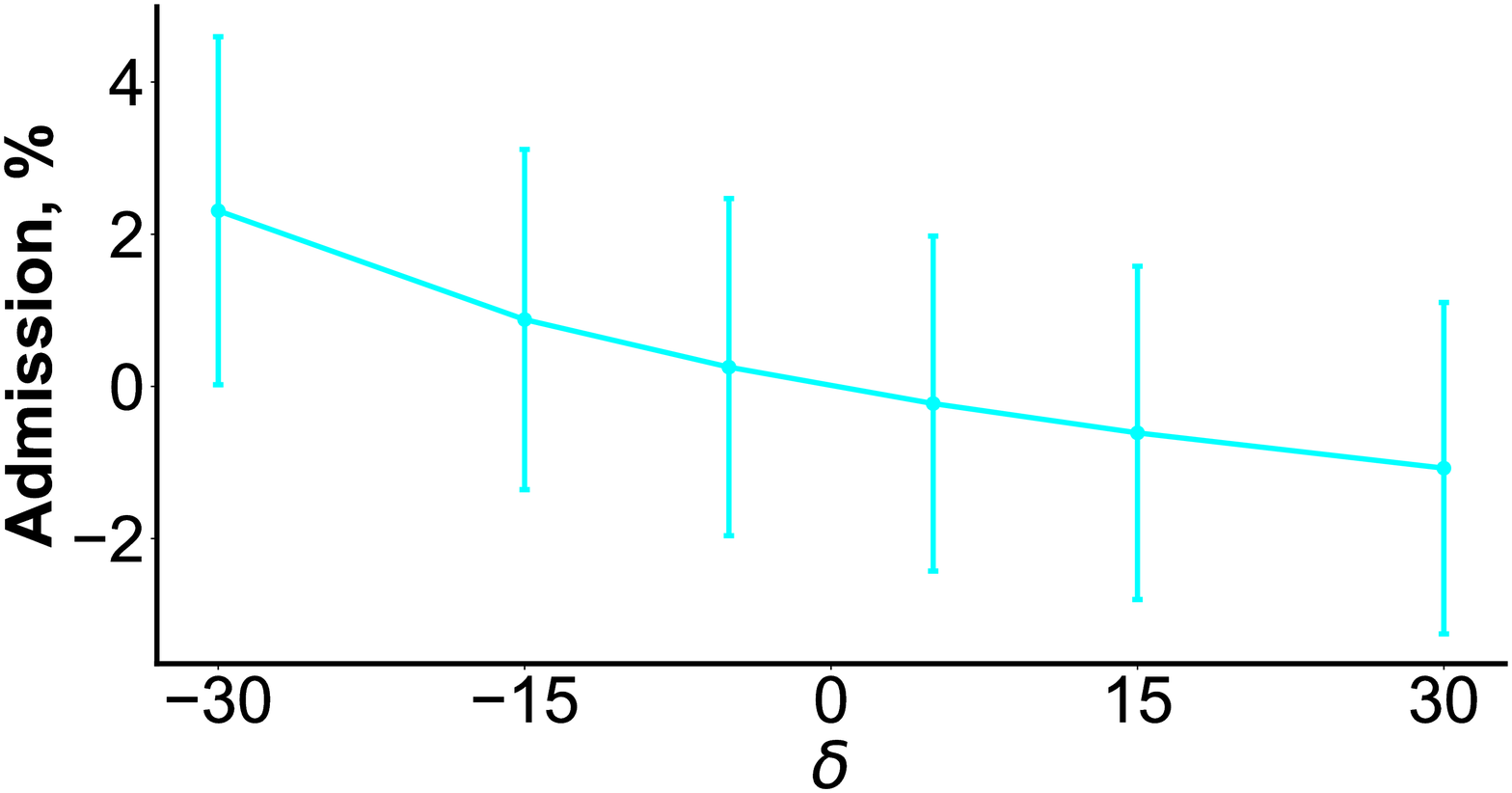}}
\caption{Shift intervention effects on admissions, revisits, and readmissions when using policy $f(T) = \max\{T+\delta, 5\}\text{ minutes}$}
\label{fig:brown_shifts} 
\end{figure}

Our final estimates are for the impact of shift interventions, which capture average responses of a keeping a patient longer or shorter \emph{relative} to their realized treatment time as opposed to keeping a patient a fixed amount of time. Specifically, we evaluated the policy $f(T) = \max\{T+\delta, 5\}\text{ mins}$ for different values $\delta \in \{-30, -15, 0, 15, 30\}$.  These are depicted in Figure \ref{fig:brown_shifts}. We find that if the time to make an admission is reduced by 30 minutes, then the admission rate is increased by 2.3\% (95\% CI: [0, 4.6]) on average. In a reciprocal manner, delaying the admission decision by 30 minutes is estimated to decrease the admission rate by 1.1\% (95\% CI: [-1.1, 3.2]) on average. On the other hand, changing the time to make an admission decision does not significantly impact the other outcomes (i.e., revisit and readmission rates).  To summarize, a premature admission decision leads to an estimated increase in the admission rate, while waiting at least 30 minutes more, leads to an estimated lower admission rate.

\section{Discussion}\label{sec:conclusion}

Admission decisions in the ED are one of the most critical, routine, and expensive decisions made in health care. These decisions are often made in the absence of clear cut clinical guidelines, leading to potentially harmful and costly variability in care. To better inform this decision, providers could potentially intervene on the duration in which they keep patients for treatment, monitoring, and assessment. Recovering evidence for these interventions is difficult, in large part due to unmeasured confounding and the intervention being continuous. Here, we presented a model-based method to measure average response curves for treatment time and applied this method to adults presenting with abdominal pain, one of the most common ED chief complaints. This paper addresses a fundamental question in many healthcare settings: if a patient were kept a little longer, could we improve their care and reduce adverse consequences? 

We rely on several innovations to address this question.  We first formalize a conceptual model of the ED admission process in a DAG. We then use the potential outcomes framework in \cite{richardson2013single} to define average potential responses of our continuous intervening variable (treatment time) and prove that these responses cannot be identified using non-parametric estimation due to unmeasured confounding. We thus turn to a parametric model that includes a latent variable to capture the unmeasured confounder. Notably, the model also includes a joint model of the admission decision and treatment time using threshold regression; threshold regression models use passage times for stochastic processes. A joint model is thought to better reflect the actual decision-making scenario than alternative time-to-event models that fail to capture an event that has two possible outcomes. We fit this model to a large EHR dataset using the well known Expectation Maximization algorithm \citep{dempster1977maximum}.


This model-based approach provided information on how long abdominal pain patients should be kept at a large tertiary teaching hospital (n=28,862). Short treatment times (i.e., one hour or less) are estimated to cause high rates of admissions, approximately 95\%, after which longer treatment times can reduce admission rates down to a steady 33\%. We stratified admissions rates down by latent health needs and found that keeping patients for at least 1 hour can dramatically reduce the risk of admitting a low-needs patient but raises the risk of discharging a high-needs patient. This additional risk, however, did not appear to translate into an increase risk of 30-day revisits and readmissions. Hence, our findings indicate that keeping abdominal pain patients for at least 2 hours (with diminishing returns beyond 2 hours) may help to reduce admission rates without raising the risk of revisits or readmissions, but may do so at the expense of more severe patients being incorrectly discharged. 

We identified similar trends for different patient and operational contexts: admission rates are dramatically reduced by treatment times of at least 1 hour with minimal impact on revisits or readmission, regardless of the patient group or level of ED congestion or APP workload.  Individuals with high admission rates overall (e.g., older, obese, with comborbid conditions) were also associated with high admission rates from short treatment times, suggesting providers may be erring more on a side of admission for these individuals when rushed to make a decision. Further, individuals who were obese were possibly at greatest risk of experiencing an increase in revisits ($\sim$2\%), though not readmissions, from longer treatments. Overall, individuals who were obese had a high risk for revisits ($\sim$30\%) and readmissions ($\sim$15\%) compared to other patient groups ($\sim$16\% and $\sim$6\%, respectively).

Estimates described above reflect a hypothetical scenario when all patients are kept the same time at the ED. We also contributed average potential response curves for shift interventions \citep{sani2020identification}, which reflect a hypothetical scenario when the realized treatment time is shifted by a prespecified amount for all patients. We estimate that keeping patients for at least 30 minutes less increased admission rates by 2\%, but keeping them for at least 30 minutes more can lower admission rates by 1\% and increase revisits and readmissions rates by only about 0.2\%. Shift interventions may be easier to translate into practice, because it still allows for variable treatment times among patients.

Our model-based approach may generalize beyond abdominal pain patients in the ED to other complaints or other healthcare settings. For example, empirical evidence is also needed for determining how long to keep patients presenting with chest pain or syncope, both are which are common ED chief complaints and associated with variable admission rates \citep{sabbatini2014reducing}. Empirical evidence is also needed for determining how long to keep patients in an in-patient unit or in post-acute care \citep{bartel2020should}.

It is important to scrutinize assumptions when applying causal inference methods. Sensitivity analysis described in the Appendix found that violations to a assumption about conditional independence of potential outcomes, which is key for identifiability, did not alter our conclusions. Alternative models were also investigated in the Appendix that did not account for either unmeasured confounding or for a joint model of treatment time and the admission decision. These alternative models yield estimates that more closely match the unadjusted estimates compared to the model presented in the main text. Further, when treatment time and admission decision were not modelled jointly, we found that adjusting for unmeasured confounding had minimal impact on estimated effects. So while estimates are sensitive to modeling assumptions, only the model in the main text did any meaningful adjustment for unmeasured confounding.

There are several limitations to consider, which may serve as a basis for future work. Since estimates are sensitive to the choice of the model, additional models may need to be tested.  Further, we have focused on abdominal pain patients, but other chief complaints, such as chest pain and syncope, may also be of interest. Last, outside environmental factors not observed in the EHR may have influenced ED treatment and admissions. These include changes in medical technology for imaging or staffing in radiology.  

In short, finding fewer admissions from longer ED treatment times without negatively impacting revisits or readmissions may support the literature on the adoption of protocols that prolong treatment times for patients presenting with abdominal pain, including observation protocols and units.  

\section{Acknowledgements}\label{sec:acknowledge}

We thank the HIP \& BerbeeWalsh Department of Emergency Medicine Collaborative Working Group at UW for their help with data acquisition and management, and their input into several aspects of this study.

\newpage
\bibliographystyle{unsrtnat}
\bibliography{references} 
\newpage

\appendix
\section{Appendix}

\subsection{Proof of proposition~\ref{thm:identifiability}} \label{appendix:proof}

Before we prove Proposition~\ref{thm:identifiability}, we introduce causal inference in a more formal setting. We use the generalized version of the Pearl's do-calculus framework presented in \cite{richardson2013single} called NPSEM-IE, also defined in \cite{malinsky2019potential}, which is compatible with the classical Neyman-Rubin framework. Suppose that we have a set of variables $\mathcal{V} = \{V_1, \dots, V_n\}$, and a set of functions $\mathcal{F} = \{f: \mathfrak{X}_{V_i} \mapsto \mathfrak{X}_{V_j}\}$, where $\mathfrak{X}_{V_i}$ represents the support of $V_i$. We assume that $(\mathcal{V}, \mathcal{F})$ can be seen as a \textit{Directed Acyclic Graph} (DAG), where $\mathcal{V}$ are the nodes and $\mathcal{F}$ are the arrows between the nodes. For example, the Figure \ref{fig:dag1} represents a DAG. In particular, we are assuming no cycles, in the sense that if $f: \mathfrak{X}_{V_i} \mapsto \mathfrak{X}_{V_j}$ in $\mathcal{F}$, then there doesn't exist maps $f_1:\mathfrak{X}_{V_j}\mapsto \mathfrak{X}_{V_{k_1}}, f_2:\mathfrak{X}_{V_{k_1}}\mapsto \mathfrak{X}_{V_{k_2}},\dots, f_m:\mathfrak{X}_{V_{k_m}}\mapsto \mathfrak{X}_{V_i}$ in $\mathcal{F}$, such that $f_m\circ \dots \circ f_1$ is in $\mathcal{F}$.

Given a causal diagram $\mathcal{G} = (\mathcal{V}, \mathcal{F})$, denote the set of parents of a variable $V\in \mathcal{V}$ as $\text{pa}_{\mathcal{G}, V} $, or $ \text{pa}_{\mathcal{G}}(V)$. We assume the existence of atomic counterfactual variables $V(\text{pa})$ where $\text{pa}$ is a realization of the random variable $\text{pa}_{\mathcal{G}}(V)$. Given a set $A\subseteq \mathcal{V}$ of variables, and a realization $a$ of $A$, we define more general counterfactuals using recursion $V(a) = V(a\cap{\text{pa}_{\mathcal{G}}(V)}, \{V^*(a)| V^*\in \text{pa}_{\mathcal{G}}(V)\setminus A\})$. Intervening by artificially making $A=a$, generates a new causal diagram $\mathcal{G}(a) = (\mathcal{V}(a), \mathcal{F}(a))$, where the occurrences of the variable $A$ in $\mathcal{F}$, are replaced by $a$.

We work with the generalized version of Pearl's do-calculus (\cite{pearl2009causal}), which asserts that the following sets of variables are mutually independent:
\begin{align*}
    \{V_1(a_1)|a_1\in\text{pa}_\mathcal{G}(V_1)\},\dots,\{V_n(a_n)|a_n\in\text{pa}_\mathcal{G}(V_n)\}.
\end{align*}
Other model with less restrictions is presented in \cite{richardson2013single}. To measure those variables in terms of estimable probabilities, we use the \textit{factorization} and \textit{consistency} properties. We say that a distribution $\PR(\mathcal{V}(a))$ factorizes with respect to a DAG $\mathcal{G}$, if $\PR(\mathcal{V}(a)) = \prod_{V\in \mathcal{V}} \PR(V(a)|\text{pa}_\mathcal{G, V}(a)\setminus A(a))$. Consistency is a consequence of the above definitions, and allows us to express probabilities of the counterfactuals in terms of the variables $\mathcal{V}$. Precisely, we have that, for any $V\in\mathcal{V}$
\begin{align*}
    \PR(V(a)|a \cap {\text{pa}_\mathcal{G}(V)}, \text{pa}_{\mathcal{G}(a)}(V(a))\setminus a) = \PR(V|a \cap {\text{pa}_\mathcal{G}(V)}, \text{pa}_\mathcal{G}(V)\setminus A).
\end{align*}
For example, if from the diagram $\mathcal{G}$ we can prove that $Y(a)\ind A|X$, then, by the consistency property we can conclude that $\PR(Y(a)|X) = \PR(Y(a)|A=a, X) = \PR(Y|A=a, X)$. 

We say that a counterfactual probability $\PR(V(a))$ is non-parametrically \textit{identifiable}, if it can be expressed uniquely in terms of estimable quantities using e.g. the properties above. It's difficult to prove non-identifiability, since it involves proving that there doesn't exist a unique formula, expressed in terms of the observed variables, that is equal to the desired quantity. \citet{shpitser2008complete} proved the following result, which allows us to disprove identifiability of the desired effects based on our conceptual understanding of the model as depicted in the DAG~\ref{fig:dag1}. It was originally developed for the do-calculus, but \cite{malinsky2019potential} proved that the axioms of do-calculus are still valid in our framework.

\begin{thm}[Shpitser and Pearl 2008, Theorem 12]\label{disproveid}
If all paths from $X$ to $Y$ in the graph $G$ are confounded by unobserved confounders, then, the causal effect from $X$ to $Y$ cannot be identified in the graph $G$.
\end{thm}

We are now ready to prove Proposition~\ref{thm:identifiability}.

\textbf{Proof of Proposition~\ref{thm:identifiability}.}
If $H$ is unobserved, then all the paths from $T$ to $Y$ and all paths from $A$ to $Y$ are confounded by unobserved confounders.  This, together with Theorem~\ref{disproveid}, imply that the DAG in Fig~\ref{fig:dag1} is not sufficient to identify the effect of $T$ on $Y$ and the effect of $T$ on $A$.

On the other hand, if $H$ is observed, then we can get explicit expressions for $\beta(t)$ and $\gamma(t)$. By the factorization property applied to the DAG of Figure \ref{fig:dag1}, we have that
\begin{align*}
    \PR(Y(t),A(t)|Z,H,X) = \PR(Y(t)|A(t),Z,H,X)\PR(A(t)|Z,H,X),
\end{align*}
and then, by the independence of $A(t)$ and $T$ conditional on $Z$, $H$, and $X$, and by consistency, we have that
\begin{align*}
    \PR(A(t)=a|Z, H, X) &= \PR(A(t)=a\,|T=t,\, Z, H, X)\\
    &= \PR(A=a\,|T=t\, Z, H, X).
\end{align*}
Similarly, by the independence of $Y$ and $T$ conditional on $A, Z, H, X$, we have that
\begin{align*}
    \PR(Y(t)|Z, H, X) &= \sum_a\PR(Y(t)|A(t)=a,Z,H,X)\PR(A(t)=a|Z,H,X) \\
    &= \sum_a\PR(Y|A=a,Z,H,X)\PR(A=a|T=t,Z,H,X).
\end{align*}
Therefore, integrating by $Z, H, X$, we get
\begin{align*}
    \gamma(t) &= \E[Y(t)] = \E[\PR(Y(t)=1|Z, H, X)]\\
    &= \E\left[\sum_a\PR(Y|A=a,Z,H,X)\PR(A=a|T=t,Z,H,X)\right]
\end{align*}
and,
\begin{align*}
    \theta(t) &= \E[A(t)] = \E[\PR(A(t)=1|Z, H, X)]\\
    &= \E[\PR(A=1|T=t, Z, H, X)]
\end{align*}

so that $\gamma(t)$ and $\theta(t)$ are identified, as desired.
\EndPf

\subsection{Proof of proposition~\ref{thm:identifiability_shift}} \label{appendix:proof_shift}

We start by formally defining the effect we want to recover. We follow the definition of shift interventions made in \cite{sani2020identification}. Given a function $f:\mathfrak{X}_{T} \mapsto \mathfrak{X}_{T}$, we want to compute the value of the mean outcome $Y$ when $T$ attains the value $f(T)$, denoted as $Y(f(T))$. As a special case, when $f(T)$ is a constant $t$, this effect is equivalent to the interventional effect defined in the Appendix \ref{appendix:proof} and can thus be viewed as a refinement of the effect considered in Appendix~\ref{appendix:proof}. Similar to the interventional case, the value of $Y(f(T))$ might not be observed unless $f(T) = T$, and we require additional assumptions to identify the average effect. More formally, we define the counterfactual outcome $Y(f(T))$ recursively as
\begin{align*}
Y(f(T)) = Y(\{Y^*=f(Y^*)| Y^*\in \text{pa}_{\mathcal{G}}(Y)\cap T\}, \{Y^*(f(T))| Y^*\in \text{pa}_{\mathcal{G}}(Y)\setminus T\}).       
\end{align*}

The same properties described in the Appendix \ref{appendix:proof}, like consistency and factorization are valid with this definition. To evaluate the effects of the policy $f$, we compare the relative difference of the outcome affected by the policy $f$ and the observed outcome $\E_Y[Y(f)-Y]=\E_{Y, T}[Y(f(T))]-\E[Y]$. We can now prove the Proposition \ref{thm:identifiability_shift}.

\textbf{Proof of Proposition~\ref{thm:identifiability_shift}.}
    First note that all the paths from $T$ to $A$ are confounded by $H$, which means that $A(T=f(T)) = A(T=f(T), H)$ varies with $H$. If $H$ is unobserved, then $P(A(T=f(T)))$ can not be uniquely expressed given a nonparametric probabilistic model of the joint distribution of the observed variables. Therefore, the effect is not nonparametrically identifiable.
    
    By the other hand, if $H$ is observed, we have that the joint distribution of the outcome, admission decision, and treatment time in the graph $\mathcal{G}_{f(T)}$, can be decomposed using the factorization property
    \begin{align*}
        \PR(Y(f(T)), A(f(T)), T|Z, H, X) &= \PR(Y(f(T))|A(f(T)),Z, H, X)\PR(A(f(T))|Z,H,X) \PR(T|Z,H,X).
    \end{align*}
    Integrating by $Y, T$, using the independence of $A(f(T))$ and $T$ conditional on $Z, H, X$, and using consistency, we get 
    \begin{align*}
        \E[A(f)|Z,H,X] &= \E_{T}\left[\PR(A(f(T))|Z, H, X)\right]\\ 
         &= \E_{T}\left[\PR(A(f(T))|T=f(T), Z, H, X)\right|Z,H,X]\\
         &= \E_{T}\left[  \right|Z,H,X].
    \end{align*}
    Also, integrating by $A, T$, by the independence of $Y$ and $T$ conditional on $Z, H, X$, and by consistency, we get
    \begin{align*}
        \E[Y(f)|Z,H,X] &= \E_{T}\left[\PR(Y(f(T))|Z, H, X)\right]\\
        &= \E_{T}\left[\sum_A \PR(Y|A, Z, H, X)\PR(A|T=f(T), Z, H, X)\bigg|Z,H,X\right].
    \end{align*}
    Finally, we estimate the effects on the admission decision and outcomes, given a function $f$,
    \begin{align*}
        \hat{\theta}(f) = \E[\E[A(f)|Z,H,X]],
    \end{align*}
    and
    \begin{align*}
        \hat{\gamma}(f) = \E[\E[Y(f)|Z,H,X]],
    \end{align*}
\EndPf

\subsection{Internal validity of model} \label{appendix:validity}

To support estimated average response curves, we evaluated internal validity of the fitted model by assessing its ability to recover certain demographic and clinical trends in empirical, or unadjusted, estimates of admission rates and average treatment times. Model-based estimates are marginalized over the latent state $H$ when compared to empirical estimates. Fitted model parameters can be found in Table~\ref{tab:brown_rev_params} when 30-day revisit is the outcome of interest and in Table~\ref{tab:brown_readm_params} when 30-day readmission is the outcome in Appendix~\ref{appendix:parameter_estimates}. Fitted parameters for the outcome 30-day revisit are similar to fitted parameters for 30-day readmission, except for the parameters specifying expected outcomes. Thus, we report model-based estimates associated with 30-day revisits, except when examining expected outcomes. 

\begin{figure}[t]
\centering
\begin{tikzpicture}
\node at (-5.25,0)
{\includegraphics[width=0.32\textwidth]{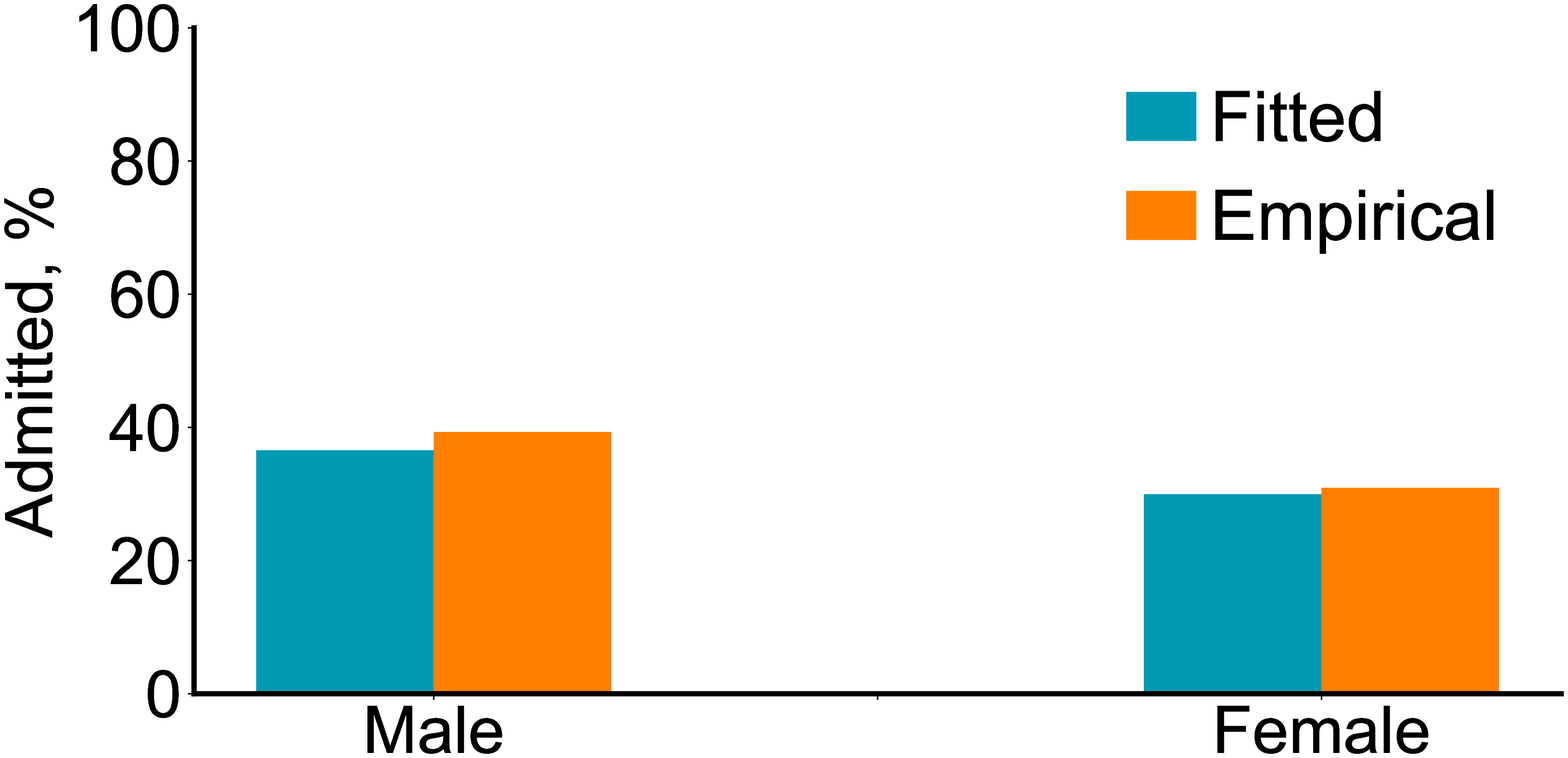}};
\node at (0,0) 
{\includegraphics[width=0.32\textwidth]{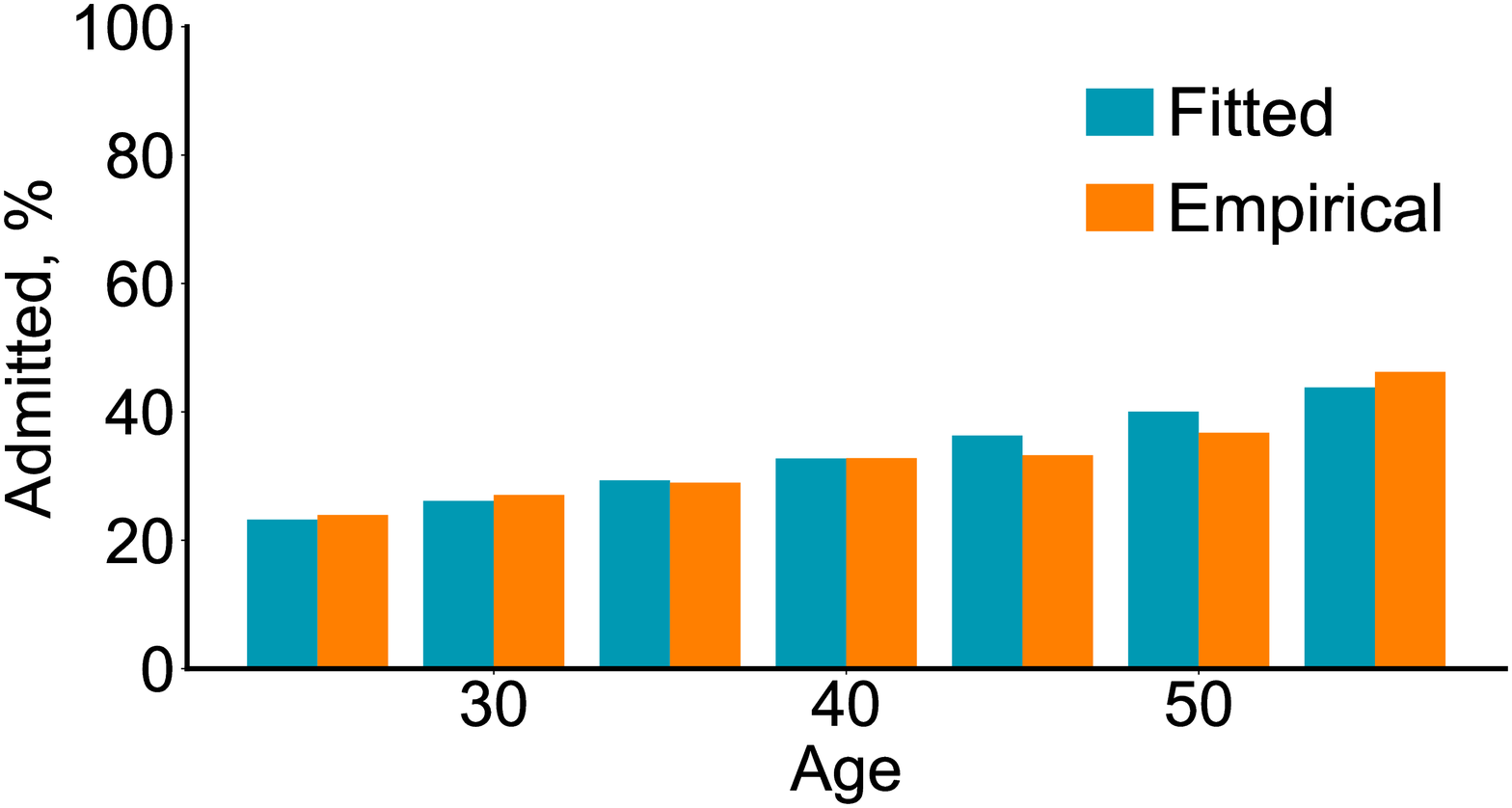}};
\node at (5.25,0)
{\includegraphics[width=0.32\textwidth]{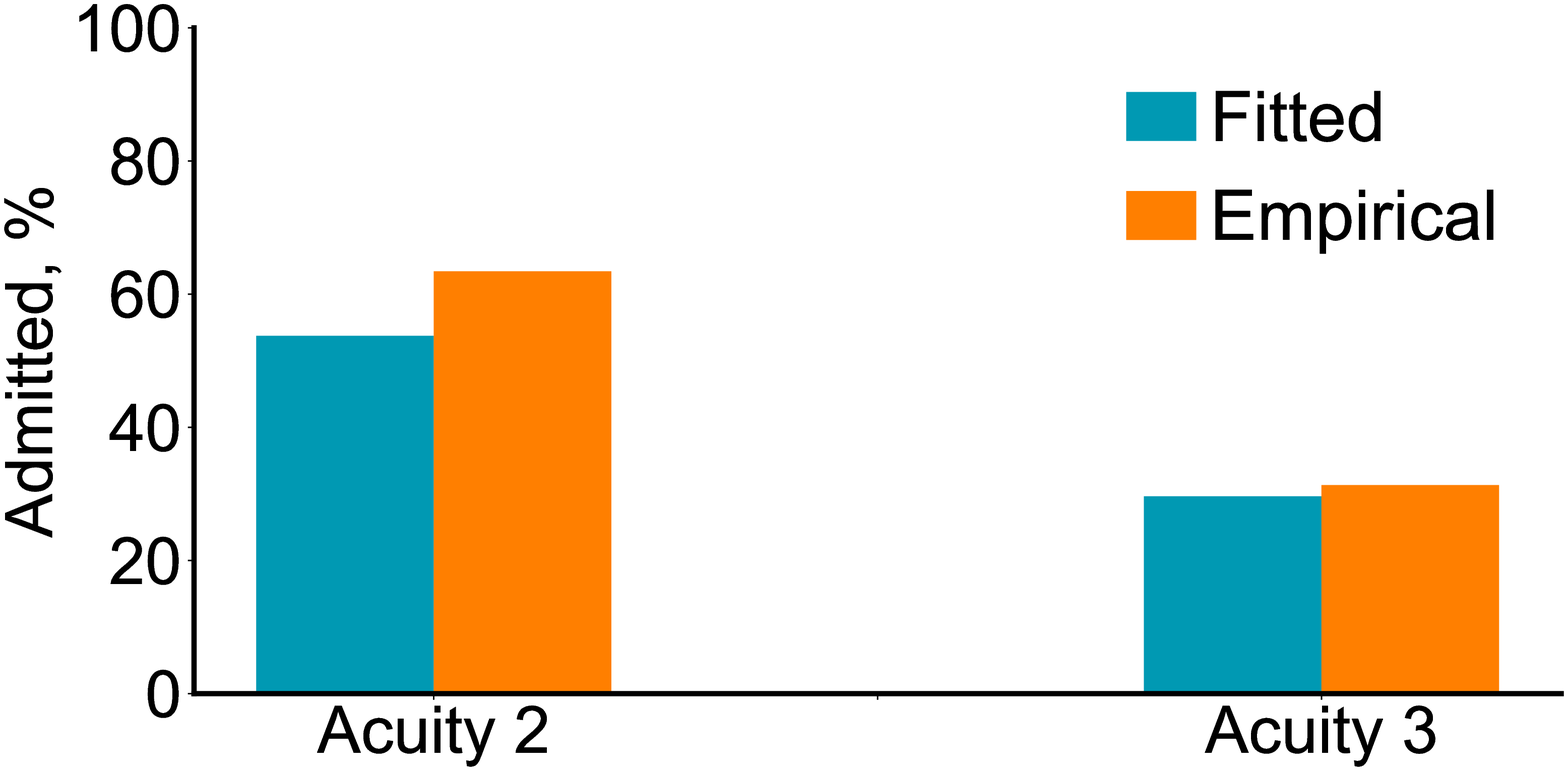}};
\node at (-7.7,1.55) {\large \textbf{\textsf{A}}};
\node at (-2.45,1.55) {\large \textbf{\textsf{B}}};
\node at (2.8,1.55) {\large \textbf{\textsf{C}}};
\end{tikzpicture}
\caption{Admissions rates by (A) gender, (B) age, and (C) acuity based on unadjusted estimates or model-based estimates.}
\label{fig:prelim-emp}
\end{figure}

We find that admissions rates among patients presenting with abdominal pain are higher for men than women, a trend which is replicated when estimating admission rates using the model (Figure~\ref{fig:prelim-emp}A). Similarly, admission rates increases with age (Figure~\ref{fig:prelim-emp}B) and, with lower acuity (Figure~\ref{fig:prelim-emp}C). Both trends are replicated when estimating rates using the model. 

\begin{figure}[htb]
\centering
\begin{tikzpicture}
\node at (-5,0)
{\includegraphics[width=0.31\textwidth]{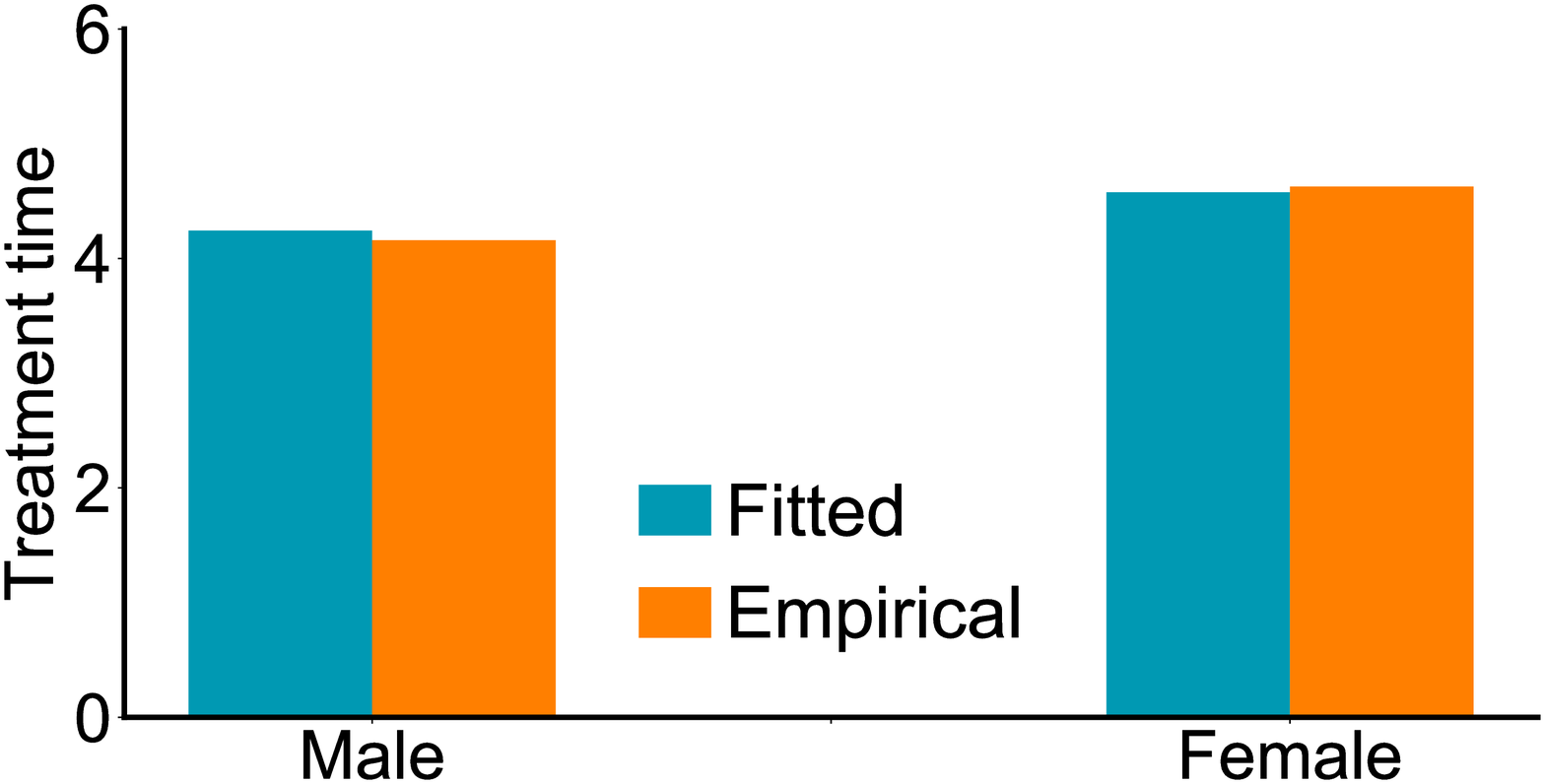}};
\node at (0,0) 
{\includegraphics[width=0.31\textwidth]{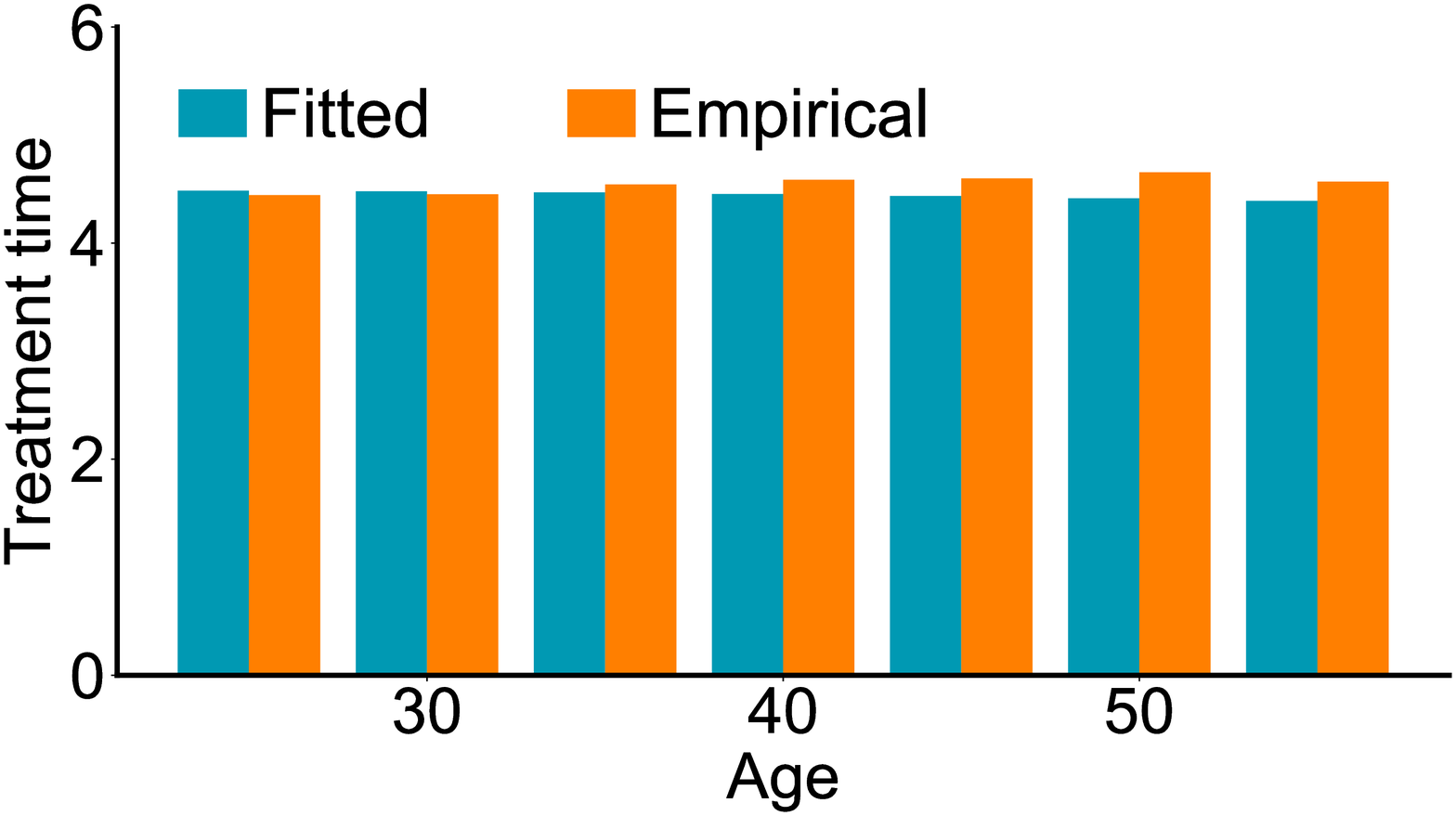}};
\node at (5,0)
{\includegraphics[width=0.31\textwidth]{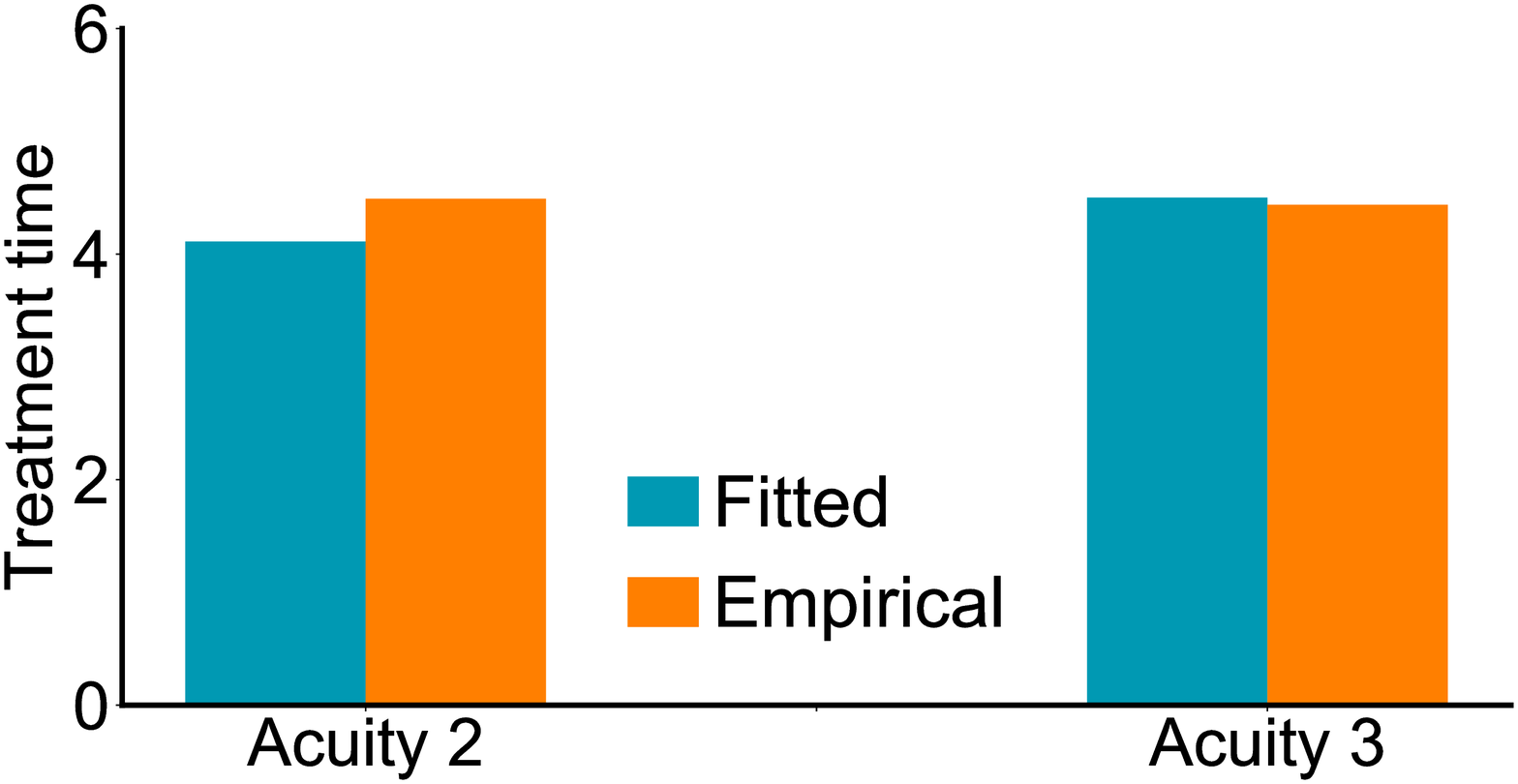}};
\node at (-5,-3) {\includegraphics[width=0.31\textwidth]{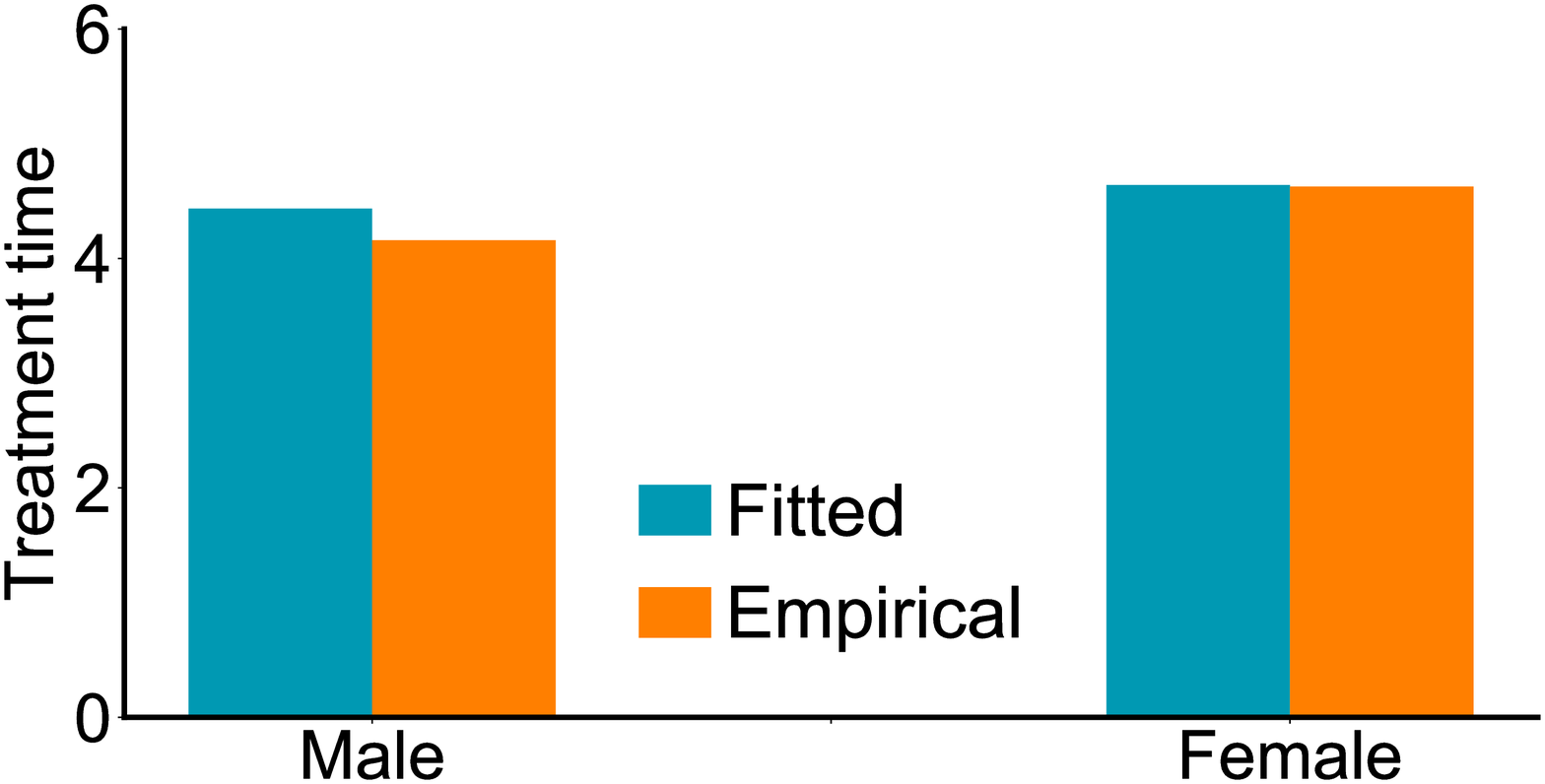}};
\node at (0,-3) {\includegraphics[width=0.31\textwidth]{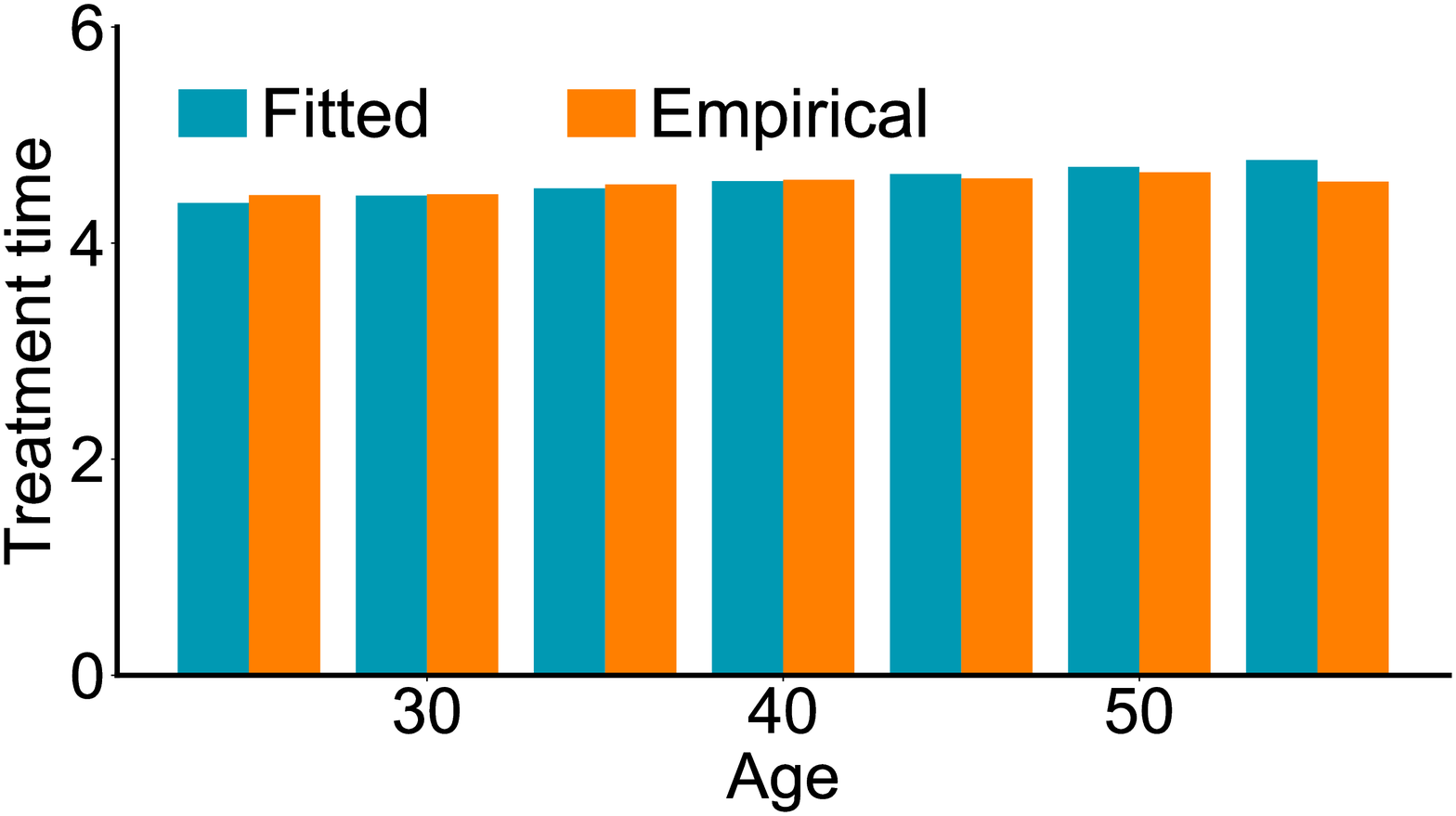}};
\node at (5,-3) {\includegraphics[width=0.31\textwidth]{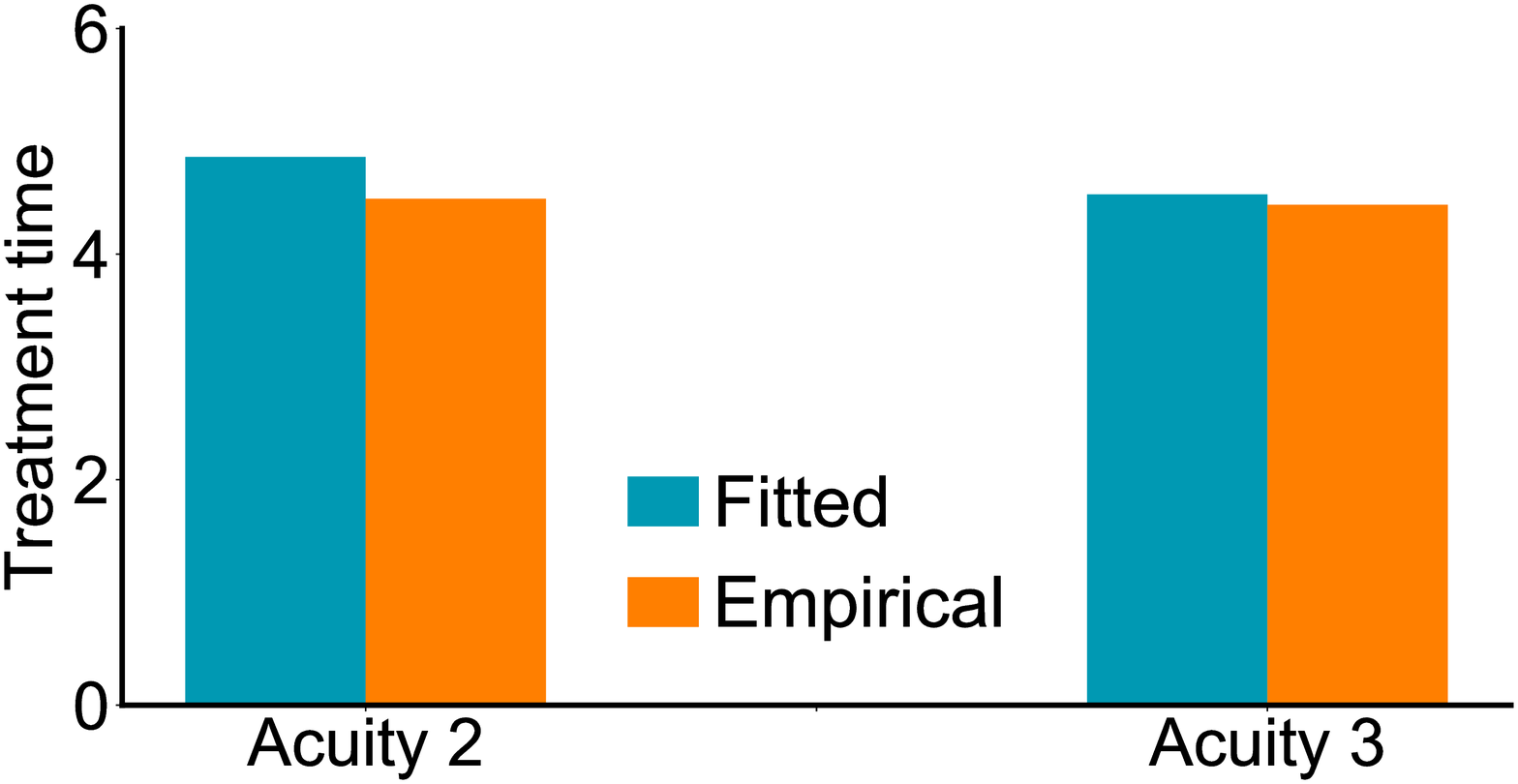}};
\node[rotate=90] at (-7.9,0.2) { \textbf{\textsf{Admitted}}};
\node[rotate=90] at (-7.9,-2.8) { \textbf{\textsf{Discharged}}};
\node at (-7.45,1.7) {\large \textbf{\textsf{A}}};
\node at (-2.20,1.7) {\large \textbf{\textsf{B}}};
\node at (2.55,1.7) {\large \textbf{\textsf{C}}};
\end{tikzpicture}
\caption{Average treatment times by admission decision and by (A) gender, (B) age, and (C) acuity based on unadjusted estimates and model-based estimates}
\label{fig:prelim-diffusion}
\end{figure}

Women have slightly longer average treatment times than men among admitted and discharged patients (Figure~\ref{fig:prelim-emp}A). Model-based estimates also show slightly longer average treatment times for women. Meanwhile, unadjusted and model-based estimates of average treatment times show a slight increase with age among admitted patients, but stay relatively consistent with age among discharged patients (Figure~\ref{fig:prelim-diffusion}B). For discharged patients, average treatment times slightly decreases with acuity (Figure~\ref{fig:prelim-diffusion}C). However, for admitted patients average treatment times slightly increases with acuity, suggesting that for these patients (i.e. those with ESI level 3) there is perhaps a higher degree of uncertainty related to their care needs. Here again, trends in model-based estimates by acuity closely match the unadjusted estimates. 

\begin{figure}[htb]
\centering
\begin{tikzpicture}
\node at (-8,0)
{\includegraphics[width=0.49\textwidth]{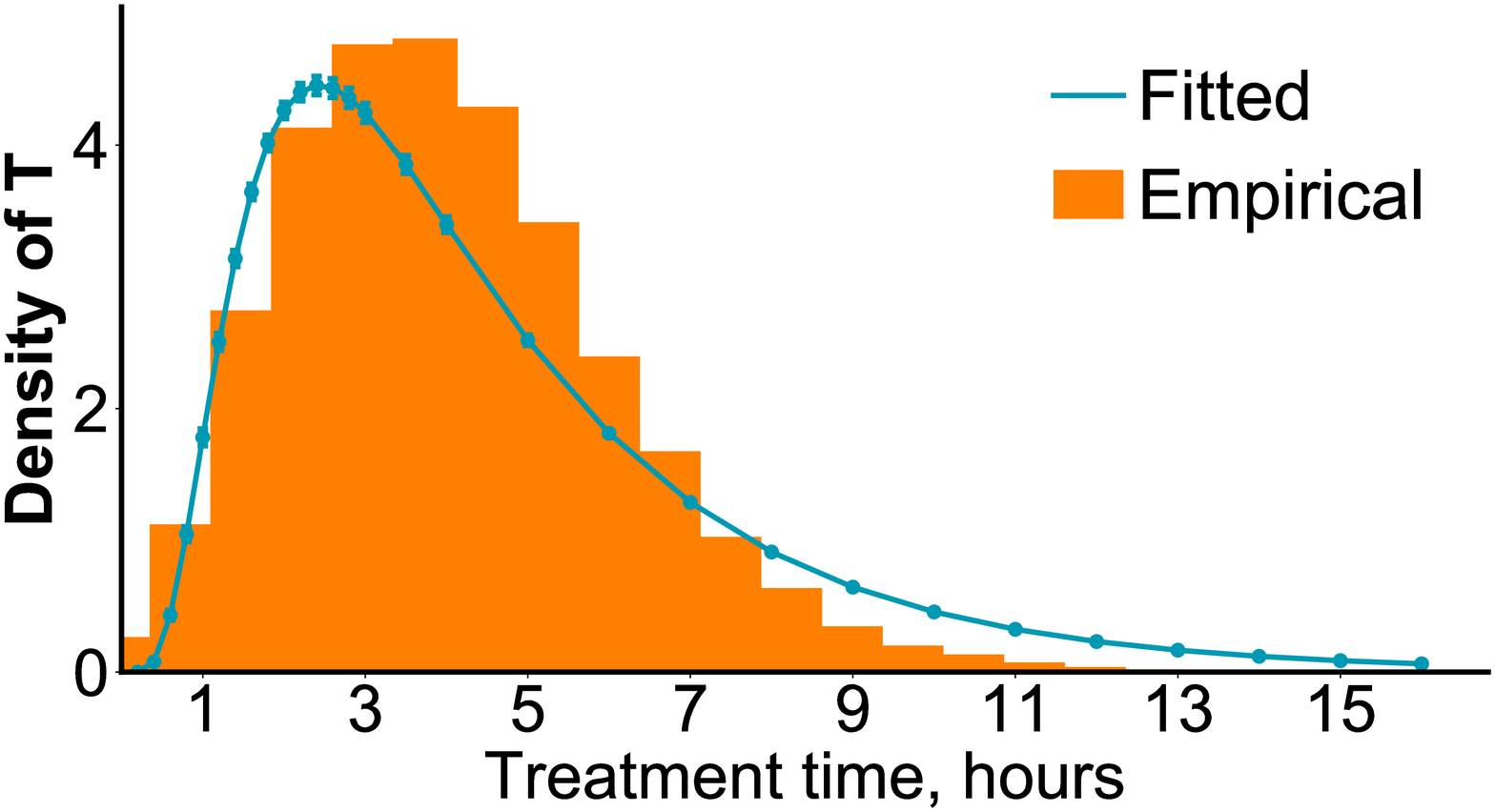}};
\node at (0,0) {\includegraphics[width=0.49\textwidth]{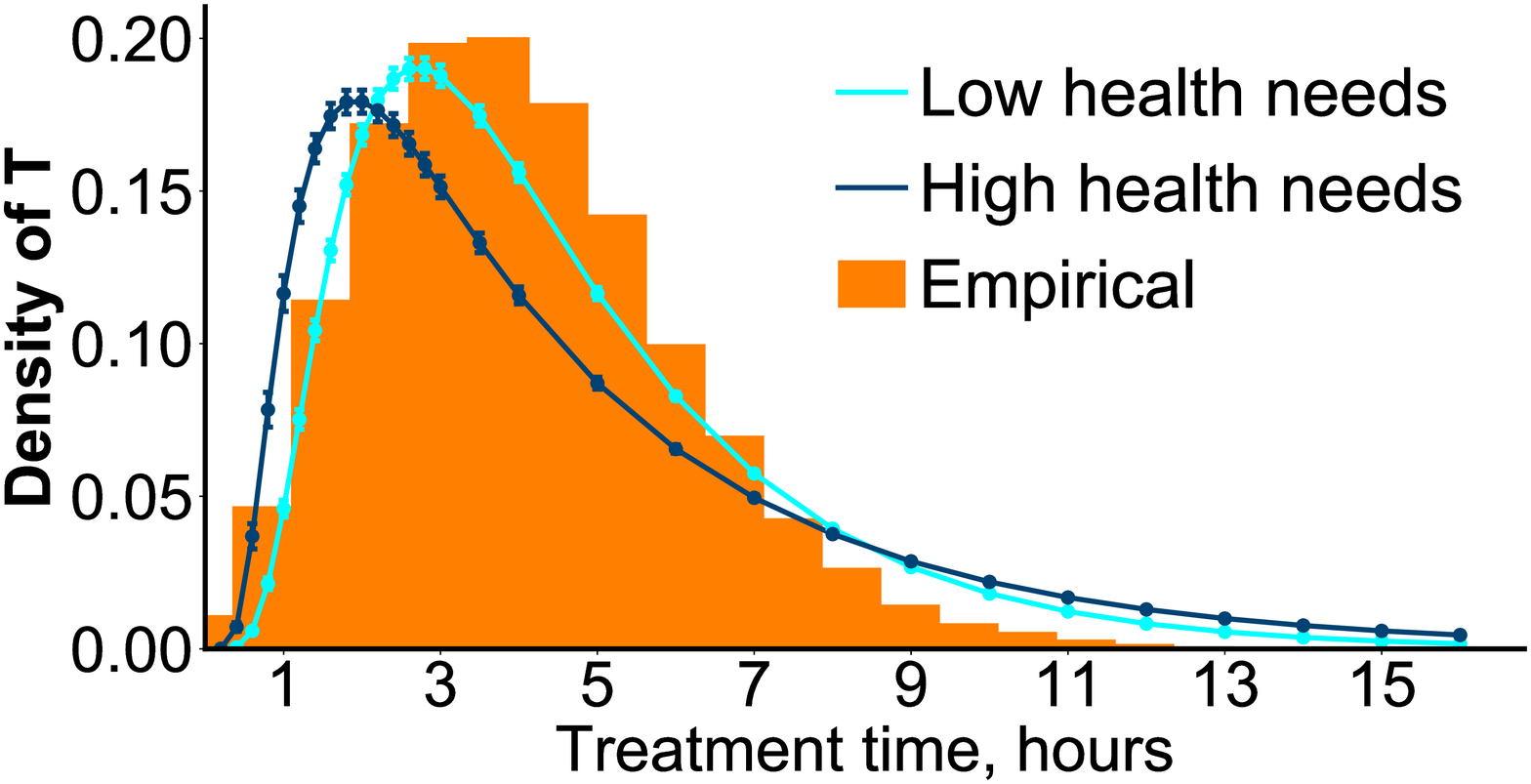}};
\node at (-11.75,2.5) {\large \textsf{\textbf{A}}};
\node at (-3.75,2.5) {\large \textsf{\textbf{B}}};
\end{tikzpicture}

\caption{Distribution of treatment time (A) overall and (B) by health needs based on unadjusted estimates and model-based (i.e. fitted) estimates.}
\label{fig:brown_los}
\end{figure}

As a final validity check, we compared the empirical distribution of ED treatment time against the fitted distribution of ED treatment time (Figure~\ref{fig:brown_los}A). We find that the fitted distribution has a similar shape to the empirical / unadjusted distribution, thought slightly larger in its right tail. Interestingly, splitting the fitted distribution by health needs shows that care providers spend more time in the ED with individuals in a lower health state ($H=0$) than a higher health state ($H=1$; Figure~\ref{fig:brown_los}B). Together, these comparisons show a degree of consistency in trends and distributional shape between fitted and unadjusted estimates, providing visual support for the internal validity of the model.   


\subsection{Sensitivity analyses} \label{appendix:sensitivitydetails}

\subsubsection{Violating a key identifiability assumption}
One of the key identifiability assumptions of the DAG depicted in Figure~\ref{fig:dag1} is that the potential response $Y(a)$ is independent of the admission decision process $(A,T)$ conditional on $H$, $Z$, and $Z$. We can systematically violate this assumption to test the impact of this particular assumption on 30-day risks of revisit and readmission. To this end, we note that the probability density function of $Y(a)$, $A$, $T$ given $H$, $X$, $Z$ decomposes as
\begin{align*}
\left[ f\left(A=1,T|H,X,Z,Y(1)\right) \PR\left(Y(1) | H, X, Z\right)\right]^A \left[ f\left(A=0,T|H,X,Z,Y(0)\right) \PR\left(Y(0) | H,X,Z\right)\right]^{1-A},
\end{align*}
where (with an abuse of notation) $f$ denotes any density function. Our identifiability assumption allows us to drop the variables $Y(a)$ from the density functions $f$, since $(A,T)$ is not influenced by $Y(a)$. If we want to violate this assumption, however, we can assume:
\begin{align*}
f\left(A,T|H,X,Z,Y(1)\right) &\propto \psi_1^{A Y(1)} f\left(A,T|H,X,Z\right); \\
f\left(A,T|H,X,Z,Y(0)\right) &\propto \psi_0^{(1-A)Y(0)} f\left(A,T|H,X,Z\right)
\end{align*}
for some scaling factors $\psi_1$ and $\psi_0$. The identifiability assumption can be violated by choosing values for $\psi_1$ and $\psi_0$ that are not both one. We thus fit the model to data for $\psi_0$ and $\psi_1$ set to either 0.95, 0.975, 1, 1.025, or 1.05 in a factorial design, resulting in 25 comparisons. 

Upon violating the identifiability assumption, estimates of the average potential response on 30-day revisits ranged from 16.2\% to 16.6\% for a treatment time of 3 hours, compared to 16.5\% if the identifiable assumption is satisfied  (Table~\ref{tab:sensitivity_revisits}). Similarly, estimated revisit rates ranged from 15.9\% to 16.5\% for a treatment time of 2 hours and from 15.5\% to 16.5\% for a treatment time of 1 hour, compared to 16.5\% and 16.4\%, respectively, if the identifiable assumption is satisfied.

\begin{table}[h!]
\centering 
\begin{tabular}{l l c c c c }
\toprule
&& \multicolumn{3}{c}{\textbf{Intervention treatment time $t$, hours}} \\
\cmidrule(lr){3-6}
    \multicolumn{1}{c}{$\psi_1$} & \multicolumn{1}{c}{$\psi_2$} & \multicolumn{1}{c}{\textbf{1/2}} & \multicolumn{1}{c}{\textbf{1}} & \multicolumn{1}{c}{\textbf{2}} & \multicolumn{1}{c}{\textbf{3}} \\
    \midrule
    0.95 & 0.95 & 7.1 (4.7,9.6) & 5.9 (5.5,6.4) & 6.1 (5.8,6.3) & 6.3 (6.0,6.6) \\ 
     & 0.975 & 6.5 (5.3,7.7) & 6.0 (5.7,6.4) & 6.2 (5.9,6.5) & 6.4 (6.1,6.6) \\ 
     & 1.0 & 5.6 (4.8,6.5) & 5.9 (5.6,6.2) & 6.2 (5.9,6.4) & 6.3 (6.0,6.6) \\ 
     & 1.025 & 5.5 (4.6,6.3) & 5.9 (5.6,6.2) & 6.2 (5.9,6.5) & 6.4 (6.1,6.6) \\ 
     & 1.05 & 8.0 (6.7,9.2) & 6.6 (6.3,7.0) & 6.4 (6.1,6.7) & 6.3 (6.1,6.6) \\ 
    0.975 & 0.95 & 4.9 (4.6,5.2) & 5.8 (5.5,6.1) & 6.2 (5.9,6.4) & 6.3 (6.0,6.6) \\ 
     & 0.975 & 7.4 (6.2,8.6) & 6.0 (5.7,6.3) & 6.1 (5.8,6.4) & 6.3 (6.0,6.6) \\ 
     & 1.0 & 5.0 (4.5,5.6) & 5.9 (5.6,6.1) & 6.2 (5.9,6.5) & 6.3 (6.1,6.6) \\ 
     & 1.025 & 6.5 (5.3,7.6) & 6.0 (5.6,6.3) & 6.1 (5.9,6.4) & 6.3 (6.0,6.6) \\ 
     & 1.05 & 4.5 (4.3,4.8) & 5.8 (5.5,6.1) & 6.2 (5.9,6.5) & 6.4 (6.1,6.6) \\ 
    1.0 & 0.95 & {7.2 (6.0,8.3)} & {6.5 (6.1,6.8)} & {6.3 (6.1,6.6)} & {6.3 (6.0,6.6)} \\ 
     & 0.975 & 6.8 (5.7,7.9) & 6.0 (5.6,6.3) & 6.1 (5.9,6.4) & 6.3 (6.1,6.6) \\ 
     & 1.0 & \textbf{6.0 (5.4,6.7)} & \textbf{6.0 (5.7,6.3)} & \textbf{6.2 (5.9,6.5)} & \textbf{6.3 (6.1,6.6)} \\ 
     & 1.025 & 5.5 (4.8,6.2) & 5.8 (5.6,6.1) & 6.1 (5.9,6.4) & 6.3 (6.0,6.6) \\ 
     & 1.05 & 6.6 (5.7,7.6) & 6.0 (5.7,6.3) & 6.1 (5.9,6.4) & 6.3 (6.0,6.6) \\ 
    1.025 & 0.95 & 6.9 (4.9,9.0) & 6.0 (5.5,6.4) & 6.1 (5.8,6.4) & 6.3 (6.0,6.6) \\ 
     & 0.975 & 5.8 (5.0,6.7) & 6.0 (5.7,6.3) & 6.2 (6.0,6.5) & 6.4 (6.1,6.6) \\ 
     & 1.0 & 5.1 (4.7,5.5) & 5.9 (5.6,6.2) & 6.2 (6.0,6.5) & 6.4 (6.1,6.6) \\ 
     & 1.025 & 5.0 (4.7,5.4) & 5.9 (5.6,6.1) & 6.2 (6.0,6.5) & 6.4 (6.1,6.6) \\ 
     & 1.05 & 7.3 (5.4,9.2) & 6.5 (6.1,6.9) & 6.4 (6.1,6.6) & 6.3 (6.0,6.6) \\ 
    1.05 & 0.95 & 6.9 (5.7,8.1) & 6.0 (5.6,6.3) & 6.2 (5.9,6.5) & 6.4 (6.1,6.7) \\ 
     & 0.975 & 7.8 (6.4,9.1) & 6.0 (5.7,6.4) & 6.1 (5.8,6.3) & 6.3 (6.0,6.6) \\ 
     & 1.0 & 8.5 (4.5,12.5) & 6.8 (5.9,7.6) & 6.4 (6.1,6.7) & 6.3 (6.0,6.6) \\ 
     & 1.025 & 7.5 (6.0,9.1) & 6.6 (6.2,7.0) & 6.4 (6.1,6.7) & 6.3 (6.1,6.6) \\ 
     & 1.05 & 7.4 (5.5,9.2) & 6.6 (6.1,7.0) & 6.4 (6.1,6.7) & 6.3 (6.1,6.6)\\
\bottomrule
\end{tabular}
\caption{Sensitivity of estimated average potential responses of treatment time on 30-day readmissions with respect to violations in one of the key identifiability assumptions. Estimates presented in main text are bolded.} \label{tab:sensitivity_readmissions}
\end{table}

For the average potential response on 30-day readmissions, estimates ranged from 6.3\% to 6.4\% for a treatment time of 3 hours, compared to 6.3\% if the identifiable assumption is satisfied (Table~\ref{tab:sensitivity_readmissions}). Similarly, estimated readmission rates ranged from 6.1\% to 6.4\% for a treatment time of 2 hours and from 5.8\% to 6.6\% for a treatment time of 1 hour, compared to 6.2\% and 6.0\%, respectively, if the identifiable assumption is satisfied.

\begin{table}[h!]
\centering
\begin{tabular}{l l c c c c }
\toprule
&& \multicolumn{3}{c}{\textbf{Intervention treatment time $t$, hours}} \\
\cmidrule(lr){3-6}
    \multicolumn{1}{c}{$\psi_1$} & \multicolumn{1}{c}{$\psi_2$} & \multicolumn{1}{c}{\textbf{1/2}} & \multicolumn{1}{c}{\textbf{1}} & \multicolumn{1}{c}{\textbf{2}} & \multicolumn{1}{c}{\textbf{3}}  \\
    \midrule 
    0.95 & 0.95 & 13.9 (12.4,15.4) & 15.7 (15.2,16.1) & 16.2 (15.8,16.6) & 16.4 (16.0,16.8) \\ 
     & 0.975 & 16.6 (14.4,18.8) & 16.5 (15.9,17.1) & 16.5 (16.1,17.0) & 16.5 (16.1,16.9) \\
     & 1.0 & 16.4 (13.4,19.3) & 16.4 (15.7,17.1) & 16.5 (16.1,16.9) & 16.5 (16.1,16.9) \\
     & 1.025 & 15.9 (13.9,17.9) & 16.3 (15.8,16.9) & 16.5 (16.1,16.9) & 16.5 (16.1,16.9) \\ 
     & 1.05 & 14.1 (12.2,15.9) & 15.5 (14.9,16.0) & 16.0 (15.6,16.5) & 16.3 (15.9,16.7) \\ 
    \midrule 
    0.975 & 0.95 & 15.2 (12.5,17.9) & 15.6 (15.0,16.2) & 16.1 (15.7,16.5) & 16.4 (15.9,16.8) \\ 
     & 0.975 & 13.9 (11.1,16.7) & 15.6 (15.0,16.2) & 16.2 (15.8,16.6) & 16.4 (16.0,16.9) \\ 
     & 1.0 & 14.8 (10.3,19.4) & 15.6 (14.7,16.5) & 16.0 (15.6,16.5) & 16.3 (15.9,16.7) \\ 
     & 1.025 & 11.6 (10.7,12.4) & 15.5 (15.1,16.0) & 16.4 (16.0,16.8) & 16.6 (16.1,17.0) \\ 
     & 1.05 & 12.8 (11.4,14.3) & 15.5 (15.1,16.0) & 16.3 (15.9,16.7) & 16.5 (16.1,16.9) \\ 
    \midrule 
    1.0 & 0.95 & 14.9 (12.7,17.1) & 15.8 (15.2,16.3) & 16.2 (15.8,16.6) & 16.4 (16.0,16.8) \\ 
     & 0.975 & 16.1 (13.7,18.6) & 15.8 (15.2,16.3) & 16.1 (15.7,16.5) & 16.3 (15.9,16.7) \\ 
     & 1.0 & \textbf{16.1 (13.7,18.4)} & \textbf{16.4 (15.8,17.0)} & \textbf{16.5 (16.1,16.9)} & \textbf{16.5 (16.1,16.9)} \\ 
     & 1.025 & 15.2 (12.0,18.5) & 15.6 (15.0,16.3) & 16.0 (15.6,16.4) & 16.3 (15.8,16.7) \\ 
     & 1.05 & 13.9 (10.7,17.1) & 15.3 (14.6,16.0) & 15.9 (15.5,16.4) & 16.2 (15.8,16.6) \\ 
    \midrule 
    1.025 & 0.95 & 15.6 (11.8,19.5) & 16.3 (15.5,17.1) & 16.4 (16.0,16.9) & 16.4 (16.0,16.9) \\ 
     & 0.975 & 13.0 (12.0,14.1) & 15.7 (15.3,16.1) & 16.4 (16.0,16.8) & 16.6 (16.2,17.0) \\ 
     & 1.0 & 15.7 (13.4,18.1) & 16.4 (15.7,17.0) & 16.5 (16.1,17.0) & 16.6 (16.1,17.0) \\ 
     & 1.025 & 13.6 (11.8,15.3) & 15.6 (15.1,16.1) & 16.3 (15.9,16.7) & 16.5 (16.1,16.9) \\ 
     & 1.05 & 14.3 (12.8,15.8) & 15.7 (15.2,16.1) & 16.2 (15.8,16.6) & 16.4 (16.0,16.8) \\ 
    \midrule 
    1.05 & 0.95 & 15.0 (13.3,16.8) & 15.7 (15.2,16.2) & 16.2 (15.7,16.6) & 16.4 (16.0,16.8) \\ 
     & 0.975 & 15.9 (12.7,19.2) & 15.7 (15.0,16.4) & 16.0 (15.6,16.4) & 16.3 (15.9,16.7) \\ 
     & 1.0 & 14.3 (12.1,16.6) & 15.5 (14.9,16.1) & 16.1 (15.6,16.5) & 16.3 (15.9,16.7) \\ 
     & 1.025 & 16.2 (12.7,19.7) & 16.4 (15.5,17.2) & 16.5 (16.1,16.9) & 16.5 (16.1,16.9) \\ 
     & 1.05 & 10.5 (10.0,11.1) & 15.4 (14.9,15.8) & 16.4 (16.0,16.8) & 16.6 (16.2,17.0)\\
\bottomrule
\end{tabular}
\caption{Sensitivity of estimated average potential responses of treatment time on 30-day revisits with respect to violations in one of the key identifiability assumptions. Estimates presented in main text are bolded.}\label{tab:sensitivity_revisits}
\end{table}

\subsubsection{Removing the latent variable from the model}

\begin{figure}
\centering
\includegraphics[width=0.49\textwidth]{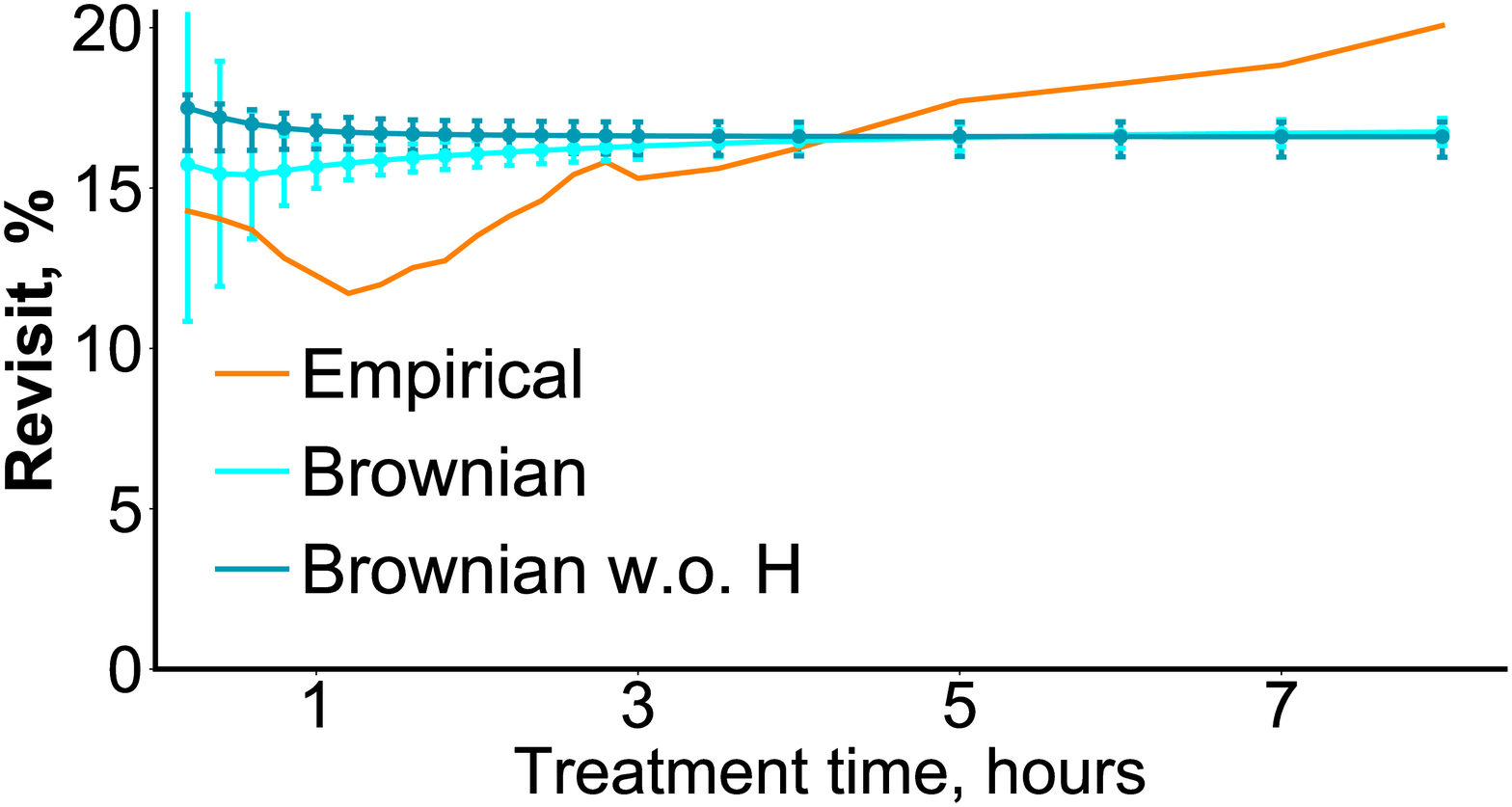}
\includegraphics[width=0.49\textwidth]{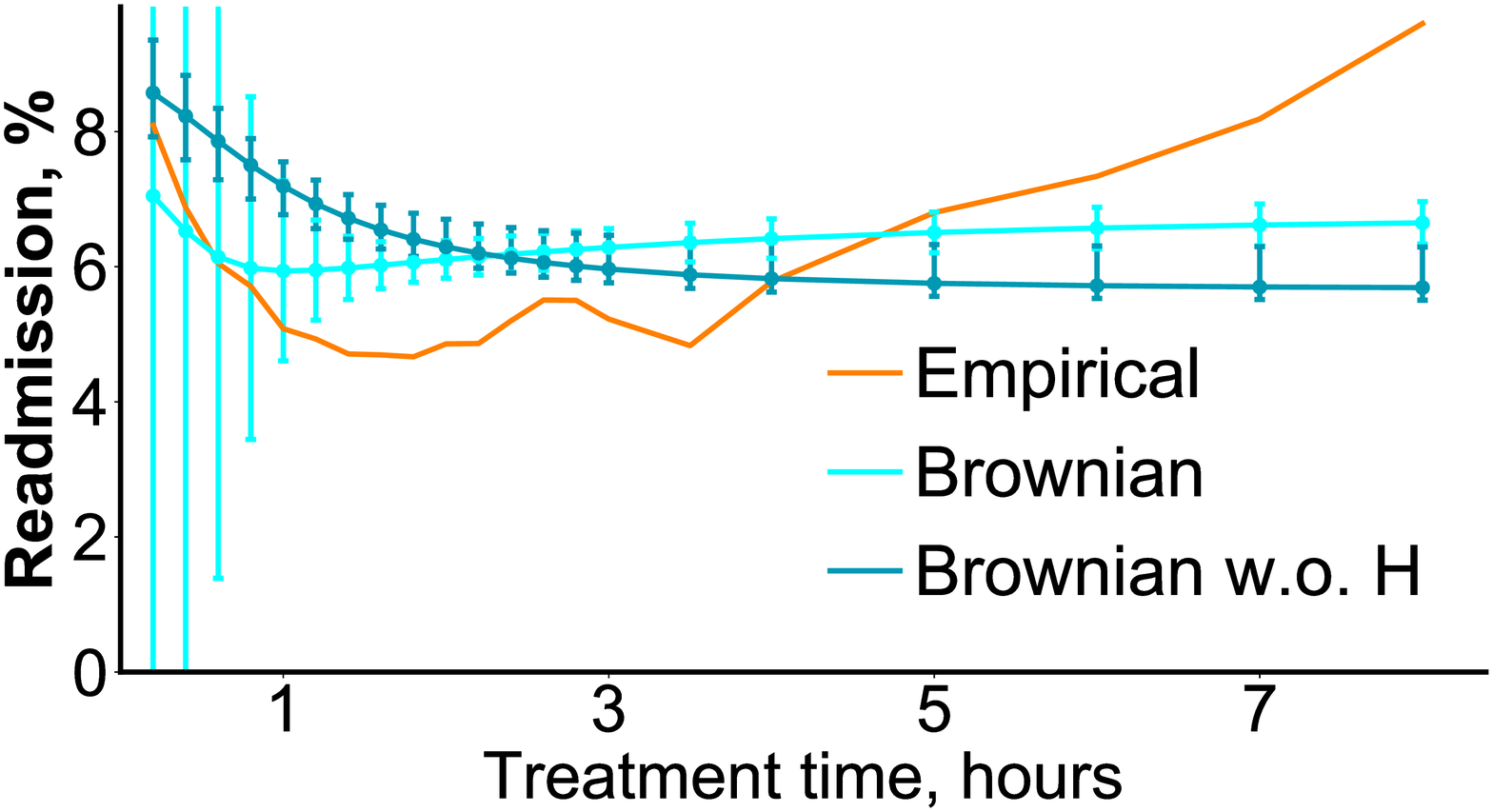} \\
\includegraphics[width=0.49\textwidth]{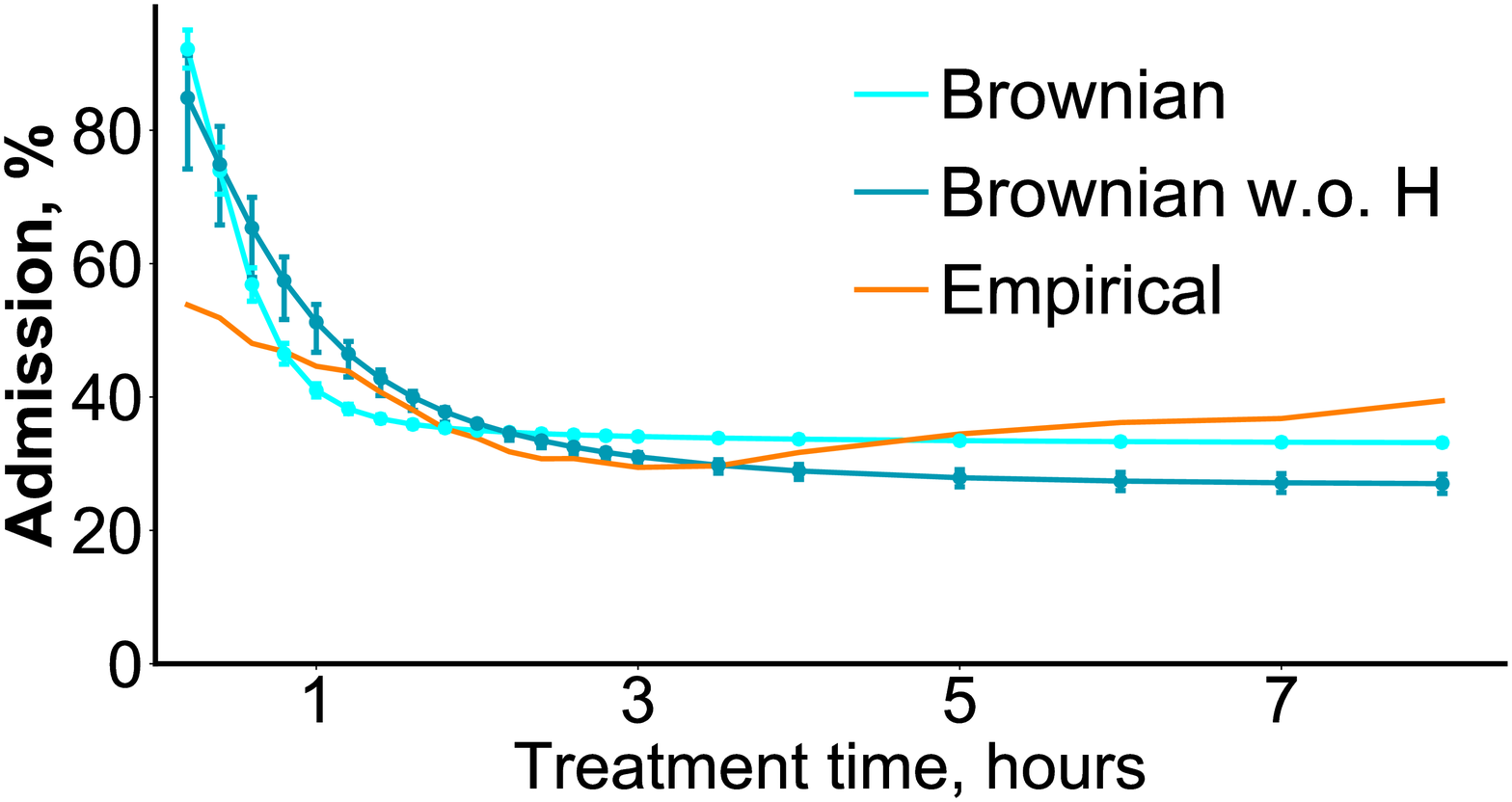}
\caption{Estimated 30-day revisit rate (top left), 30-day readmission rate (top right), and admission rate (bottom) comparison between drift diffusion model with and without $H$ . } \label{fig:brown_woh_outcomes_comp} 
\end{figure}

To show the importance of controlling for unobserved confounders \emph{via} the latent variable $H$, we removed $H$ from the original model in the main text. This model is referred to as ``Brownian" in the Appendix to respect our consideration of other models here. The parameters are summarized in Tables~\ref{tab:brown_woh_rev_params} and~\ref{tab:brown_woh_readm_params} in Appendix~\ref{appendix:parameter_estimates}. We depicted the admission, revisit and readmission risks in Figure \ref{fig:brown_woh_outcomes_comp}. The most notable difference when ignoring latent health needs $H$ can be seen in the admission rates, as opposed to revisit and readmission rates which have overlapping errors bars between the models with and without $H$. Specifically, the model without $H$ draws a similar conclusion as the empirical, or unadjusted, model in terms of suggesting the third hour of treatment time is much more critical to reducing admission rates than the model with $H$. One explanation is that individuals with treatment times around 3 hours tend to have less severe health needs and hence are less likely to be admitted, which if not adjusted for, can make the average potential response of a 3-hour treatment time on admission rates appear lower than it actually is. 


\subsubsection{Using general propensity score (GPS) without a latent variable}\label{app:gps}

In another exploration into the role of adjusting for unmeasured confounding, we define a new estimator by adapting GPS to our causal diagram. Recall that GPS only adjusts for measured confounding, not unmeasured confounding. Thus, we need to rely on stronger assumptions that ignore the potential latent confounder $H$:  $Y^t$ is independent of $T$ conditional on $A$ and $X$, and $A^t$ is independent of $T$ conditional on $X$ and $Z$.  Next, let $f(t|x,z)$ be the conditional density of treatment time $T$ given the observed characteristics $X$, and initial assessment of the vitals $Z$. The GPS is given by $R = f(T|X, Z)$. The most relevant property of the GPS is that the potential outcome is independent of the treatment conditional on $R$ and assuming that $X$ and $Z$ are the only confounders.  In our case, we have that $A^t\ind T\,|\,r(t, X, Z)$ \emph{and} $Y^t \ind T\,|\,A, r(t, X, Z)$. Define
\begin{align*}
 l(t,r) = \sum_{a=0, 1}P(Y=1|T=t, R=r, A=a)P(A=a|T=t, R=r).
\end{align*}
Theorem 2 in~\citet{hirano2004propensity} implies that
\begin{align*}
 l(t, r) = \sum_{a=0, 1}P(Y^t=1|A=a, r(t, X, Z)=r)P(A^t=a|r(t, X, Z)=r),
\end{align*}
and that
\begin{align*}
    \E[Y^t] = \E[l(t, r(t, X, Z))].
\end{align*}
To estimate effects, we propose a parametric model and assume that
\begin{align*}
 T | X, Z \sim \text{Lognormal}(\mu, \sigma)
\end{align*}
and that
\begin{align*}
 W = \log(\E[T| X, Z]) = \beta_0 + \beta_1'X + \beta_2'Z.
\end{align*}
As a result,
\begin{align*}
 R = \frac{1}{T\sigma\sqrt{2\pi}}\exp\left(-\frac{(\log(T)-W-\frac{\sigma^2}{2})^2}{2\sigma^2}\right).
\end{align*}
For $A$ and $Y$, define $\phi(y|a, t, r) = P(Y=y|A=a, T=t, R=r)$, and $\psi(a|t, r) = P(A=a|T=t, R=r)$. We assume that the logit of $\phi$ and $\psi$ are linear in its dependant variables
\begin{align*}
    \text{logit}(\phi(1|a, t, r)) = \gamma_0 + \gamma_1 r + \gamma_2 r^2 +  \gamma_3 t + \gamma_3 t^2 + \gamma_4 a  + \gamma_5 ar + \gamma_6 ar,
\end{align*}
and that
\begin{align*}
    \text{logit}(\psi(1|t, r)) = \alpha_0 + \alpha_1 r + \alpha_2 r^2  + \alpha_3 t + \alpha_3 t^2.
\end{align*}

We estimate parameters for the model above using maximum likelihood estimation for the parameters defined above using the joint distribution $\phi(y|a, t, r)\psi(a|t, r)$ and present an estimator of $l$
\begin{align*}
    \hat{l}(t, r) &= \sum_a \hat{\phi}(1|a,t, r)\hat{\psi}(a|t, r)
\end{align*}
This estimator is consistent. Finally, to estimate the average potential outcome of $T$ on $Y$, we let
\begin{align*}
    \widehat{\E[Y^t]} = \frac{1}{n}\sum_i \hat{l}(t, r(t, x_i, z_i)).
\end{align*}

\begin{figure}
\centering
\begin{tikzpicture}
\node at (8,0) {\includegraphics[width=0.49\textwidth]{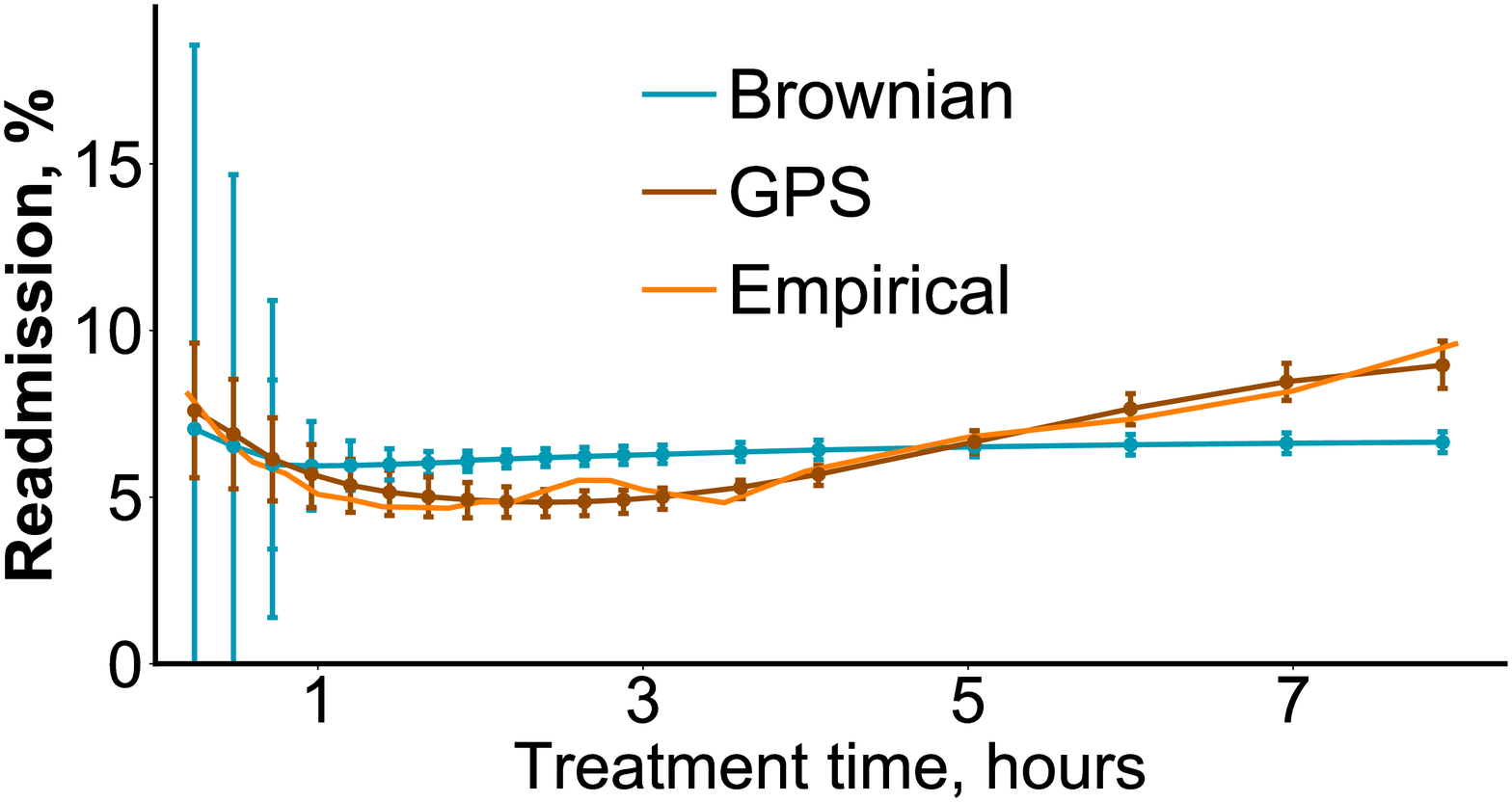}};
\node at (0,0) {\includegraphics[width=0.49\textwidth]{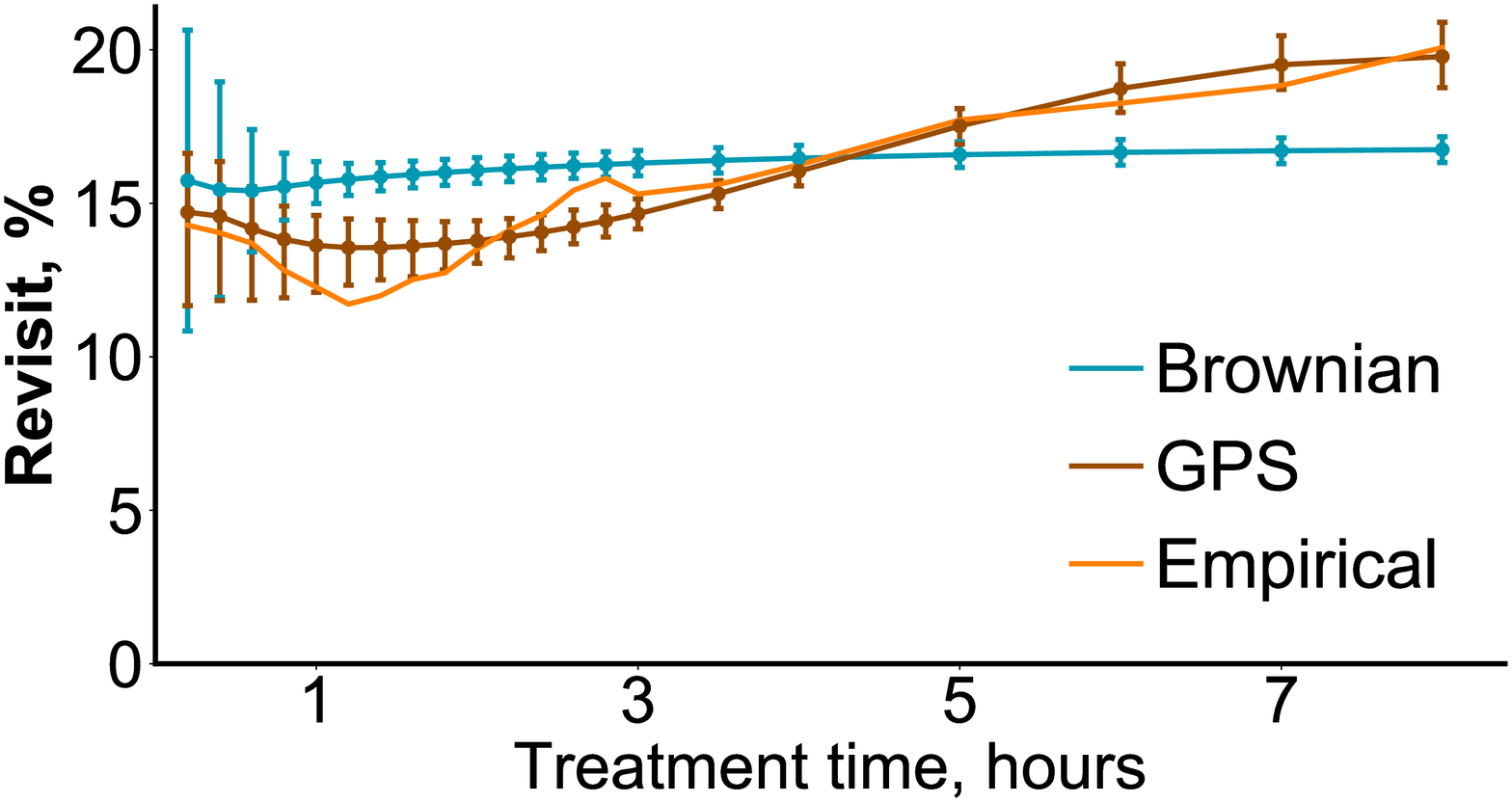}};
\node at (4,-5) {\includegraphics[width=0.49\textwidth]{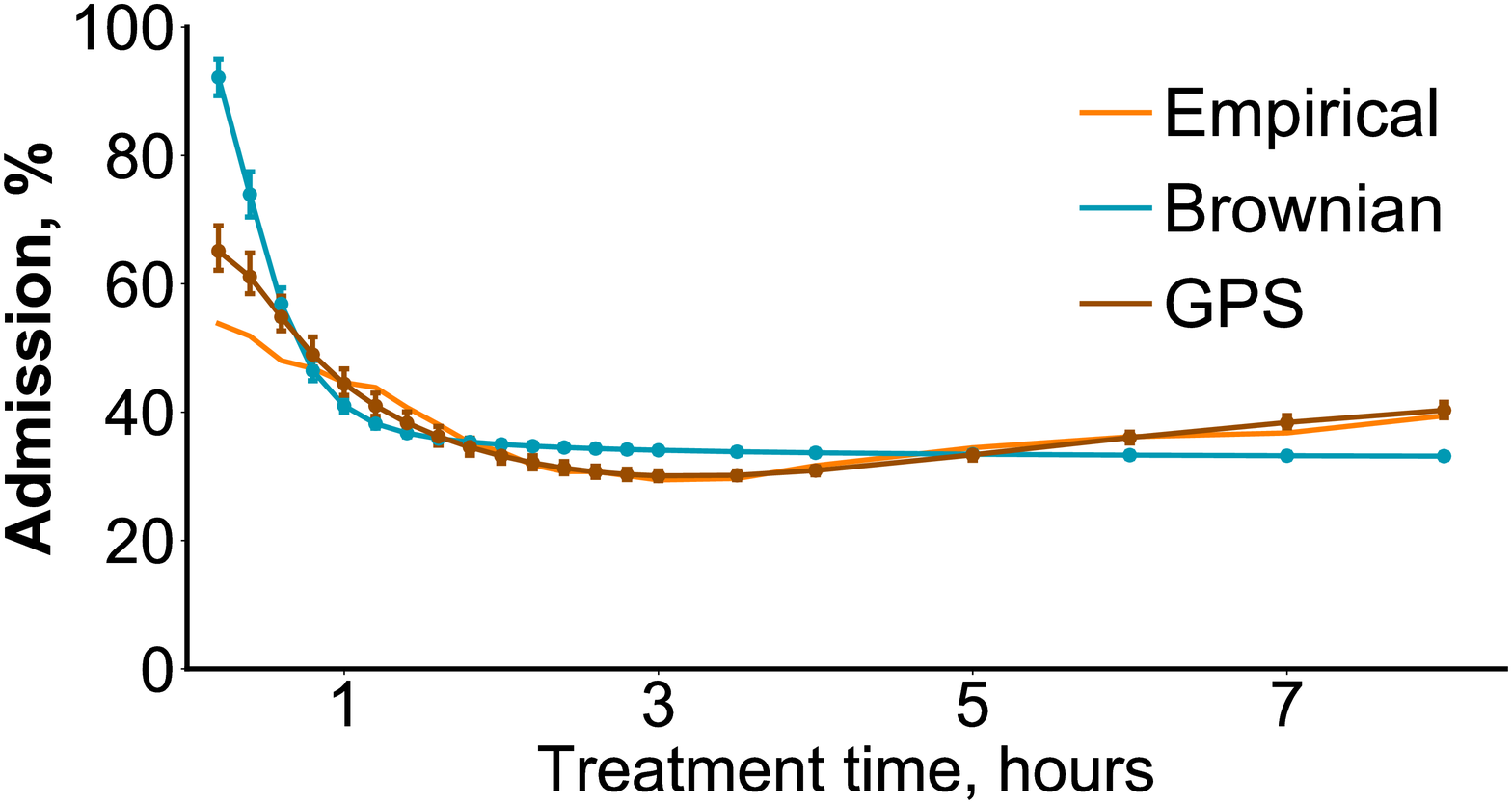}};
\node at (-3.7,2.4) {\large \textbf{\textsf{A}}};
\node at (4.15,2.4) {\large \textbf{\textsf{B}}};
\node at (0.3,-2.6) {\large \textbf{\textsf{C}}};
\end{tikzpicture}
\caption{Estimated (A) 30-day revisit rates, (B) 30-day readmission rates, and (C) admission rates compared between latent variable model based on Brownian model and GPS. Unadjusted estimates are also provided.} \label{fig:brown_gps_outcomes_comp}
\end{figure}

Estimated model parameters for the 30-day revisit and readmission outcomes using the GPS parameters are summarized in Tables~\ref{tab:gps_rev_params} and~\ref{tab:gps_readm_params} in Appendix~\ref{appendix:parameter_estimates}. Using the confounded GPS approach, we have reconstructed the prospect counterfactual curves for 30-day revisit, 30-day readmission, and admission rates in Figure~\ref{fig:brown_gps_outcomes_comp}. Notably, GPS follows the empirical curves more closely than the original model (``Brownian") for each variables. Thus even though GPS is controlling for possible measured confounding, it does not seem to translate into meaningful differences from the empirical, or unadjusted estimates.



\subsubsection{Modifying the parametric model of the admission process}

We present a different parametric model presented in the main text by modifying the joint distribution of $A$ and $T$. The original joint distribution of $A$ and $T$, which is based on Brownian motion, was designed to reflect how a provider accumulates evidence on latent health needs $H$ to help make their admission decision. To examine a more traditional regression approach, we assume that $T$ conditional on $X$, $Z$, and $H$ follows a lognormal distribution with mean given by \begin{align*}
 \log(\E[T\,|\,X,H,Z]) = \beta_1'X + \beta_2'Z +\beta_3H +\beta_4(1-H).
\end{align*}
It follows that the conditional density is given by
\begin{align*}
 f_{T|X, Z, H}(t|x, z, h) = \frac{1}{t\sigma\sqrt{2\pi}}\exp\left(-\frac{(\log(t)-\eta-\frac{\sigma^2}{2})^2}{2\sigma^2}\right),
\end{align*}
where $\eta = \beta_1'x + \beta_2'z +\beta_3h +\beta_4(1-h)$ and $\sigma$ is an unknown parameter. We also assume that the logit of the conditional probability of $A=1$ given $T$, $H$, $Z$ and $X$ is linear in its arguments, 
\begin{align*}
    \text{logit}(\PR(A=1|T, H, Z, X)) = \nu_1'X+\nu_2'Z+\nu_3\log(T) + \nu_4H + \nu_5(1-H).
\end{align*}
The remaining distributions are the same as in the main text, and the EM algorithm was again used for estimation.

\begin{figure}
\centering
\includegraphics[width=0.49\textwidth]{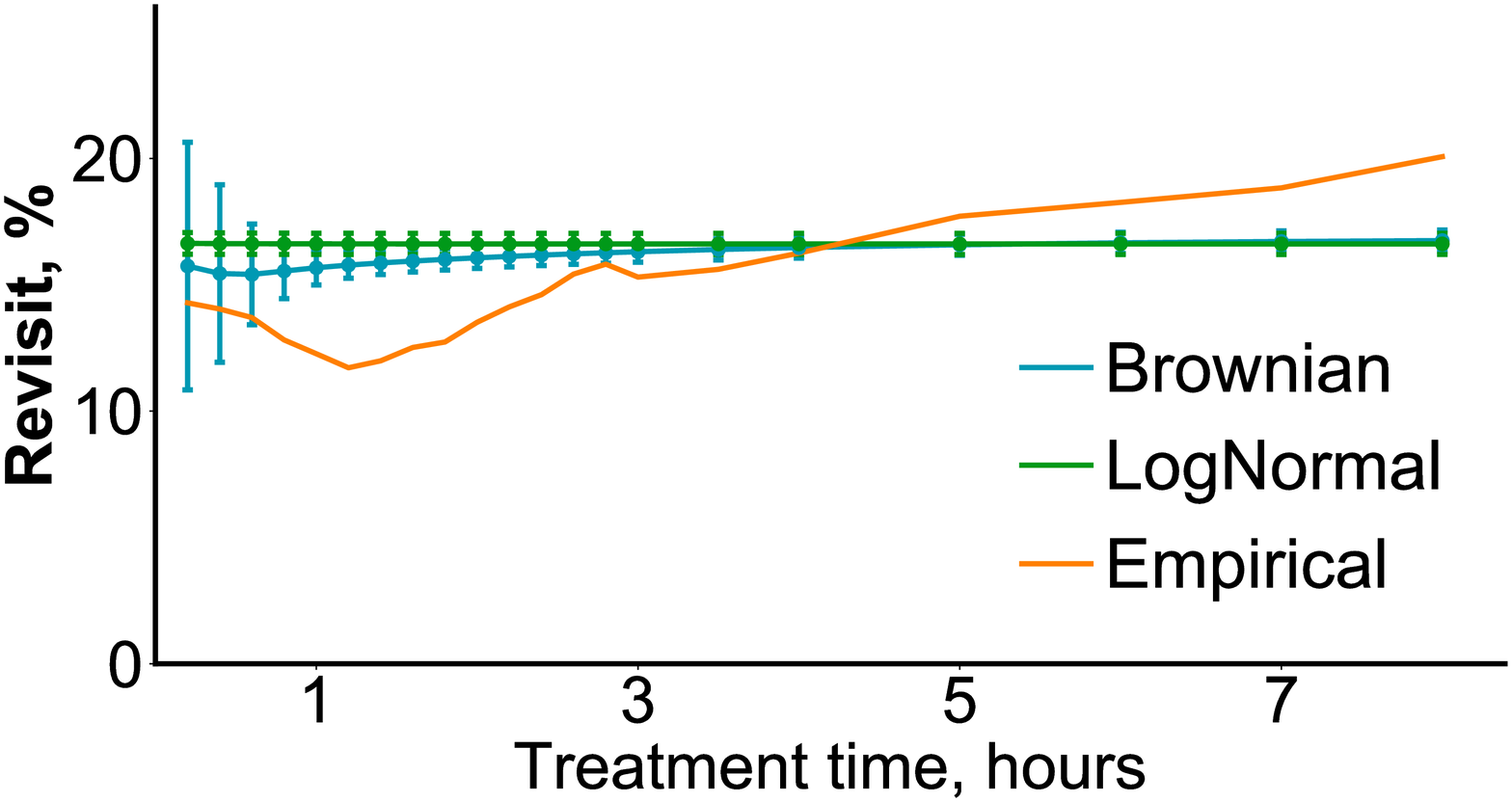}
\includegraphics[width=0.49\textwidth]{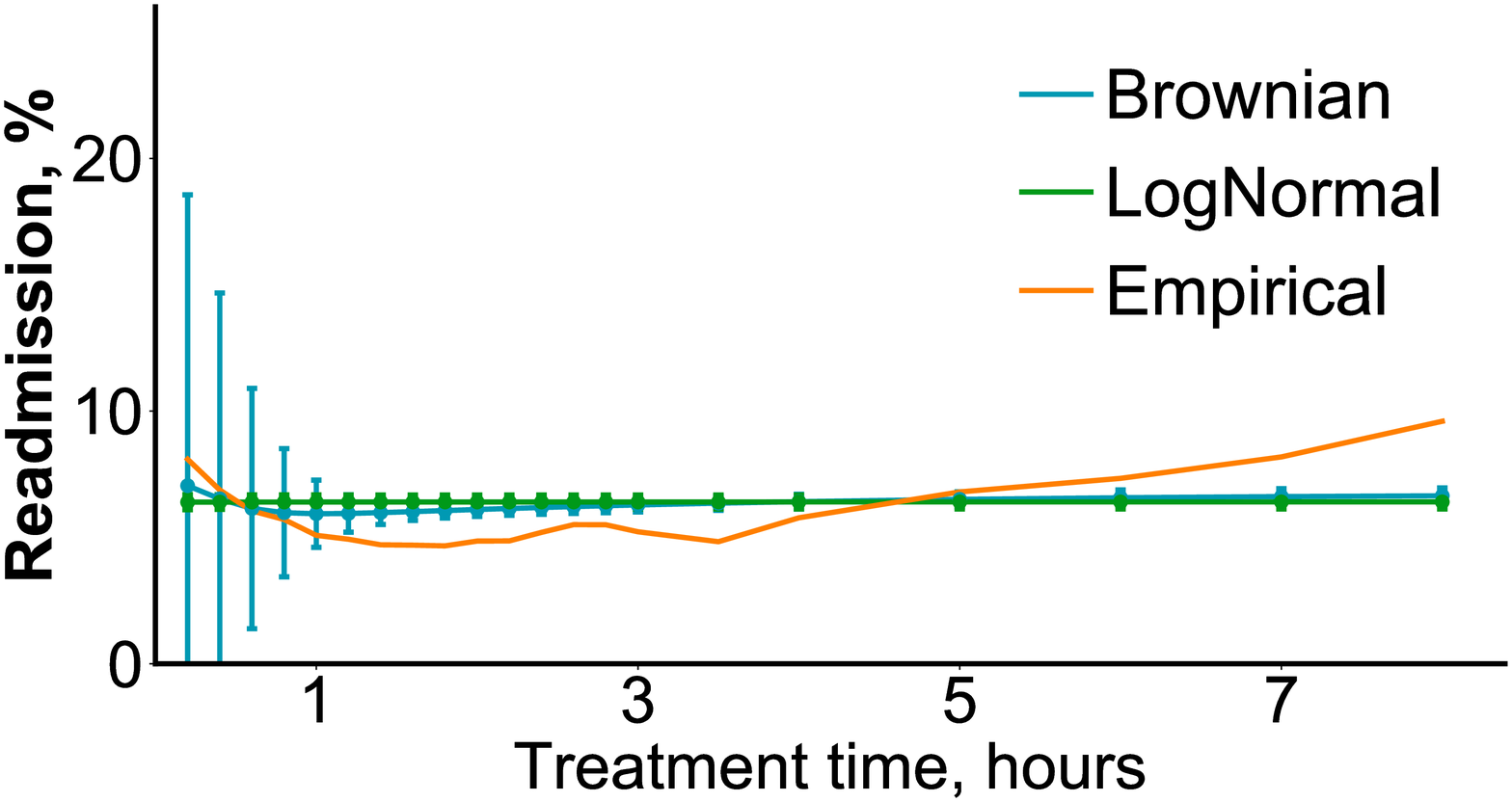}
\includegraphics[width=0.49\textwidth]{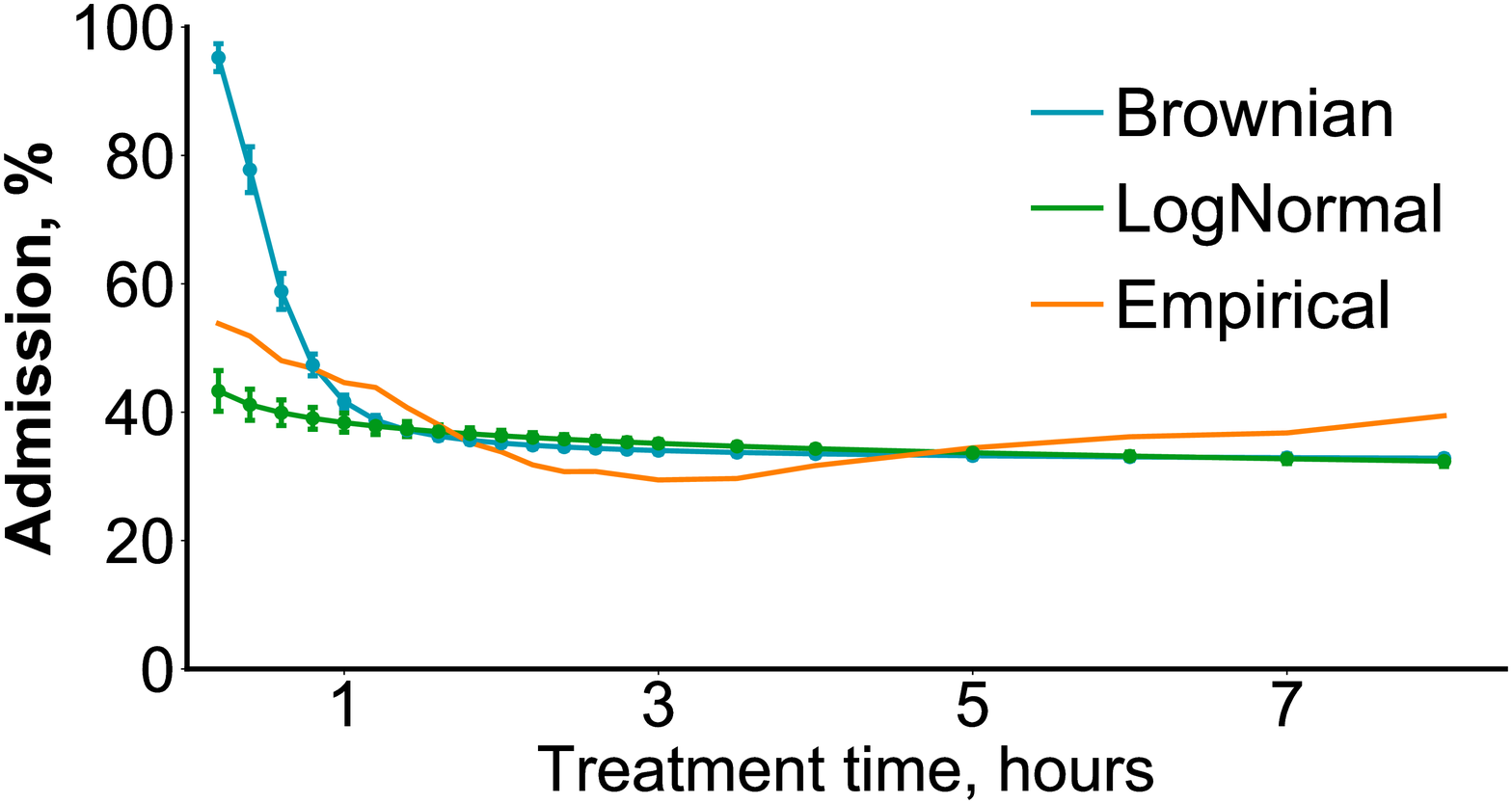}
\caption{Estimated 30-day revisit rate (top left), 30-day readmission rate (top right), and admission decision (bottom) comparison between latent variable models based on drift diffusion versus lognormal. } \label{fig:brown_logn_outcomes} 
\end{figure}

The estimated parameters using the latent-variable approach with lognormal distribution for the treatment time for the 30-day revisit and readmission risk, respectively, are presented in Tables~\ref{tab:logn_rev_params} and~\ref{tab:logn_read_params}.  The aggregated outcomes are compared in Figure~\ref{fig:brown_logn_outcomes}. We note that for treatment times longer than one hour, this alternative model based on a log-normal distribution (``LogNormal") agrees with original model (``Brownian") with respect to the estimated readmission, revisit, and admission rates. There is however a slight difference between models for admission rates under treatment times less than 1-hour, whereby the LogNormal model estimates that the admission rates are lower than the estimated by the Brownian motion model during the first hour. This difference is greater than 50\%. 

\begin{figure}
\centering
\includegraphics[width=0.49\textwidth]{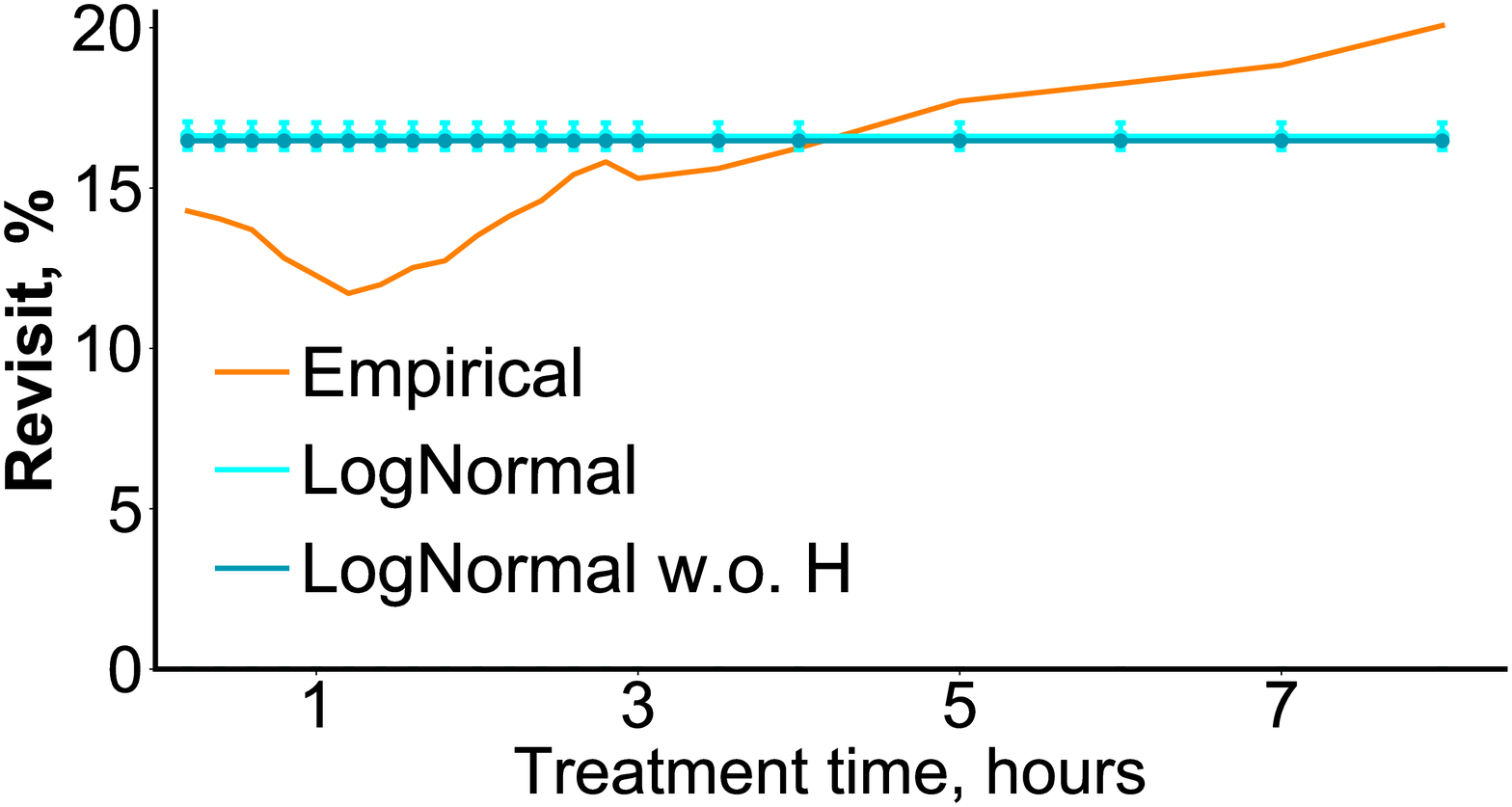}
\includegraphics[width=0.49\textwidth]{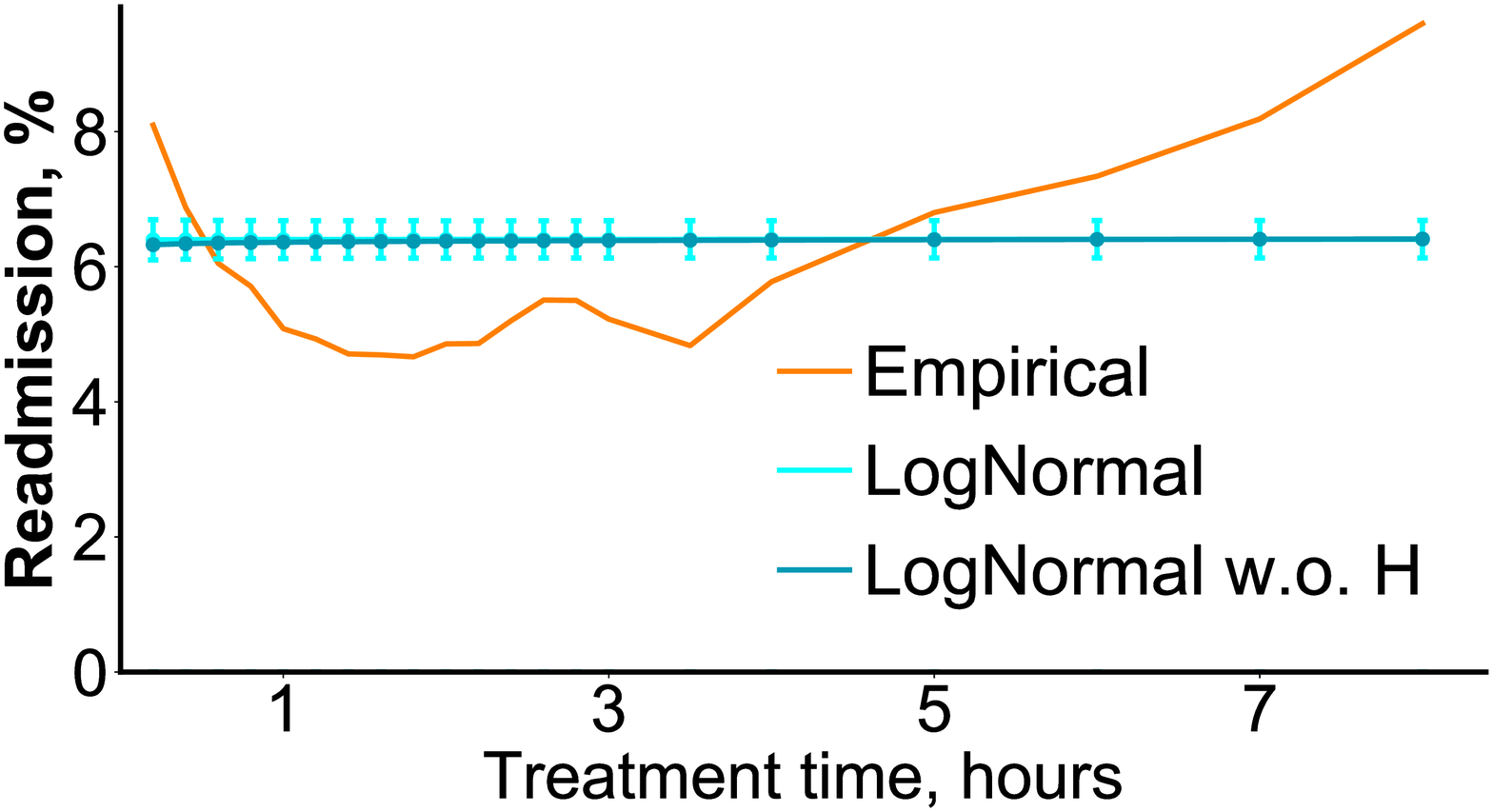} \\
\includegraphics[width=0.49\textwidth]{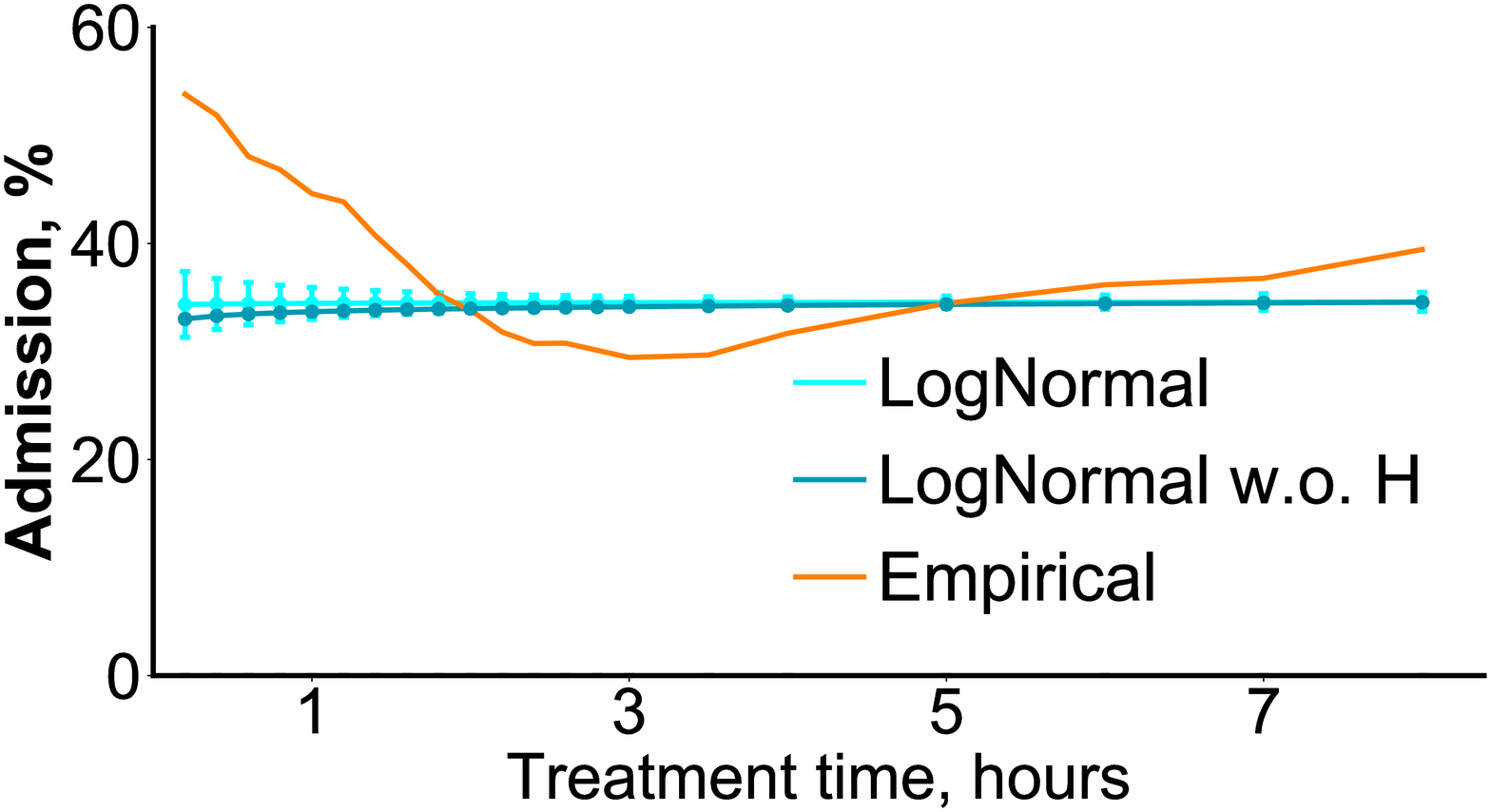}
\caption{Estimated 30-day revisit rate (top left), 30-day readmission rate (top right), and admission rate (bottom) comparison between LogNormal model with $H$ and LogNormal model without $H$ } \label{fig:logn_woh_outcomes_comp} 
\end{figure}

One potential reason for the discrepancy for models is that the Brownian motion model is better able to capture how latent health needs influence the admission decision. For example, latent health needs influence treatment time and admission decisions linearly in the LogNormal model. To explore this point, we also recovered estimates from the model that uses a LogNormal distribution for treatment time but also removes the latent variable $H$. Parameter estimates are summarized in Tables~\ref{tab:lognwoh_rev_params} and~\ref{tab:lognwoh_readm_params} in Appendix~\ref{appendix:parameter_estimates}. The admission, revisit and readmission risks are depicted in Figure \ref{fig:logn_woh_outcomes_comp}. We note that there is no difference between the versions with and without the health needs, which means that the LogNormal model fails to recover any possible unobserved confounder that might be present in the model. Thus, the latent health needs $H$ plays a larger role in the original Brownian model to correct estimates, which could mean that the Brownian motion model is better able to capture how latent health needs influence the admission decision.

\newpage

\subsection{Estimated model parameters}\label{appendix:parameter_estimates}

\subsubsection{Model based on Brownian motion}

\mbox{}
\vspace{-1em}

\begin{center}
\begin{longtable}{l l c c c }
\caption{Estimated coefficients for 30-day revisit outcome using latent variable approach based on Brownian motion. The link between input variables and output is described in the main text.}\label{tab:brown_rev_params}\\
\toprule
\endfirsthead

\multicolumn{5}{c}%
{{\bfseries \tablename\ \thetable{} -- Continued from previous page}} \\
\toprule

    \multicolumn{2}{c}{} & \multicolumn{1}{c}{\textbf{}} & \multicolumn{2}{c}{\textbf{95\% CI}} \\
    \multicolumn{1}{l}{\textbf{Variable}} &
    \multicolumn{1}{l}{\textbf{Model term}} & \multicolumn{1}{c}{\textbf{Estimate}} & \multicolumn{1}{c}{\textbf{Lower}} & \multicolumn{1}{c}{\textbf{Upper}} \\
    \midrule
\endhead

\hline \multicolumn{5}{r}{{Continued on next page}} \\ \hline
\endfoot

\hline
\endlastfoot

    \multicolumn{2}{c}{} & \multicolumn{1}{c}{\textbf{}} & \multicolumn{2}{c}{\textbf{95\% CI}} \\
    \multicolumn{1}{l}{\textbf{Variable}} &
    \multicolumn{1}{l}{\textbf{Model term}} & \multicolumn{1}{c}{\textbf{Estimate}} & \multicolumn{1}{c}{\textbf{Lower}} & \multicolumn{1}{c}{\textbf{Upper}} \\
    \midrule    Health State $H$ & Intercept & -0.755 & -0.828 & -0.682 \\ 
     & Age & 0.474 & 0.439 & 0.51 \\ 
     & Female & -0.298 & -0.357 & -0.239 \\ 
     & Non-White & -0.324 & -0.399 & -0.249 \\ 
     & Workers Comp & 0.075 & 0.012 & 0.137 \\ 
     & Income & -0.097 & -0.127 & -0.067 \\ 
     & High Comorbidity & 0.26 & 0.237 & 0.283 \\ 
     & Hypertension & -0.029 & -0.108 & 0.05 \\ 
     & Obesity & -0.331 & -0.435 & -0.228 \\ 
    Acuity $Z$ & $H=0$ & -2.868 & -2.976 & -2.76 \\ 
     & $H=1$ & -1.828 & -1.936 & -1.72 \\ 
     & Age & 0.345 & 0.292 & 0.398 \\ 
     & Female & -0.214 & -0.299 & -0.129 \\ 
     & Non-White & 0.004 & -0.108 & 0.116 \\ 
     & Workers Comp & -0.093 & -0.183 & -0.002 \\ 
     & Income & -0.008 & -0.051 & 0.036 \\ 
     & High Comorbidity & 0.138 & 0.11 & 0.165 \\ 
     & Hypertension & -0.019 & -0.132 & 0.093 \\ 
     & Obesity & -0.003 & -0.153 & 0.147 \\ 
     & Temperature & -0.013 & -0.066 & 0.039 \\ 
     & Heart rate & 0.33 & 0.285 & 0.375 \\ 
     & Blood pressure & 0.001 & -0.041 & 0.043 \\ 
     & Respiration rate & -0.0 & -0.045 & 0.045 \\ 
    Temperature & $H=0$ & -0.012 & -0.035 & 0.011 \\ 
     & $H=1$ & 0.163 & 0.137 & 0.19 \\ 
     & Age & -0.154 & -0.165 & -0.144 \\ 
     & Female & -0.0 & -0.018 & 0.018 \\ 
     & Non-White & 0.001 & -0.022 & 0.023 \\ 
     & Workers Comp & -0.064 & -0.084 & -0.044 \\ 
     & Income & 0.002 & -0.007 & 0.011 \\ 
     & High Comorbidity & -0.008 & -0.016 & -0.0 \\ 
     & Hypertension & 0.059 & 0.032 & 0.086 \\ 
     & Obesity & -0.026 & -0.059 & 0.007 \\ 
     & $H=0$ & -0.051 & -0.081 & -0.021 \\ 
     & $H=1$ & 0.302 & 0.268 & 0.337 \\ 
     & Age & -0.352 & -0.366 & -0.337 \\ 
     & Female & 0.005 & -0.018 & 0.029 \\ 
     & Non-White & -0.163 & -0.192 & -0.134 \\ 
     & Workers Comp & -0.168 & -0.194 & -0.142 \\ 
     & Income & 0.0 & -0.012 & 0.012 \\ 
     & High Comorbidity & 0.066 & 0.056 & 0.077 \\ 
     & Hypertension & 0.047 & 0.012 & 0.082 \\ 
     & Obesity & 0.009 & -0.034 & 0.052 \\ 
    Heart rate & $H=0$ & -0.023 & -0.05 & 0.003 \\ 
     & $H=1$ & 0.006 & -0.025 & 0.036 \\ 
     & Age & 0.391 & 0.378 & 0.403 \\ 
     & Female & 0.0 & -0.021 & 0.021 \\ 
     & Non-White & 0.0 & -0.025 & 0.026 \\ 
     & Workers Comp & -0.0 & -0.023 & 0.023 \\ 
     & Income & -0.021 & -0.031 & -0.01 \\ 
     & High Comorbidity & -0.066 & -0.075 & -0.057 \\ 
     & Hypertension & 0.383 & 0.352 & 0.415 \\ 
     & Obesity & 0.077 & 0.039 & 0.115 \\ 
     & $H=0$ & -0.001 & -0.028 & 0.026 \\ 
     & $H=1$ & 0.219 & 0.187 & 0.25 \\ 
     & Age & -0.243 & -0.256 & -0.23 \\ 
     & Female & -0.083 & -0.105 & -0.062 \\ 
     & Non-White & -0.0 & -0.026 & 0.026 \\ 
     & Workers Comp & -0.14 & -0.163 & -0.116 \\ 
     & Income & 0.013 & 0.003 & 0.024 \\ 
     & High Comorbidity & 0.03 & 0.02 & 0.039 \\ 
     & Hypertension & 0.131 & 0.099 & 0.163 \\ 
     & Obesity & -0.001 & -0.039 & 0.038 \\ 
    Blood pressure & $H=0$ & 0.103 & 0.086 & 0.12 \\ 
     & $H=1$ & 0.014 & -0.013 & 0.041 \\ 
     & Age & -0.021 & -0.055 & 0.012 \\ 
     & Female & 0.0 & -0.028 & 0.029 \\ 
     & Non-White & -0.029 & -0.043 & -0.015 \\ 
     & Workers Comp & -0.0 & -0.011 & 0.011 \\ 
     & Income & -0.091 & -0.13 & -0.053 \\ 
     & High Comorbidity & -0.002 & -0.047 & 0.043 \\ 
     & Hypertension & 0.009 & -0.008 & 0.026 \\ 
     & Obesity & 0.052 & 0.038 & 0.066 \\ 
     & $H=0$ & -0.047 & -0.061 & -0.032 \\ 
     & $H=1$ & -0.029 & -0.043 & -0.014 \\ 
     & Age & 0.052 & 0.024 & 0.079 \\ 
     & Female & 0.051 & 0.039 & 0.064 \\ 
     & Non-White & 0.123 & 0.103 & 0.144 \\ 
     & Workers Comp & 0.12 & 0.092 & 0.147 \\ 
     & Income & -0.009 & -0.031 & 0.014 \\ 
     & High Comorbidity & -0.023 & -0.033 & -0.013 \\ 
     & Hypertension & 0.048 & 0.038 & 0.058 \\ 
     & Obesity & -0.001 & -0.033 & 0.031 \\ 
    Respiration rate & $H=0$ & 0.022 & -0.016 & 0.059 \\ 
     & $H=1$ & 1.043 & 1.028 & 1.058 \\ 
     & Age & 0.654 & 0.626 & 0.683 \\ 
     & Female & -2.031 & -2.144 & -1.917 \\ 
     & Non-White & 2.632 & 2.207 & 3.057 \\ 
     & Workers Comp & -1.879 & -2.557 & -1.2 \\ 
     & Income & -1.655 & -1.797 & -1.514 \\ 
     & High Comorbidity & -0.012 & -0.035 & 0.011 \\ 
     & Hypertension & 0.163 & 0.137 & 0.19 \\ 
     & Obesity & -0.154 & -0.165 & -0.144 \\ 
     & $H=0$ & -0.0 & -0.018 & 0.018 \\ 
     & $H=1$ & 0.001 & -0.022 & 0.023 \\ 
     & Age & -0.064 & -0.084 & -0.044 \\ 
     & Female & 0.002 & -0.007 & 0.011 \\ 
     & Non-White & -0.008 & -0.016 & -0.0 \\ 
     & Workers Comp & 0.059 & 0.032 & 0.086 \\ 
     & Income & -0.026 & -0.059 & 0.007 \\ 
     & High Comorbidity & -0.051 & -0.081 & -0.021 \\ 
     & Hypertension & 0.302 & 0.268 & 0.337 \\ 
     & Obesity & -0.352 & -0.366 & -0.337 \\ 
    Initial Evidence $c(X, Z)$ & Age & 0.005 & -0.018 & 0.029 \\ 
     & Female & -0.163 & -0.192 & -0.134 \\ 
     & Non-White & -0.168 & -0.194 & -0.142 \\ 
     & Workers Comp & 0.0 & -0.012 & 0.012 \\ 
     & Income & 0.066 & 0.056 & 0.077 \\ 
     & High Comorbidity & 0.047 & 0.012 & 0.082 \\ 
     & Hypertension & 0.009 & -0.034 & 0.052 \\ 
     & Obesity & -0.023 & -0.05 & 0.003 \\ 
     & Temperature & 0.006 & -0.025 & 0.036 \\ 
     & Heart rate & 0.391 & 0.378 & 0.403 \\ 
     & Blood pressure & 0.0 & -0.021 & 0.021 \\ 
     & Respiration rate & 0.0 & -0.025 & 0.026 \\ 
    Evidence Threshold $b(X)$ & Intercept & -0.0 & -0.023 & 0.023 \\ 
     & Age & -0.021 & -0.031 & -0.01 \\ 
     & Female & -0.066 & -0.075 & -0.057 \\ 
     & Non-White & 0.383 & 0.352 & 0.415 \\ 
     & Workers Comp & 0.077 & 0.039 & 0.115 \\ 
     & Income & -0.001 & -0.028 & 0.026 \\ 
     & High Comorbidity & 0.219 & 0.187 & 0.25 \\ 
     & Hypertension & -0.243 & -0.256 & -0.23 \\ 
     & Obesity & -0.083 & -0.105 & -0.062 \\ 
    Drift Rate $d(H)$ & $H=0$ & -0.0 & -0.026 & 0.026 \\ 
     & $H=1$ & -0.14 & -0.163 & -0.116 \\ 
    $\mu_1$ & Intercept & 0.013 & 0.003 & 0.024 \\ 
    $\mu_2$ & Intercept & 0.03 & 0.02 & 0.039 \\ 
    $\mu_3$ & Intercept & 0.131 & 0.099 & 0.163 \\ 
    $\mu_4$ & Intercept & -0.001 & -0.039 & 0.038\\
\end{longtable}
\end{center}

\begin{center}
\begin{longtable}{l l c c c }
\caption{Estimated coefficients for 30-day readmission outcome using latent-variable approach based on Brownian motion. The link between input variables and output is described in the main text.}\label{tab:brown_readm_params}\\
\toprule
\endfirsthead

\multicolumn{5}{c}%
{{\bfseries \tablename\ \thetable{} -- Continued from previous page}} \\
\toprule

    \multicolumn{2}{c}{} & \multicolumn{1}{c}{\textbf{}} & \multicolumn{2}{c}{\textbf{95\% CI}} \\
    \multicolumn{1}{l}{\textbf{Variable}} &
    \multicolumn{1}{l}{\textbf{Model term}} & \multicolumn{1}{c}{\textbf{Estimate}} & \multicolumn{1}{c}{\textbf{Lower}} & \multicolumn{1}{c}{\textbf{Upper}} \\
    \midrule
\endhead

\hline \multicolumn{5}{r}{{Continued on next page}} \\ \hline
\endfoot

\hline
\endlastfoot

    \multicolumn{2}{c}{} & \multicolumn{1}{c}{\textbf{}} & \multicolumn{2}{c}{\textbf{95\% CI}} \\
    \multicolumn{1}{l}{\textbf{Variable}} &
    \multicolumn{1}{l}{\textbf{Model term}} & \multicolumn{1}{c}{\textbf{Estimate}} & \multicolumn{1}{c}{\textbf{Lower}} & \multicolumn{1}{c}{\textbf{Upper}} \\
    \midrule    Health State $H$ & Intercept & -0.705 & -0.775 & -0.635 \\ 
     & Age & 0.452 & 0.418 & 0.486 \\ 
     & Female & -0.315 & -0.373 & -0.257 \\ 
     & Non-White & -0.381 & -0.456 & -0.306 \\ 
     & Workers Comp & 0.057 & -0.005 & 0.119 \\ 
     & Income & -0.104 & -0.134 & -0.074 \\ 
     & High Comorbidity & 0.258 & 0.235 & 0.281 \\ 
     & Hypertension & -0.027 & -0.105 & 0.051 \\ 
     & Obesity & -0.354 & -0.458 & -0.249 \\ 
    Acuity $Z$ & $H=0$ & -2.819 & -2.935 & -2.702 \\ 
     & $H=1$ & -1.642 & -1.756 & -1.527 \\ 
     & Age & -0.0 & -0.054 & 0.054 \\ 
     & Female & -0.132 & -0.219 & -0.044 \\ 
     & Non-White & -0.218 & -0.331 & -0.105 \\ 
     & Workers Comp & -0.281 & -0.377 & -0.184 \\ 
     & Income & 0.003 & -0.042 & 0.047 \\ 
     & High Comorbidity & 0.183 & 0.155 & 0.211 \\ 
     & Hypertension & 0.0 & -0.121 & 0.121 \\ 
     & Obesity & 0.0 & -0.154 & 0.154 \\ 
     & Temperature & -0.003 & -0.056 & 0.049 \\ 
     & Heart rate & -0.005 & -0.051 & 0.042 \\ 
     & Blood pressure & -0.001 & -0.044 & 0.043 \\ 
     & Respiration rate & 0.333 & 0.293 & 0.373 \\ 
    Temperature & $H=0$ & -0.011 & -0.035 & 0.012 \\ 
     & $H=1$ & 0.158 & 0.132 & 0.184 \\ 
     & Age & -0.146 & -0.157 & -0.135 \\ 
     & Female & 0.0 & -0.018 & 0.018 \\ 
     & Non-White & 0.004 & -0.019 & 0.026 \\ 
     & Workers Comp & -0.058 & -0.078 & -0.038 \\ 
     & Income & 0.003 & -0.006 & 0.012 \\ 
     & High Comorbidity & 0.0 & -0.008 & 0.008 \\ 
     & Hypertension & -0.0 & -0.027 & 0.027 \\ 
     & Obesity & -0.018 & -0.051 & 0.015 \\ 
    Heart rate & $H=0$ & -0.131 & -0.16 & -0.102 \\ 
     & $H=1$ & 0.219 & 0.186 & 0.251 \\ 
     & Age & -0.352 & -0.365 & -0.338 \\ 
     & Female & 0.124 & 0.102 & 0.147 \\ 
     & Non-White & -0.16 & -0.187 & -0.132 \\ 
     & Workers Comp & -0.163 & -0.187 & -0.138 \\ 
     & Income & -0.005 & -0.016 & 0.006 \\ 
     & High Comorbidity & 0.07 & 0.06 & 0.08 \\ 
     & Hypertension & 0.048 & 0.015 & 0.082 \\ 
     & Obesity & 0.037 & -0.004 & 0.077 \\ 
    Blood pressure & $H=0$ & -0.0 & -0.027 & 0.027 \\ 
     & $H=1$ & 0.006 & -0.024 & 0.036 \\ 
     & Age & 0.395 & 0.383 & 0.407 \\ 
     & Female & -0.143 & -0.165 & -0.122 \\ 
     & Non-White & 0.089 & 0.063 & 0.114 \\ 
     & Workers Comp & 0.034 & 0.011 & 0.057 \\ 
     & Income & -0.018 & -0.029 & -0.008 \\ 
     & High Comorbidity & -0.055 & -0.064 & -0.046 \\ 
     & Hypertension & 0.388 & 0.357 & 0.42 \\ 
     & Obesity & 0.106 & 0.068 & 0.144 \\ 
    Respiration rate & $H=0$ & -0.022 & -0.051 & 0.006 \\ 
     & $H=1$ & 0.187 & 0.156 & 0.219 \\ 
     & Age & -0.238 & -0.251 & -0.226 \\ 
     & Female & -0.0 & -0.022 & 0.022 \\ 
     & Non-White & -0.004 & -0.031 & 0.022 \\ 
     & Workers Comp & -0.162 & -0.186 & -0.138 \\ 
     & Income & 0.015 & 0.004 & 0.026 \\ 
     & High Comorbidity & 0.027 & 0.018 & 0.037 \\ 
     & Hypertension & 0.094 & 0.062 & 0.126 \\ 
     & Obesity & 0.028 & -0.011 & 0.067 \\ 
    Initial Evidence $c(X, Z)$ & Age & 0.112 & 0.095 & 0.129 \\ 
     & Female & 0.033 & 0.007 & 0.059 \\ 
     & Non-White & -0.027 & -0.061 & 0.006 \\ 
     & Workers Comp & 0.001 & -0.027 & 0.029 \\ 
     & Income & -0.017 & -0.031 & -0.003 \\ 
     & High Comorbidity & 0.006 & -0.004 & 0.016 \\ 
     & Hypertension & -0.135 & -0.172 & -0.098 \\ 
     & Obesity & 0.195 & 0.15 & 0.241 \\ 
     & Temperature & 0.038 & 0.022 & 0.055 \\ 
     & Heart rate & 0.044 & 0.03 & 0.058 \\ 
     & Blood pressure & -0.041 & -0.055 & -0.027 \\ 
     & Respiration rate & -0.03 & -0.044 & -0.016 \\ 
    Evidence Threshold $b(X)$ & Intercept & 0.136 & 0.108 & 0.164 \\ 
     & Age & 0.011 & -0.002 & 0.023 \\ 
     & Female & 0.11 & 0.089 & 0.131 \\ 
     & Non-White & -0.002 & -0.029 & 0.025 \\ 
     & Workers Comp & -0.061 & -0.083 & -0.038 \\ 
     & Income & -0.036 & -0.046 & -0.026 \\ 
     & High Comorbidity & 0.058 & 0.049 & 0.067 \\ 
     & Hypertension & 0.005 & -0.025 & 0.036 \\ 
     & Obesity & 0.048 & 0.012 & 0.084 \\ 
    Drift Rate $d(H)$ & $H=0$ & 1.054 & 1.04 & 1.069 \\ 
     & $H=1$ & 0.712 & 0.687 & 0.736 \\ 
    $\mu_1$ & Intercept & -4.057 & -4.336 & -3.777 \\ 
    $\mu_2$ & Intercept & 2.646 & 1.682 & 3.61 \\ 
    $\mu_3$ & Intercept & -3.049 & -6.812 & 0.714 \\ 
    $\mu_4$ & Intercept & -2.537 & -2.903 & -2.172\\
\end{longtable}
\end{center}

\subsubsection{Model based on Brownian motion without latent variable}

\begin{table}[H]
\centering
\begin{tabular}{l l c c c }
\toprule
    \multicolumn{2}{c}{} & \multicolumn{1}{c}{\textbf{}} & \multicolumn{2}{c}{\textbf{95\% CI}} \\
    \multicolumn{1}{l}{\textbf{Variable}} &
    \multicolumn{1}{l}{\textbf{Model term}} & \multicolumn{1}{c}{\textbf{Estimate}} & \multicolumn{1}{c}{\textbf{Lower}} & \multicolumn{1}{c}{\textbf{Upper}} \\
    \midrule    Initial Evidence $c(X, Z)$ & $Z=0$ & 0.268 & 0.157 & 0.321 \\ 
     & $Z=1$ & 0.147 & -0.003 & 0.221 \\ 
     & Age & 0.175 & 0.152 & 0.22 \\ 
     & Female & -0.065 & -0.126 & 0.005 \\ 
     & Non-White & -0.093 & -0.123 & 0.001 \\ 
     & Workers Comp & 0.03 & -0.023 & 0.104 \\ 
     & Income & -0.04 & -0.062 & -0.009 \\ 
     & High Comorbidity & 0.038 & 0.015 & 0.062 \\ 
     & Hypertension & -0.0 & -0.145 & 0.001 \\ 
     & Obesity & 0.001 & -0.049 & 0.112 \\ 
     & Temperature & 0.036 & 0.016 & 0.065 \\ 
     & Heart rate & 0.091 & 0.046 & 0.135 \\ 
     & Blood pressure & -0.034 & -0.075 & 0.0 \\ 
     & Respiration rate & -0.001 & -0.0 & 0.0 \\ 
    Evidence Threshold $b(X)$ & Intercept & -0.127 & -0.149 & -0.11 \\ 
     & Age & -0.0 & -0.0 & 0.025 \\ 
     & Female & 0.065 & 0.045 & 0.087 \\ 
     & Non-White & 0.03 & 0.003 & 0.04 \\ 
     & Workers Comp & -0.014 & -0.026 & 0.002 \\ 
     & Income & -0.0 & -0.011 & 0.003 \\ 
     & High Comorbidity & 0.019 & 0.013 & 0.024 \\ 
     & Hypertension & 0.016 & -0.003 & 0.036 \\ 
     & Obesity & 0.021 & -0.003 & 0.043 \\ 
     & Temperature & 0.0 & -0.015 & 0.001 \\ 
     & Heart rate & 0.006 & -0.005 & 0.02 \\ 
     & Blood pressure & 0.007 & -0.002 & 0.017 \\ 
     & Respiration rate & -0.007 & -0.013 & 0.001 \\ 
    Drift Rate $d(H)$ & $H=0$ & -1.244 & -1.34 & -1.17 \\ 
     & $H=1$ & 0.352 & 0.196 & 0.536 \\ 
    $\mu_1$ & Intercept & -1.444 & -1.527 & -1.276 \\ 
    $\mu_2$ & Intercept & -1.365 & -1.558 & -1.127 \\ 
    $\mu_3$ & Intercept & -0.001 & -0.053 & 0.004 \\ 
    $\mu_4$ & Intercept & 0.102 & 0.071 & 0.181\\
\bottomrule
\end{tabular}
\caption{Estimated coefficients for 30-day revisit outcome using the Brownian motion model without the latent variable. The link between input variables and output is described in the Appendix.}\label{tab:brown_woh_rev_params}
\end{table}

\begin{table}[H]
\centering
\begin{tabular}{l l c c c }
\toprule

    \multicolumn{2}{c}{} & \multicolumn{1}{c}{\textbf{}} & \multicolumn{2}{c}{\textbf{95\% CI}} \\
    \multicolumn{1}{l}{\textbf{Variable}} &
    \multicolumn{1}{l}{\textbf{Model term}} & \multicolumn{1}{c}{\textbf{Estimate}} & \multicolumn{1}{c}{\textbf{Lower}} & \multicolumn{1}{c}{\textbf{Upper}} \\
    \midrule    Initial Evidence $c(X, Z)$ & $Z=0$ & 0.212 & 0.151 & 0.323 \\ 
     & $Z=1$ & 0.086 & -0.004 & 0.224 \\ 
     & Age & 0.186 & 0.153 & 0.218 \\ 
     & Female & -0.0 & -0.124 & 0.003 \\ 
     & Non-White & -0.0 & -0.116 & 0.001 \\ 
     & Workers Comp & 0.041 & -0.011 & 0.108 \\ 
     & Income & -0.035 & -0.062 & -0.005 \\ 
     & High Comorbidity & 0.037 & 0.015 & 0.062 \\ 
     & Hypertension & 0.001 & -0.142 & 0.001 \\ 
     & Obesity & 0.015 & -0.05 & 0.113 \\ 
     & Temperature & 0.037 & 0.016 & 0.067 \\ 
     & Heart rate & 0.092 & 0.047 & 0.136 \\ 
     & Blood pressure & -0.038 & -0.075 & 0.001 \\ 
     & Respiration rate & 0.0 & -0.0 & 0.0 \\ 
    Evidence Threshold $b(X)$ & Intercept & -0.132 & -0.149 & -0.111 \\ 
     & Age & 0.021 & -0.0 & 0.024 \\ 
     & Female & 0.064 & 0.045 & 0.087 \\ 
     & Non-White & 0.032 & 0.001 & 0.04 \\ 
     & Workers Comp & -0.0 & -0.026 & 0.003 \\ 
     & Income & -0.006 & -0.011 & 0.003 \\ 
     & High Comorbidity & 0.016 & 0.013 & 0.025 \\ 
     & Hypertension & 0.014 & -0.004 & 0.036 \\ 
     & Obesity & 0.0 & -0.004 & 0.044 \\ 
     & Temperature & -0.007 & -0.015 & 0.001 \\ 
     & Heart rate & 0.012 & -0.005 & 0.019 \\ 
     & Blood pressure & 0.001 & -0.002 & 0.017 \\ 
     & Respiration rate & -0.004 & -0.012 & 0.001 \\ 
    Drift Rate $d(H)$ & $Z=0$ & -1.255 & -1.338 & -1.167 \\ 
     & $Z=1$ & 0.369 & 0.186 & 0.519 \\ 
    $\mu_1$ & Intercept & -2.7 & -3.036 & -2.629 \\ 
    $\mu_2$ & Intercept & -2.141 & -2.664 & -1.946 \\ 
    $\mu_3$ & Intercept & -0.0 & -0.012 & 0.062 \\ 
    $\mu_4$ & Intercept & 0.083 & 0.001 & 0.278\\
\bottomrule
\end{tabular}
\caption{Estimated coefficients for 30-day readmission outcome using the Brownian motion model without the latent variable. The link between input variables and output is described in the Appendix.}
\label{tab:brown_woh_readm_params}
\end{table}

\subsubsection{Model based on GPS without latent variable}

\begin{table}[H]
\centering
\begin{tabular}{l l c c c }
\toprule

    \multicolumn{2}{c}{} & \multicolumn{1}{c}{\textbf{}} & \multicolumn{2}{c}{\textbf{95\% CI}} \\
    \multicolumn{1}{l}{\textbf{Variable}} &
    \multicolumn{1}{l}{\textbf{Model term}} & \multicolumn{1}{c}{\textbf{Estimate}} & \multicolumn{1}{c}{\textbf{Lower}} & \multicolumn{1}{c}{\textbf{Upper}} \\
    \midrule    GPS $(R)$ & Intercept & -2.106 & -2.826 & -0.943 \\ 
     & Age & -0.038 & -0.236 & 0.785 \\ 
     & Female & 0.164 & -0.766 & 0.886 \\ 
     & Non-White & 0.021 & -0.75 & 1.12 \\ 
     & Workers Comp & -0.036 & -1.014 & 0.464 \\ 
     & Income & -0.004 & -0.177 & 0.39 \\ 
     & High Comorbidity & -0.155 & -0.368 & 0.154 \\ 
     & Hypertension & 0.537 & -1.463 & 1.107 \\ 
     & Obesity & -0.126 & -1.076 & 2.034 \\ 
     & Temperature & 0.008 & -0.175 & 0.391 \\ 
     & Heart rate & 0.246 & -0.206 & 0.51 \\ 
     & Blood pressure & 0.183 & -0.3 & 0.358 \\ 
     & Respiration rate & -0.017 & -0.229 & 0.38 \\ 
     & Acuity & 0.861 & -1.04 & 1.672 \\ 
     & $\sqrt{2}\sigma$ & -0.017 & -0.255 & 0.483 \\ 
    Decision $A$ & Intercept & 0.665 & 0.531 & 0.852 \\ 
     & $R$ & -0.514 & -0.584 & -0.457 \\ 
     & $R^2$ & 0.045 & 0.035 & 0.056 \\ 
     & $T$ & -2.63 & -3.783 & -1.432 \\ 
     & $T^2$ & 2.567 & -0.266 & 5.078 \\ 
    Outcome $Y$ & Intercept & -1.942 & -2.316 & -1.712 \\ 
     & $R$ & -0.158 & -0.257 & -0.048 \\ 
     & $R^2$ & 0.015 & 0.0 & 0.03 \\ 
     & $T$ & 4.382 & 3.285 & 6.141 \\ 
     & $T^2$ & -7.118 & -10.756 & -4.358 \\ 
     & $A$ & 0.234 & -0.069 & 0.656 \\ 
     & $AR$ & 0.049 & -0.036 & 0.106 \\ 
     & $AT$ & -0.741 & -1.936 & 0.211\\
\bottomrule
\end{tabular}
\caption{Estimated coefficients for 30-day Revisit outcome using the GPS model.}
\label{tab:gps_rev_params}
\end{table}

\begin{table}[H]
\centering
\begin{tabular}{l l c c c }
\toprule

    \multicolumn{2}{c}{} & \multicolumn{1}{c}{\textbf{}} & \multicolumn{2}{c}{\textbf{95\% CI}} \\
    \multicolumn{1}{l}{\textbf{Variable}} &
    \multicolumn{1}{l}{\textbf{Model term}} & \multicolumn{1}{c}{\textbf{Estimate}} & \multicolumn{1}{c}{\textbf{Lower}} & \multicolumn{1}{c}{\textbf{Upper}} \\
    \midrule    GPS $(R)$ & Intercept & -1.499 & -2.593 & -0.652 \\ 
     & Age & 0.168 & -0.318 & 0.715 \\ 
     & Female & -0.704 & -0.872 & 0.636 \\ 
     & Non-White & 0.071 & -1.045 & 0.747 \\ 
     & Workers Comp & -0.027 & -1.038 & 0.595 \\ 
     & Income & -0.06 & -0.17 & 0.265 \\ 
     & High Comorbidity & -0.102 & -0.493 & 0.048 \\ 
     & Hypertension & 0.285 & -0.953 & 1.326 \\ 
     & Obesity & -0.199 & -0.894 & 1.568 \\ 
     & Temperature & -0.003 & -0.186 & 0.249 \\ 
     & Heart rate & 0.05 & -0.126 & 0.426 \\ 
     & Blood pressure & -0.014 & -0.207 & 0.432 \\ 
     & Respiration rate & 0.016 & -0.217 & 0.252 \\ 
     & Acuity & 0.6 & -1.171 & 1.701 \\ 
     & $\sqrt{2}\sigma$ & -0.059 & -0.367 & 0.417 \\ 
    Decision $A$ & Intercept & 0.666 & 0.487 & 0.826 \\ 
     & $R$ & -0.514 & -0.574 & -0.461 \\ 
     & $R^2$ & 0.045 & 0.037 & 0.055 \\ 
     & $T$ & -2.638 & -3.483 & -1.358 \\ 
     & $T^2$ & 2.587 & -0.198 & 4.423 \\ 
    Outcome $Y$ & Intercept & -3.799 & -4.286 & -3.352 \\ 
     & $R$ & -0.298 & -0.423 & -0.145 \\ 
     & $R^2$ & 0.037 & 0.02 & 0.055 \\ 
     & $T$ & 7.985 & 5.177 & 11.328 \\ 
     & $T^2$ & -10.258 & -17.348 & -4.421 \\ 
     & $A$ & 1.625 & 1.0 & 2.198 \\ 
     & $AR$ & 0.021 & -0.094 & 0.124 \\ 
     & $AT$ & -3.506 & -5.468 & -1.748\\
\bottomrule
\end{tabular}
\caption{Estimated coefficients for 30-day readmission outcome using the GPS model.}
\label{tab:gps_readm_params}
\end{table}

\subsubsection{Model based on log-normal distribution for treatment time}

\mbox{}
\vspace{-1em}

\begin{center}
\begin{longtable}{l l c c c }
\caption{Estimated coefficients for 30-day revisit outcome with log-normal distribution for the treatment time.}\label{tab:logn_rev_params}\\
\toprule
\endfirsthead

\multicolumn{5}{c}%
{{\bfseries \tablename\ \thetable{} -- Continued from previous page}} \\
\toprule

    \multicolumn{2}{c}{} & \multicolumn{1}{c}{\textbf{}} & \multicolumn{2}{c}{\textbf{95\% CI}} \\
    \multicolumn{1}{l}{\textbf{Variable}} &
    \multicolumn{1}{l}{\textbf{Model term}} & \multicolumn{1}{c}{\textbf{Estimate}} & \multicolumn{1}{c}{\textbf{Lower}} & \multicolumn{1}{c}{\textbf{Upper}} \\
    \midrule
\endhead

\hline \multicolumn{5}{r}{{Continued on next page}} \\ \hline
\endfoot

\hline
\endlastfoot

    \multicolumn{2}{c}{} & \multicolumn{1}{c}{\textbf{}} & \multicolumn{2}{c}{\textbf{95\% CI}} \\
    \multicolumn{1}{l}{\textbf{Variable}} &
    \multicolumn{1}{l}{\textbf{Model term}} & \multicolumn{1}{c}{\textbf{Estimate}} & \multicolumn{1}{c}{\textbf{Lower}} & \multicolumn{1}{c}{\textbf{Upper}} \\
    \midrule    Health State $H$ & Intercept & -3.022 & -3.186 & -2.858 \\ 
     & Age & -0.855 & -1.01 & -0.7 \\ 
     & Female & -0.266 & -0.393 & -0.139 \\ 
     & Non-White & 0.021 & -0.121 & 0.164 \\ 
     & Workers Comp & -0.22 & -0.358 & -0.081 \\ 
     & Income & -0.01 & -0.073 & 0.054 \\ 
     & High Comorbidity & 0.032 & -0.026 & 0.091 \\ 
     & Hypertension & 0.547 & 0.313 & 0.78 \\ 
     & Obesity & -0.427 & -0.72 & -0.135 \\ 
    Acuity $Z$ & $H=0$ & -2.534 & -2.635 & -2.434 \\ 
     & $H=1$ & -1.753 & -2.013 & -1.494 \\ 
     & Age & 0.51 & 0.458 & 0.563 \\ 
     & Female & -0.218 & -0.303 & -0.134 \\ 
     & Non-White & -0.112 & -0.223 & -0.001 \\ 
     & Workers Comp & -0.053 & -0.144 & 0.037 \\ 
     & Income & -0.065 & -0.109 & -0.021 \\ 
     & High Comorbidity & 0.179 & 0.151 & 0.206 \\ 
     & Hypertension & 0.026 & -0.082 & 0.135 \\ 
     & Obesity & -0.214 & -0.364 & -0.064 \\ 
     & Temperature & -0.097 & -0.165 & -0.028 \\ 
     & Heart rate & 0.272 & 0.225 & 0.319 \\ 
     & Blood pressure & -0.015 & -0.057 & 0.028 \\ 
     & Respiration rate & 0.301 & 0.256 & 0.346 \\ 
    Temperature & $H=0$ & -0.11 & -0.13 & -0.09 \\ 
     & $H=1$ & 1.839 & 1.686 & 1.992 \\ 
     & Age & -0.065 & -0.077 & -0.054 \\ 
     & Female & 0.05 & 0.034 & 0.067 \\ 
     & Non-White & 0.011 & -0.009 & 0.032 \\ 
     & Workers Comp & -0.027 & -0.045 & -0.009 \\ 
     & Income & -0.002 & -0.01 & 0.006 \\ 
     & High Comorbidity & 0.0 & -0.007 & 0.007 \\ 
     & Hypertension & 0.025 & 0.001 & 0.05 \\ 
     & Obesity & -0.014 & -0.043 & 0.016 \\ 
     & $H=0$ & -0.109 & -0.138 & -0.08 \\ 
     & $H=1$ & 1.57 & 1.502 & 1.637 \\ 
     & Age & -0.26 & -0.278 & -0.241 \\ 
     & Female & 0.126 & 0.104 & 0.147 \\ 
     & Non-White & -0.188 & -0.214 & -0.162 \\ 
     & Workers Comp & -0.145 & -0.17 & -0.121 \\ 
     & Income & 0.002 & -0.009 & 0.013 \\ 
     & High Comorbidity & 0.086 & 0.076 & 0.095 \\ 
     & Hypertension & 0.017 & -0.016 & 0.05 \\ 
     & Obesity & 0.019 & -0.02 & 0.059 \\ 
    Heart rate & $H=0$ & 0.069 & 0.044 & 0.095 \\ 
     & $H=1$ & -0.018 & -0.077 & 0.04 \\ 
     & Age & 0.392 & 0.38 & 0.405 \\ 
     & Female & -0.175 & -0.196 & -0.154 \\ 
     & Non-White & 0.067 & 0.041 & 0.092 \\ 
     & Workers Comp & -0.001 & -0.024 & 0.021 \\ 
     & Income & -0.021 & -0.031 & -0.01 \\ 
     & High Comorbidity & -0.062 & -0.071 & -0.053 \\ 
     & Hypertension & 0.376 & 0.345 & 0.407 \\ 
     & Obesity & 0.106 & 0.068 & 0.144 \\ 
     & $H=0$ & 0.001 & -0.03 & 0.031 \\ 
     & $H=1$ & 1.3 & 1.139 & 1.461 \\ 
     & Age & -0.176 & -0.196 & -0.156 \\ 
     & Female & -0.086 & -0.108 & -0.065 \\ 
     & Non-White & -0.025 & -0.051 & 0.0 \\ 
     & Workers Comp & -0.126 & -0.15 & -0.102 \\ 
     & Income & 0.012 & 0.002 & 0.023 \\ 
     & High Comorbidity & 0.039 & 0.03 & 0.048 \\ 
     & Hypertension & 0.096 & 0.064 & 0.128 \\ 
     & Obesity & 0.049 & 0.01 & 0.087 \\ 
    Blood pressure & $H=0$ & 0.069 & 0.061 & 0.077 \\ 
     & $H=1$ & 0.115 & 0.103 & 0.128 \\ 
     & Age & 0.004 & -0.012 & 0.019 \\ 
     & Female & 0.02 & 0.007 & 0.034 \\ 
     & Non-White & -0.001 & -0.007 & 0.005 \\ 
     & Workers Comp & 0.027 & 0.021 & 0.032 \\ 
     & Income & -0.0 & -0.019 & 0.019 \\ 
     & High Comorbidity & -0.003 & -0.026 & 0.019 \\ 
     & Hypertension & 0.007 & -0.003 & 0.017 \\ 
     & Obesity & 0.01 & 0.003 & 0.017 \\ 
     & $H=0$ & 0.017 & 0.01 & 0.024 \\ 
     & $H=1$ & 0.001 & -0.007 & 0.008 \\ 
     & Age & -2.04 & -2.055 & -2.024 \\ 
     & Female & -2.141 & -2.191 & -2.091 \\ 
     & Non-White & 0.444 & 0.408 & 0.479 \\ 
     & Workers Comp & -0.303 & -0.356 & -0.249 \\ 
     & Income & -0.33 & -0.399 & -0.262 \\ 
     & High Comorbidity & 0.068 & 0.01 & 0.126 \\ 
     & Hypertension & 0.056 & 0.03 & 0.083 \\ 
     & Obesity & 0.227 & 0.204 & 0.251 \\ 
    Respiration rate & $H=0$ & -0.002 & -0.08 & 0.075 \\ 
     & $H=1$ & -0.308 & -0.407 & -0.209 \\ 
     & Age & -0.008 & -0.056 & 0.039 \\ 
     & Female & 0.124 & 0.093 & 0.155 \\ 
     & Non-White & 0.001 & -0.028 & 0.029 \\ 
     & Workers Comp & -0.01 & -0.046 & 0.026 \\ 
     & Income & -1.092 & -1.204 & -0.98 \\ 
     & High Comorbidity & 0.452 & 0.242 & 0.661 \\ 
     & Hypertension & -1.833 & -1.931 & -1.735 \\ 
     & Obesity & -1.217 & -1.437 & -0.996 \\ 
     & $H=0$ & -1.654 & -1.774 & -1.534 \\ 
     & $H=1$ & -1.796 & -2.034 & -1.559 \\ 
     & Age & -0.11 & -0.13 & -0.09 \\ 
     & Female & 1.839 & 1.686 & 1.992 \\ 
     & Non-White & -0.065 & -0.077 & -0.054 \\ 
     & Workers Comp & 0.05 & 0.034 & 0.067 \\ 
     & Income & 0.011 & -0.009 & 0.032 \\ 
     & High Comorbidity & -0.027 & -0.045 & -0.009 \\ 
     & Hypertension & -0.002 & -0.01 & 0.006 \\ 
     & Obesity & 0.0 & -0.007 & 0.007 \\ 
    Treatment time $T$ & Age & 0.025 & 0.001 & 0.05 \\ 
     & Female & -0.014 & -0.043 & 0.016 \\ 
     & Non-White & -0.109 & -0.138 & -0.08 \\ 
     & Workers Comp & 1.57 & 1.502 & 1.637 \\ 
     & Income & -0.26 & -0.278 & -0.241 \\ 
     & High Comorbidity & 0.126 & 0.104 & 0.147 \\ 
     & Hypertension & -0.188 & -0.214 & -0.162 \\ 
     & Obesity & -0.145 & -0.17 & -0.121 \\ 
     & Temperature & 0.002 & -0.009 & 0.013 \\ 
     & Heart rate & 0.086 & 0.076 & 0.095 \\ 
     & Blood pressure & 0.017 & -0.016 & 0.05 \\ 
     & Respiration rate & 0.019 & -0.02 & 0.059 \\ 
     & $H=0$ & 0.069 & 0.044 & 0.095 \\ 
     & $H=1$ & -0.018 & -0.077 & 0.04 \\ 
    Admission Decision $A$ & Age & 0.392 & 0.38 & 0.405 \\ 
     & Female & -0.175 & -0.196 & -0.154 \\ 
     & Non-White & 0.067 & 0.041 & 0.092 \\ 
     & Workers Comp & -0.001 & -0.024 & 0.021 \\ 
     & Income & -0.021 & -0.031 & -0.01 \\ 
     & High Comorbidity & -0.062 & -0.071 & -0.053 \\ 
     & Hypertension & 0.376 & 0.345 & 0.407 \\ 
     & Obesity & 0.106 & 0.068 & 0.144 \\ 
     & Temperature & 0.001 & -0.03 & 0.031 \\ 
     & Heart rate & 1.3 & 1.139 & 1.461 \\ 
     & Blood pressure & -0.176 & -0.196 & -0.156 \\ 
     & Respiration rate & -0.086 & -0.108 & -0.065 \\ 
     & $H=0$ & -0.025 & -0.051 & 0.0 \\ 
     & $H=1$ & -0.126 & -0.15 & -0.102 \\ 
    $\mu_1$ & Intercept & 0.012 & 0.002 & 0.023 \\ 
    $\mu_2$ & Intercept & 0.039 & 0.03 & 0.048 \\ 
    $\mu_3$ & Intercept & 0.096 & 0.064 & 0.128 \\ 
    $\mu_4$ & Intercept & 0.049 & 0.01 & 0.087\\
\end{longtable}
\end{center}

\begin{center}
\begin{longtable}{l l c c c }
\caption{Estimated coefficients for 30-day readmission outcome using log-normal distribution for the treatment time.}\label{tab:logn_read_params}\\
\toprule
\endfirsthead

\multicolumn{5}{c}%
{{\bfseries \tablename\ \thetable{} -- Continued from previous page}} \\
\toprule

    \multicolumn{2}{c}{} & \multicolumn{1}{c}{\textbf{}} & \multicolumn{2}{c}{\textbf{95\% CI}} \\
    \multicolumn{1}{l}{\textbf{Variable}} &
    \multicolumn{1}{l}{\textbf{Model term}} & \multicolumn{1}{c}{\textbf{Estimate}} & \multicolumn{1}{c}{\textbf{Lower}} & \multicolumn{1}{c}{\textbf{Upper}} \\
    \midrule
\endhead

\hline \multicolumn{5}{r}{{Continued on next page}} \\ \hline
\endfoot

\hline
\endlastfoot

    \multicolumn{2}{c}{} & \multicolumn{1}{c}{\textbf{}} & \multicolumn{2}{c}{\textbf{95\% CI}} \\
    \multicolumn{1}{l}{\textbf{Variable}} &
    \multicolumn{1}{l}{\textbf{Model term}} & \multicolumn{1}{c}{\textbf{Estimate}} & \multicolumn{1}{c}{\textbf{Lower}} & \multicolumn{1}{c}{\textbf{Upper}} \\
    \midrule    Health State $H$ & Intercept & -3.622 & -3.892 & -3.352 \\ 
     & Age & -1.439 & -2.271 & -0.606 \\ 
     & Female & -0.227 & -0.448 & -0.005 \\ 
     & Non-White & 0.04 & -0.113 & 0.194 \\ 
     & Workers Comp & -0.272 & -0.624 & 0.079 \\ 
     & Income & -0.0 & -0.068 & 0.068 \\ 
     & High Comorbidity & 0.044 & -0.025 & 0.112 \\ 
     & Hypertension & 0.882 & 0.308 & 1.457 \\ 
     & Obesity & -0.437 & -0.792 & -0.083 \\ 
    Acuity $Z$ & $H=0$ & -2.507 & -2.614 & -2.399 \\ 
     & $H=1$ & -2.092 & -2.556 & -1.628 \\ 
     & Age & 0.54 & 0.487 & 0.592 \\ 
     & Female & -0.235 & -0.321 & -0.149 \\ 
     & Non-White & -0.106 & -0.22 & 0.008 \\ 
     & Workers Comp & -0.026 & -0.122 & 0.07 \\ 
     & Income & -0.069 & -0.113 & -0.026 \\ 
     & High Comorbidity & 0.182 & 0.154 & 0.21 \\ 
     & Hypertension & -0.066 & -0.179 & 0.046 \\ 
     & Obesity & -0.17 & -0.324 & -0.016 \\ 
     & Temperature & -0.028 & -0.133 & 0.078 \\ 
     & Heart rate & 0.297 & 0.248 & 0.346 \\ 
     & Blood pressure & -0.024 & -0.067 & 0.018 \\ 
     & Respiration rate & 0.318 & 0.268 & 0.368 \\ 
    Temperature & $H=0$ & -0.101 & -0.128 & -0.075 \\ 
     & $H=1$ & 1.728 & 0.675 & 2.781 \\ 
     & Age & -0.049 & -0.059 & -0.039 \\ 
     & Female & 0.048 & 0.029 & 0.068 \\ 
     & Non-White & 0.017 & -0.004 & 0.037 \\ 
     & Workers Comp & -0.015 & -0.04 & 0.01 \\ 
     & Income & -0.004 & -0.012 & 0.005 \\ 
     & High Comorbidity & 0.002 & -0.005 & 0.009 \\ 
     & Hypertension & 0.017 & -0.008 & 0.042 \\ 
     & Obesity & -0.01 & -0.044 & 0.024 \\ 
     & $H=0$ & -0.107 & -0.165 & -0.048 \\ 
     & $H=1$ & 1.71 & 1.6 & 1.821 \\ 
     & Age & -0.232 & -0.291 & -0.174 \\ 
     & Female & 0.126 & 0.094 & 0.157 \\ 
     & Non-White & -0.191 & -0.218 & -0.165 \\ 
     & Workers Comp & -0.133 & -0.181 & -0.085 \\ 
     & Income & -0.017 & -0.028 & -0.005 \\ 
     & High Comorbidity & 0.084 & 0.075 & 0.094 \\ 
     & Hypertension & 0.008 & -0.036 & 0.052 \\ 
     & Obesity & -0.003 & -0.043 & 0.036 \\ 
    Heart rate & $H=0$ & 0.077 & 0.05 & 0.104 \\ 
     & $H=1$ & 0.087 & -0.046 & 0.22 \\ 
     & Age & 0.395 & 0.381 & 0.409 \\ 
     & Female & -0.178 & -0.199 & -0.157 \\ 
     & Non-White & 0.063 & 0.037 & 0.088 \\ 
     & Workers Comp & -0.012 & -0.035 & 0.011 \\ 
     & Income & -0.019 & -0.029 & -0.008 \\ 
     & High Comorbidity & -0.064 & -0.073 & -0.055 \\ 
     & Hypertension & 0.374 & 0.343 & 0.406 \\ 
     & Obesity & 0.106 & 0.067 & 0.144 \\ 
     & $H=0$ & -0.0 & -0.136 & 0.135 \\ 
     & $H=1$ & 1.748 & 0.671 & 2.825 \\ 
     & Age & -0.14 & -0.265 & -0.016 \\ 
     & Female & -0.091 & -0.137 & -0.044 \\ 
     & Non-White & -0.024 & -0.049 & 0.001 \\ 
     & Workers Comp & -0.122 & -0.194 & -0.049 \\ 
     & Income & 0.009 & -0.003 & 0.021 \\ 
     & High Comorbidity & 0.035 & 0.025 & 0.046 \\ 
     & Hypertension & 0.072 & 0.001 & 0.142 \\ 
     & Obesity & 0.052 & 0.003 & 0.101 \\ 
    Blood pressure & $H=0$ & 0.067 & 0.059 & 0.075 \\ 
     & $H=1$ & 0.115 & 0.103 & 0.128 \\ 
     & Age & 0.005 & -0.01 & 0.021 \\ 
     & Female & 0.021 & 0.008 & 0.035 \\ 
     & Non-White & -0.002 & -0.008 & 0.004 \\ 
     & Workers Comp & 0.026 & 0.021 & 0.031 \\ 
     & Income & 0.015 & -0.003 & 0.034 \\ 
     & High Comorbidity & 0.034 & 0.011 & 0.056 \\ 
     & Hypertension & 0.005 & -0.005 & 0.015 \\ 
     & Obesity & 0.017 & 0.009 & 0.024 \\ 
     & $H=0$ & 0.016 & 0.009 & 0.022 \\ 
     & $H=1$ & -0.02 & -0.031 & -0.009 \\ 
     & Age & -2.047 & -2.063 & -2.031 \\ 
     & Female & -2.086 & -2.158 & -2.013 \\ 
     & Non-White & 0.547 & 0.51 & 0.583 \\ 
     & Workers Comp & -0.358 & -0.412 & -0.303 \\ 
     & Income & -0.29 & -0.36 & -0.22 \\ 
     & High Comorbidity & 0.154 & 0.093 & 0.215 \\ 
     & Hypertension & -0.029 & -0.056 & -0.003 \\ 
     & Obesity & 0.205 & 0.181 & 0.228 \\ 
    Respiration rate & $H=0$ & -0.062 & -0.14 & 0.015 \\ 
     & $H=1$ & -0.334 & -0.433 & -0.236 \\ 
     & Age & 0.045 & -0.115 & 0.205 \\ 
     & Female & 0.241 & 0.197 & 0.286 \\ 
     & Non-White & -0.025 & -0.055 & 0.004 \\ 
     & Workers Comp & 0.045 & 0.011 & 0.079 \\ 
     & Income & -0.776 & -0.891 & -0.661 \\ 
     & High Comorbidity & 0.082 & -0.816 & 0.98 \\ 
     & Hypertension & -3.374 & -3.586 & -3.161 \\ 
     & Obesity & -1.914 & -2.503 & -1.325 \\ 
     & $H=0$ & -2.522 & -2.676 & -2.369 \\ 
     & $H=1$ & -2.556 & -2.86 & -2.251 \\ 
     & Age & -0.101 & -0.128 & -0.075 \\ 
     & Female & 1.728 & 0.675 & 2.781 \\ 
     & Non-White & -0.049 & -0.059 & -0.039 \\ 
     & Workers Comp & 0.048 & 0.029 & 0.068 \\ 
     & Income & 0.017 & -0.004 & 0.037 \\ 
     & High Comorbidity & -0.015 & -0.04 & 0.01 \\ 
     & Hypertension & -0.004 & -0.012 & 0.005 \\ 
     & Obesity & 0.002 & -0.005 & 0.009 \\ 
    Treatment Time $T$ & Age & 0.017 & -0.008 & 0.042 \\ 
     & Female & -0.01 & -0.044 & 0.024 \\ 
     & Non-White & -0.107 & -0.165 & -0.048 \\ 
     & Workers Comp & 1.71 & 1.6 & 1.821 \\ 
     & Income & -0.232 & -0.291 & -0.174 \\ 
     & High Comorbidity & 0.126 & 0.094 & 0.157 \\ 
     & Hypertension & -0.191 & -0.218 & -0.165 \\ 
     & Obesity & -0.133 & -0.181 & -0.085 \\ 
     & Temperature & -0.017 & -0.028 & -0.005 \\ 
     & Heart rate & 0.084 & 0.075 & 0.094 \\ 
     & Blood pressure & 0.008 & -0.036 & 0.052 \\ 
     & Respiration rate & -0.003 & -0.043 & 0.036 \\ 
     & $H=0$ & 0.077 & 0.05 & 0.104 \\ 
     & $H=1$ & 0.087 & -0.046 & 0.22 \\ 
    Admission Decision $A$ & Age & 0.395 & 0.381 & 0.409 \\ 
     & Female & -0.178 & -0.199 & -0.157 \\ 
     & Non-White & 0.063 & 0.037 & 0.088 \\ 
     & Workers Comp & -0.012 & -0.035 & 0.011 \\ 
     & Income & -0.019 & -0.029 & -0.008 \\ 
     & High Comorbidity & -0.064 & -0.073 & -0.055 \\ 
     & Hypertension & 0.374 & 0.343 & 0.406 \\ 
     & Obesity & 0.106 & 0.067 & 0.144 \\ 
     & Temperature & -0.0 & -0.136 & 0.135 \\ 
     & Heart rate & 1.748 & 0.671 & 2.825 \\ 
     & Blood pressure & -0.14 & -0.265 & -0.016 \\ 
     & Respiration rate & -0.091 & -0.137 & -0.044 \\ 
     & $H=0$ & -0.024 & -0.049 & 0.001 \\ 
     & $H=1$ & -0.122 & -0.194 & -0.049 \\ 
    $\mu_1$ & Intercept & 0.009 & -0.003 & 0.021 \\ 
    $\mu_2$ & Intercept & 0.035 & 0.025 & 0.046 \\ 
    $\mu_3$ & Intercept & 0.072 & 0.001 & 0.142 \\ 
    $\mu_4$ & Intercept & 0.052 & 0.003 & 0.101\\
\end{longtable}
\end{center}

\subsubsection{Model based on log-normal distribution without latent variable}

\begin{table}[H]
\centering
\begin{tabular}{l l c c c }
\toprule

    \multicolumn{2}{c}{} & \multicolumn{1}{c}{\textbf{}} & \multicolumn{2}{c}{\textbf{95\% CI}} \\
    \multicolumn{1}{l}{\textbf{Variable}} &
    \multicolumn{1}{l}{\textbf{Model term}} & \multicolumn{1}{c}{\textbf{Estimate}} & \multicolumn{1}{c}{\textbf{Lower}} & \multicolumn{1}{c}{\textbf{Upper}} \\
    \midrule    Treatment Time $T$ & Age & 0.072 & 0.072 & 0.072 \\ 
     & Female & -0.0 & -0.0 & -0.0 \\ 
     & Non-White & 0.015 & 0.015 & 0.015 \\ 
     & Workers Comp & 0.0 & 0.0 & 0.0 \\ 
     & Income & -0.001 & -0.001 & -0.001 \\ 
     & High Comorbidity & 0.023 & 0.023 & 0.023 \\ 
     & Hypertension & 0.012 & 0.012 & 0.012 \\ 
     & Obesity & -0.002 & -0.002 & -0.002 \\ 
     & Temperature & -0.001 & -0.001 & -0.001 \\ 
     & Heart rate & 0.012 & 0.012 & 0.012 \\ 
     & Blood pressure & 0.0 & 0.0 & 0.0 \\ 
     & Respiration rate & -0.02 & -0.02 & -0.02 \\ 
     & Intercept & -1.964 & -1.964 & -1.964 \\ 
    Admission Decision $A$ & Age & 0.409 & 0.409 & 0.409 \\ 
     & Female & -0.0 & -0.0 & -0.0 \\ 
     & Non-White & -0.0 & -0.0 & -0.0 \\ 
     & Workers Comp & -0.002 & -0.002 & -0.002 \\ 
     & Income & 0.0 & 0.0 & 0.0 \\ 
     & High Comorbidity & 0.222 & 0.222 & 0.222 \\ 
     & Hypertension & 0.0 & 0.0 & 0.0 \\ 
     & Obesity & -0.0 & -0.0 & -0.0 \\ 
     & Temperature & 0.223 & 0.223 & 0.223 \\ 
     & Heart rate & 0.0 & 0.0 & 0.0 \\ 
     & Blood pressure & 0.006 & 0.006 & 0.006 \\ 
     & Respiration rate & 0.001 & 0.001 & 0.001 \\ 
     & Intercept & -0.985 & -0.985 & -0.985 \\ 
    $\mu_1$ & Intercept & -1.326 & -1.326 & -1.326 \\ 
    $\mu_2$ & Intercept & -1.256 & -1.256 & -1.256 \\ 
    $\mu_3$ & Intercept & -0.002 & -0.002 & -0.002 \\ 
    $\mu_4$ & Intercept & 0.072 & 0.072 & 0.072\\
\bottomrule
\end{tabular}
\caption{Estimated coefficients for 30-day revisit outcome using the LogNormal model without the latent variable.}
\label{tab:lognwoh_rev_params}
\end{table}

\begin{table}[H]
\centering
\begin{tabular}{l l c c c }
\toprule

    \multicolumn{2}{c}{} & \multicolumn{1}{c}{\textbf{}} & \multicolumn{2}{c}{\textbf{95\% CI}} \\
    \multicolumn{1}{l}{\textbf{Variable}} &
    \multicolumn{1}{l}{\textbf{Model term}} & \multicolumn{1}{c}{\textbf{Estimate}} & \multicolumn{1}{c}{\textbf{Lower}} & \multicolumn{1}{c}{\textbf{Upper}} \\
    \midrule    Treatment Time $T$ & Age & 0.072 & 0.072 & 0.072 \\ 
     & Female & 0.116 & 0.116 & 0.116 \\ 
     & Non-White & -0.002 & -0.002 & -0.002 \\ 
     & Workers Comp & 0.006 & 0.006 & 0.006 \\ 
     & Income & 0.002 & 0.002 & 0.002 \\ 
     & High Comorbidity & 0.003 & 0.003 & 0.003 \\ 
     & Hypertension & 0.01 & 0.01 & 0.01 \\ 
     & Obesity & 0.013 & 0.013 & 0.013 \\ 
     & Temperature & 0.0 & 0.0 & 0.0 \\ 
     & Heart rate & -0.0 & -0.0 & -0.0 \\ 
     & Blood pressure & 0.017 & 0.017 & 0.017 \\ 
     & Respiration rate & -0.0 & -0.0 & -0.0 \\ 
     & Intercept & -2.018 & -2.018 & -2.018 \\ 
    Admission Decision $A$ & Age & -0.0 & -0.0 & -0.0 \\ 
     & Female & -0.024 & -0.024 & -0.024 \\ 
     & Non-White & 0.003 & 0.003 & 0.003 \\ 
     & Workers Comp & 0.001 & 0.001 & 0.001 \\ 
     & Income & -0.0 & -0.0 & -0.0 \\ 
     & High Comorbidity & -0.0 & -0.0 & -0.0 \\ 
     & Hypertension & 0.0 & 0.0 & 0.0 \\ 
     & Obesity & -0.006 & -0.006 & -0.006 \\ 
     & Temperature & -0.001 & -0.001 & -0.001 \\ 
     & Heart rate & -0.003 & -0.003 & -0.003 \\ 
     & Blood pressure & 0.182 & 0.182 & 0.182 \\ 
     & Respiration rate & 0.0 & 0.0 & 0.0 \\ 
     & Intercept & -0.738 & -0.738 & -0.738 \\ 
    $\mu_1$ & Intercept & -2.979 & -2.979 & -2.979 \\ 
    $\mu_2$ & Intercept & -2.361 & -2.361 & -2.361 \\ 
    $\mu_3$ & Intercept & 0.0 & 0.0 & 0.0 \\ 
    $\mu_4$ & Intercept & 0.067 & 0.067 & 0.067\\
\bottomrule
\end{tabular}
\caption{Estimated coefficients for 30-day readmission outcome using the LogNormal model without the latent variable.}
\label{tab:lognwoh_readm_params}
\end{table}


\end{document}